\documentclass[11pt]{article}

\usepackage{jheppubmod}
\usepackage[utf8]{inputenc}
\usepackage{amsmath,amsfonts,amstext,amssymb,mathrsfs,mathtools}
\usepackage{graphicx}
\usepackage{comment}
\usepackage{dutchcal}
\usepackage{feynmp-auto}
\usepackage{color}
\usepackage{cancel}
\usepackage{multicol}
\usepackage{multirow}
\usepackage{graphicx}
\usepackage[sorting=none]{biblatex}
\newcommand{\secref}[1]{Section \ref{#1}}
\newcommand{\appref}[1]{Appendix \ref{#1}}
\newcommand{\figref}[1]{Fig. \ref{#1}}

\newcommand{\ed}{\text{d}}

\newcommand{\lpl}{\ell_{\text{Pl}}}
\newcommand{\up}{x}
\newcommand{\um}{y}

\newcommand{\gap}{\text{ }}
\newcommand{\sgap}{\gap\gap}
\newcommand{\mgap}{\sgap\sgap}
\newcommand{\lgap}{\gap\gap\gap\gap\gap\gap}
\newcommand{\p}{\partial}

\newcommand{\proj}{\text{p}}

\begin{document}

\title{Soft graviton exchange and the information paradox}
\author{Nava Gaddam and}
\author{Nico Groenenboom}
\emailAdd{gaddam@uu.nl}
\emailAdd{n.groenenboom@uu.nl}
\affiliation{Institute for Theoretical Physics and Center for Extreme Matter and Emergent Phenomena, Utrecht University, 3508 TD Utrecht, The Netherlands.}
\date{\today}

\abstract{We show that there is a remarkable phase in quantum gravity where gravitational scattering amplitudes mediated by virtual gravitons can be calculated explicitly in effective field theory, when the impact parameter $b$ satisfies $L_{Pl}\ll b \lesssim R_S$, with $R_S$ being the Schwarzschild radius. This phase captures collisions with energies satisfying $\sqrt{s}\gg \gamma M_{Pl}$ (with $\gamma \sim M_{Pl}/M_{BH}$) near the horizon. We call this the black hole eikonal phase, in contrast to its flat space analogue where collisions are trans-Planckian. Hawking's geometric optics approximation neglects gravitational interactions near the horizon, and results in thermal occupation numbers in the Bogoliubov coefficients. We show that these interactions are mediated by graviton exchange in $2 \rightarrow 2$ scattering near the horizon, and explicitly calculate the S-matrix non-perturbatively in $M_{Pl}/M_{BH}$. This involves a re-summation of infinitely many ladder diagrams near the horizon, all mediated by virtual soft gravitons. The S-matrix turns out to be a pure phase upon this re-summation and is agnostic of Planckian physics and any specific ultraviolet completion. In contrast to the flat space eikonal limit, the black hole eikonal phase captures collisions of extremely low energy near the horizon.}

\maketitle

\section{Introduction and a summary of results}

Hawking argued that semi-classical black hole physics is well approximated by the propagation of free quantum fields in the presence of a fixed classical background, and therefore that they reliably describe the state of radiation \cite{Hawking:1974sw, Hawking:1976ra}. The supposition is therefore that the gravitational effect of quantum radiation is described by an adiabatic change of the background. In stark contrast to this picture, 't Hooft has long argued that strong gravitational interactions between the infalling matter and outgoing Hawking quanta dramatically change the observations of the external observer \cite{tHooft:1990ang, Stephens:1993an, tHooft:1996rdg}; the claim therefore being that the radiation is not sufficiently described by quantum radiation on a fixed background and that quantum coherence is in fact preserved. If this claim is to be true, the natural question to then ask is, how must Hawking's calculation be modified to include these strong gravitational interactions? An educated guess might be that this must be via the inclusion of graviton exchange between the in and out going quanta near the horizon; after all, gravitational interactions dominate all others in this region. In this paper, we explicitly compute an infinite number of such graviton exchange diagrams near the horizon, non-perturbatively re-sum them in a controlled approximation, and show that the result provides for compelling evidence in support of the said claim.

Consider a spherical shell of matter that is energetic enough to collapse into a black hole. For as long as the matter fields are far away from the eventual Schwarzschild radius of the horizon that is to be formed, they may safely be taken to be propagating on flat space. At such large impact parameters, gravitational interactions can largely be ignored if the energy of collisions is small. However, if the energy of collisions becomes Planckian, graviton exchange between the modes starts to dominate \cite{tHooft:1987vrq}. In the so-called eikonal approximation, these processes can be reliably calculated non-perturbatively by summing an infinite number of ladder diagrams \cite{Kabat:1992tb}; this is the regime of negligible momentum transfer. Considerable effort has gone into studying trans-Planckian scattering (when impact parameters $b \gg G_N \sqrt{s}$, where $\sqrt{s}$ measures the centre of mass energy of the collision), both within semi-classical gravity and string theory \cite{Amati:1987wq, Amati:1987uf, Amati:1992zb,Lousto:1996zk, DiVecchia:2020ymx}. However, when the impact parameters reach a certain critical value $b \sim G_N \sqrt{s}$, the eikonal result diverges \cite{Veneziano:2004er, Amati:2007ak}. This suggests the production of an intermediate state. In equal measure, it also prevents a study of scattering at impact parameters smaller than the said critical scale. Notwithstanding this difficulty, various attempts have been made to capture black hole formation and evaporation \cite{Veneziano:2008zb, Veneziano:2008xa, Ciafaloni:2008dg, Dvali:2014ila, Addazi:2016ksu, Ciafaloni:2017ort, Ciafaloni:2018uwe}. Nevertheless, it is fair to say that no universally accepted picture of information retrieval has emerged so far. Any hope to repair this situation seems to be hidden in physics near the only intrinsic ultra-violet scale available, when impact parameters are comparable to Planck length; a regime where large momentum transfers and microscopic black hole formation dominates. A regime where little to nothing is known.

\begin{figure}[!htbp]
    \centering
    \includegraphics[scale=0.45]{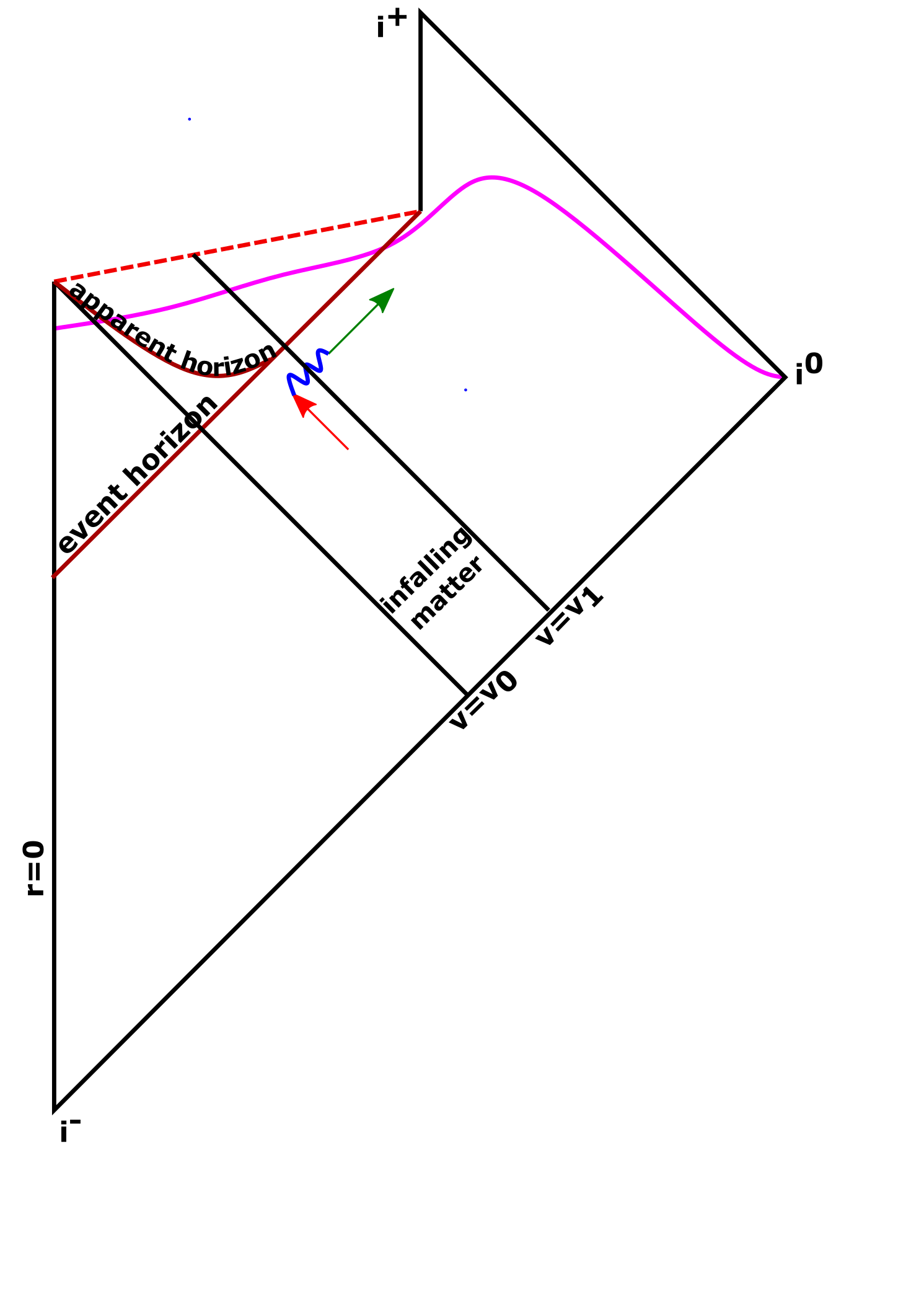}
    \caption{In this figure, a schematic picture of black hole evolution is shown. It is formed by a collapsing shell of null matter of width $v_{1} - v_{0}$. Already at a classical level, an apparent horizon forms long before the entire collapse has occurred. For most of the lifetime of the black hole, however, the apparent horizon is essentially indistinguishable from the event horizon. In fact, already at the final stages of the collapse, the apparent horizon begins to resemble the event horizon very closely. Therefore, except during the very dynamical phase of the collapse, all infalliing matter essentially appears to fall into an already formed horizon. The infalling mode is drawn in red, the outgoing one in green, while the symbolic exchange of gravitons is shown by the blue wavy line. The central conclusion of this paper is that the radiation receives information about the infalling matter from the mediated virtual soft gravitons; this is manifest after a re-summation of infinitely many ladder diagrams. Our calculations are valid so long as $\gamma = \frac{M_{Pl}}{M_{BH}}$ is small. Consequently, the radiation imprints this information on the Cauchy slice drawn in yellow, allowing the external observer to retrieve it.}
    \label{fig:penrosediagram}
\end{figure}

Gravity has the remarkable property that strong quantum effects arise not only in scenarios with large momentum transfer, but also via the emergence of strong gravitational effects at emergent scales (much larger than Planck length). A case in point being the Schwarzschild radius. In fact, in the collapsing spherical shell scenario under consideration, an apparent horizon opens up long before the shell has fallen past the Schwarzschild radius of the eventual horizon to be formed. This is depicted in \figref{fig:penrosediagram}. 

Of course, in the very early stages of the opening up of the horizon, it is Planckian in size and all processes are dominated by large momentum transfers. However, as the horizon grows to be larger than Planck size, impact parameters of collisions on the horizon are of the order of the Schwarzschild radius or less, but larger than Planck size: $L_{Pl} \ll b \leq R_S$. Consequently, momentum transfer effects are suppressed. Nevertheless, in this regime, the physics is necessarily dominated by scattering in the presence of the horizon that has opened up due to the collapse \cite{Banks:1999gd}. The primary difficulty with carrying out this calculation is that unlike in flat space, the graviton propagator on such a background is analytically intractable, rendering the scattering process difficult to study.

When there is a separation of scales of the kind $L_{Pl} \ll b \leq R_S$, it is known that the transverse directions of the horizon can be integrated out. Such an expectation arose from arguments due to the Verlindes \cite{Verlinde:1991iu, Verlinde:1993mi}. It was anticipated in their work that single graviton exchange in this approximation would already modify the state of Hawking radiation. Carrying this integration out explicitly allows for a partial wave basis, in which the scattering problem becomes analytically tractable near the horizon.

In this article, we calculate a four point correlator of matter fields in the said partial wave basis, near the black hole horizon. Owing to a sub-dominant transverse momentum transfer, we may choose the external momentum to be ingoing to the black hole for the infalling modes and outgoing from the horizon for the Hawking quanta. These modes exchange soft gravitons on the horizon. The interactions are governed by the universal three-point vertex coupling the graviton with the stress tensor of the matter fields. Upon integrating the transverse sphere out, the strength of the interactions is dictated by the emergent dimensionless parameter: $\gamma = \kappa/R_S \sim M_{Pl}/M_{BH}$. The main result of this article is that, for every partial wave $\ell$, the four point function is given by
\begin{align}\label{eqn:fourpoint}
    \langle\phi\phi\phi\phi\rangle ~ &= ~ 4 p_{\text{in}} p_{\text{out}} \left[\exp\left(i \dfrac{\gamma^2 R^2_S}{\hbar \left(\ell^2 + \ell + 2\right)} p_{\text{in}} p_{\text{out}}\right) - 1\right] \nonumber \\
    &= ~ 4 p_{\text{in}} p_{\text{out}} \left[\exp\left(i \dfrac{8 \pi G_N}{\hbar\left(\ell^2 + \ell + 2\right)} p_{\text{in}} p_{\text{out}}\right) - 1\right] \, ,
\end{align}
where $p_{in}$ and $p_{out}$ are the in and out going momenta of the infalling matter and Hawking quanta respectively. This is a result of a resummation of an infinite number of ladder diagrams on the horizon. Therefore, it is non-perturbative\footnote{It is non-perturbative in the sense that, for small $\gamma$, effects that are exponentially suppressed in $\gamma$ are captured. But it may also be called `perturbatively exact' in the sense that Planckian (large momentum transfer) effects that arise from the $\gamma \sim 1$ regime are not captured.} in $\gamma$ and $\hbar$. Exponential behaviour in such four point functions has received renewed interest owing to their close connection to quantum chaos \cite{Shenker:2014cwa, Maldacena:2015waa}. Scattering near the horizon has also been argued to be related to chaos \cite{Shenker:2013pqa, Polchinski:2015cea}. This may be seen to arise from \eqref{eqn:fourpoint}, when one moves to exponentiated coordinates near the horizon. In fact, the two-dimensional Dray-'t Hooft scattering amplitude was shown to exactly agree with the semi-classical limit of the four point function of the Schwarzian quantum mechanics that describes the collective infra-red mode of the SYK model \cite{Jensen:2016pah, Maldacena:2016upp, Engelsoy:2016xyb, Kitaev:2017awl, Lam:2018pvp}\footnote{In contrast to the expression (1.1) in reference \cite{Lam:2018pvp}, we find the extra factor arising from the partial wave $\ell$. This factor is of importance, as we explain below \eqref{eqn:shapiro} and discuss further in \secref{sec:discrepancy}.}.

However, as far as four dimensional black holes are concerned, the importance of the emergent scale and the new dimensionless coupling $\gamma$ cannot be overstated. Based on intuition from the eikonal approximation in flat space, one may have feared that near-horizon scattering may result in firewalls \cite{Almheiri:2012rt, Almheiri:2013hfa, Marolf:2013dba, Polchinski:2015cea}; it is indeed true that soft graviton exchange in flat space requires collision energies that are trans-Planckian. So, an infalling observer may worry about encountering highly energetic outgoing modes. However, as we will show in the present article, this amplitude \eqref{eqn:fourpoint} actually arises at leading order in the approximation
\begin{equation}\label{eqn:approx}
    \sqrt{s} ~ \gg ~ \gamma \, M_{Pl} \,  .
\end{equation}
We call this the \textit{black hole eikonal phase} of quantum gravity. This is in contrast to the eikonal approximation in flat space for which $s \gg M^2_{Pl}$. For an earth mass black hole (with $R_s \sim 1 cm \gg L_{Pl}$), where our calculation is valid, we see that \eqref{eqn:approx} implies that $s \gg 10^{-64} M^2_{Pl}$. The calculation captures collisions of extremely low energy because the physics is captured by zero momentum transfer effects for the $2\rightarrow 2$ amplitudes under consideration.\footnote{We expect that momentum transfer effects will play a significant role in general $2\rightarrow N$ amplitudes. Moreover, the time scale associated to those scattering processes is also expected to be longer. Therefore, all external legs are better interpreted as asymptotic states at future and past null infinity; therefore conclusions about firewalls would require further care in general.} Nevertheless, given that we perform an asymptotic scattering amplitude calculation, the experience of an infalling observer is not directly addressed. The effective field theory we set up in this paper, may be useful to this end using the in-in formalism. Moreover, in contrast to scattering in  flat space in a partial wave basis (which for a fixed impact parameter and collision energy is dominated by large $\ell$ modes \cite{Verlinde:1991iu}), scattering near the horizon is evidently dominated by the low $\ell$ modes, as can be seen from \eqref{eqn:fourpoint}. The importance of metric perturbations for a resolution of the information paradox has previously been emphasised \cite{Brustein:2013qma,  Alberte:2015cxa, Goldberger:2019sya, Goldberger:2020geb}.

A rather important consequence of \eqref{eqn:fourpoint} is the fact that upon a Fourier transform of the right hand side of that equation, we see that the four point function is non-zero only when
\begin{equation}\label{eqn:shapiro}
    y_{out} ~ = ~ \dfrac{8 \pi G_N}{\ell^2 + \ell + 2} p_{in} \, .
\end{equation}
This relation is very close to the Shapiro delay derived from the backreaction calculation of \cite{Dray:1984ha} in a first quantised formalism \cite{Hooft:2015jea, Hooft:2016itl, Betzios:2016yaq}. The curious discrepancy is in the extra factor in $\ell^2 + \ell + 2$ instead of the $\ell^2 + \ell + 1$ that was found in those references. As we will show in \secref{sec:discrepancy}, both results are correct in their own right and the present one must be seen as a second quantised generalisation that includes arbitrary off-shell fluctuations in the path integral. It is worth noting that our results are consistent with the expectation that quantum chaos is non-perturbatively realised; our calculation shows which parameters are required to be non-perturbatively treated (namely $\gamma$ and $\hbar$), and which perturbatively ($\sqrt{s}/\gamma M_{Pl}$), for this realisation. Our effective field theory setup allows us to see that while the eikonal expectation \cite{Shenker:2014cwa} that ladder diagrams yield the shock wave geometry is indeed true, the `classicalisation' also allows for a different near-horizon approximation where an alternative black hole eikonal amplitude emerges, as we show in \secref{sec:discrepancy}. Moreover, our effective field theory is a natural framework that can capture exponentially suppressed amplitudes, and exponentially many of them, in the form of general $2\rightarrow N$ amplitudes in addition to the $2\rightarrow 2$ amplitudes we study in this paper.

That scattering amplitudes capture non-linearities in Einstein's equations is increasingly being appreciated \cite{Dafermos:2016uzj, Damour:2016gwp, Damour:2017zjx, Bjerrum-Bohr:2018xdl, Cheung:2018wkq, Kosower:2018adc, Bern:2019nnu, Bern:2019crd, Masaood:2020uhi}. In fact, the classical tree-level three point vertex is sufficient \cite{Amati:2007ak, Cristofoli:2020hnk} to completely determine the Aichelberg-Sexl solution \cite{Bonnor:1969, Aichelburg:1970dh, Penrose:1973}. In the case of black holes, under consideration in the present article, multiple near-horizon approximations emerge, one of which is related to the shockwave, that contribute to the non-perturbative amplitudes\footnote{See \cite{Mogull:2020sak} for a recent perspective on classical scattering of a pair of black holes.}. It is interesting that the different approximations result in differences of numerical prefactors\footnote{The change in prefactor is compared to what appears in the non-linear Dray-'t Hooft solution; see eq. (2.9) of \cite{Hooft:2016itl} for instance. We discuss this further in \secref{sec:discrepancy}.} in \eqref{eqn:shapiro}. This may implore one to ask whether quantum chaotic behaviour can also be seen to emerge in the first quantised formalism of \cite{Hooft:2015jea, Hooft:2016itl, Betzios:2016yaq}. As it turns out, appropriate boundary conditions that respect the scattering algebra near the central causal diamond do indeed generate a rich and chaotic spectrum; those of the zeroes of the Riemann zeta and Dirichlet beta functions \cite{Betzios:2020wcv}. This boundary condition can be motivated to arise from the expectation that CPT is a gauge symmetry in a putative UV complete theory of quantum gravity, as was also argued in \cite{Betzios:2020wcv}. It is to be seen as a generalisation of the antipodal identification that has gained some traction in the context of black holes \cite{Sanchez:1986qn, Hooft:2016itl, Betzios:2016yaq, Sanchez:2018lnb, Bzowski:2018aiq, Sanchez:2018lad, Strauss:2020rpb} and wormholes \cite{Betzios:2017krj, Bzowski:2020umc}.

\paragraph{Organisation of the paper} We begin with a quick review of the eikonal regime in flat space in \secref{sec:minkEikonal}. In \secref{sec:SoftQG}, we set up the theory of quantum gravity coupled to a massless scalar field in a certain soft limit, in the background field method. We first begin with arbitrary backgrounds, and then specialise to spherically symmetric ones, integrate the sphere out, and formulate an effective two-dimensional theory. Thereafter, in \secref{sec:horizon}, we move to physics near the horizon and derive the propagators and Feynman rules governing the interactions. In \secref{sec:scattering}, we compute the advertised four point function. We end with a discussion on the implications of our results for various existing proposals for a resolution of the information paradox and an outlook, in \secref{sec:discussion}. The sections and subsections have been so titled to allow for an understanding of the flow of the paper, merely by reading the table of contents.

\section{A review of eikonal physics in flat space}
\label{sec:minkEikonal}
Certain features of Planckian scattering in flat space are dominated by virtual gravitons. The significance of soft gravitons in quantum gravity was perhaps first noted in this classic example. In the eikonal limit, elastic forward scattering of massive scalar particles can be calculated exactly \cite{tHooft:1987vrq, Kabat:1992tb}. It is instructive to review this example. We begin with the flat space metric in Cartesian coordinates in four dimensions
\begin{eqnarray}
\ed s^2 ~ = ~ -\ed t^2+\ed x^2+\ed y^2+\ed z^2 \, .
\end{eqnarray}
Of the four external particles, the two ingoing ones are taken to carry momenta $p_1$ and $p_2$ while the outgoing momenta are labelled by $p_3$ and $p_4$. The Mandelstam variables of interest are 
\begin{equation}
    s ~ \coloneqq ~ - \left(p_1 + p_2\right)^2 \qquad \text{and} \qquad t ~ \coloneqq ~ - \left(p_1 - p_3\right)^2 \, ,
\end{equation}
and we focus on the limit $s \gg t$. The eikonal limit consists of discarding effects that are sub-dominant in $s/m^2$, where $m$ is the mass of the scalar field. Moreover, to avoid large transverse momentum transfer, the impact parameter is taken to be large; in flat space, the only availables scales to compare the impact parameter with are the Planck length, i.e. $b\gg \lpl$, and the scale associated to the centre of mass energy of the collisions, i.e. $b \gg G_{N} \sqrt{s}$. Therefore, the two scattering particles maintain most of their  momentum in the scattering direction which we call longitudinal, i.e. $p_1^{\parallel}\approx p_2^{\parallel}$. The two particles do however exchange momentum in the remaining two directions; the transverse directions, such that $p_1^{\perp}\neq p_2^{\perp}$. Nevertheless, for all particles, we take $p_i^{\parallel}\gg p_i^{\perp}$. This will later allow us to consider diagrams where there is no exchange in four momentum, all virtual gravitons will be `soft'.

\begin{figure}[h!]
    \centering
	\includegraphics[scale=0.6]{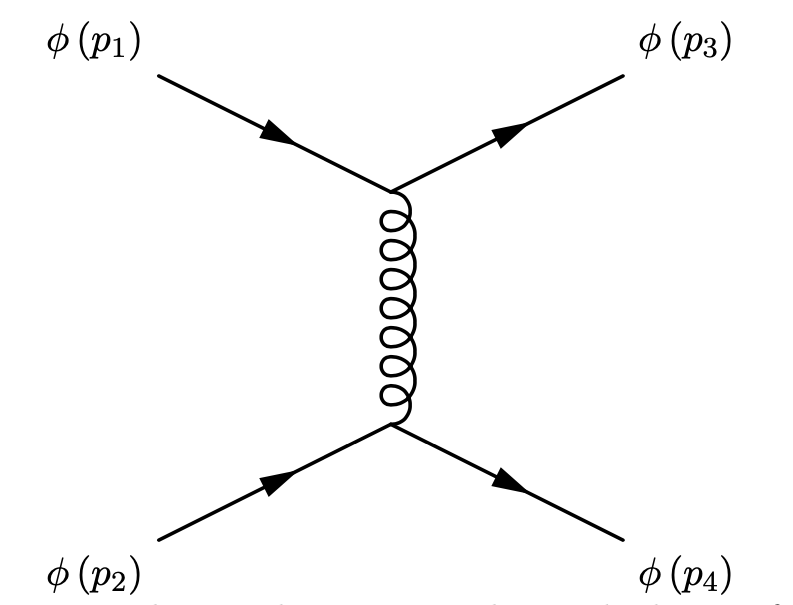}
    \caption{Tree level diagram with virtual graviton exchange, built out of two three-point vertices.}
    \label{fig:minkTree}
\end{figure}

The Feynman rules of interest include the familiar flat space scalar and graviton propagators that are well known and relatively straight forward to calculate\\
\begin{fmffile}{feyn_minkpropagators}
\begin{align}
  \begin{fmfgraph*}(130,0)
    \fmfleft{i1}
    \fmfright{i2}
    \fmf{phantom}{i1,i2}
    \fmfi{fermion,label=$\phi(p)$,label.side=left}{vpath (__i1,__i2)}
  \end{fmfgraph*} ~~~ &= ~ \dfrac{-i}{p^2 + m^2 - i \epsilon} \nonumber \\
  \begin{fmfgraph*}(100,0)
    \fmfleft{i5}
    \fmfright{i6}
    \fmf{phantom}{i5,i6}
    \fmflabel{$h_{\mu\nu}$}{i5}
    \fmflabel{$h_{\rho\sigma}$}{i6}
    \fmfi{curly,label=$k$,label.side=left}{vpath (__i5,__i6)}
  \end{fmfgraph*} ~ \qquad &= ~ \frac{-2 i \kappa^2}{k^2 - i \epsilon} \left(\eta^{\mu\rho}\eta^{\nu\sigma} + \eta^{\mu\sigma}\eta^{\nu\rho} - \eta^{\mu\nu}\eta^{\rho\sigma}\right) \, , \nonumber
  \end{align}
\end{fmffile}

\noindent and the interactions are governed by a three point vertex that arises from the coupling of the stress tensor to the metric fluctuations. It is of the form:\\
\begin{fmffile}{feyn_minkvertex}
\begin{align}
  \begin{fmfgraph*}(100,100)
    \fmfleft{i8,i7}
    \fmfright{o2}
    \fmf{gluon}{i7,v1}
    \fmf{fermion}{i8,v1}
    \fmf{fermion}{v1,o2}
    \fmflabel{$h_{\mu\nu}$}{i7}
    \fmflabel{$p_2$}{i8}
    \fmflabel{$p_1 ~ .$}{o2}
    \fmfdot{v1}
    \fmfv{label=$i ~ p^1_\mu ~ p^2_\nu$,label.angle=60,label.dist=20}{v1}
  \end{fmfgraph*} \nonumber \, .
\end{align}
\end{fmffile}
\noindent The scalar propagator is the familiar Klein-Gordon propagator and the graviton propagator is written in the harmonic gauge with $\kappa^2 = 8 \pi G_N$. The vertex in principle contains another term
\begin{equation}
    p_{\mu} p_{\nu} - \dfrac{1}{2} \eta_{\mu\nu} \left(p^2 + m^2\right) \, ,
\end{equation}
but the second recoil term of the matter field can be neglected for eikonal scattering. Similarly, for large external momentum $p$, internal scalar propagators can be approximated as:
\begin{equation}
    \dfrac{1}{\left(p + k\right)^2 + m^2 - i \epsilon} ~ \approx ~ \dfrac{1}{2 p \cdot k - i \epsilon} \, .
\end{equation}
The tree level contribution to the four point function $\langle\phi\left(p_1\right)\phi\left(p_2\right)\phi\left(p_3\right)\phi\left(p_4\right)\rangle$ arises from the Feynman diagram shown in \figref{fig:minkTree}. \\

\begin{figure}[h!]
\centering
	\includegraphics[scale=0.5]{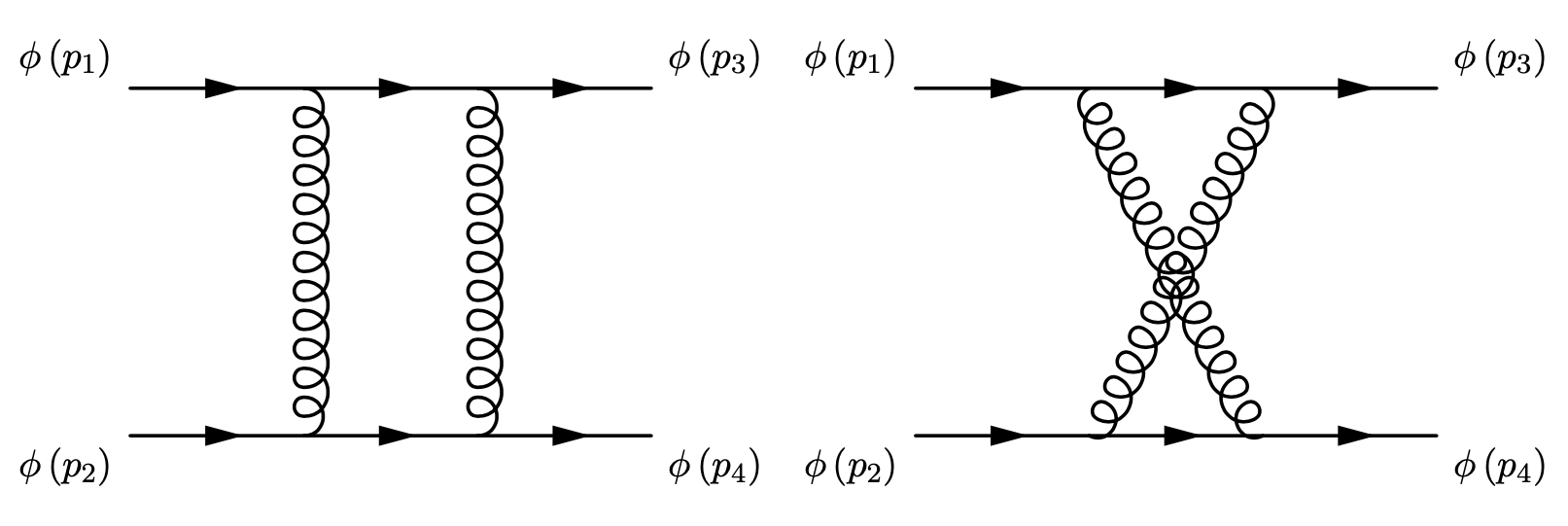}
    \caption{The one-loop ladder diagrams contributing to the eikonal scattering amplitude. The two graviton lines crossing each other in the diagram on the right do not interact, as graviton self-interactions are sub-leading in the eikonal limit. The ladder diagrams give all leading contributions in $s/m^2$.}
    \label{fig:minkOneLoop}
\end{figure}

This tree level diagram \figref{fig:minkTree} is easily shown to evaluate to 
\begin{equation}\label{eq:minkowskitree}
    i \mathcal{M} ~ = ~ \dfrac{2 i \kappa^2 j(s)}{-t} \qquad \text{with} \qquad j(s) ~ \coloneqq ~ \dfrac{1}{2} \left(\left(s - 2 m^2\right)^2 - 2 m^4 \right) \, .
\end{equation}
In the eikonal limit, the Feynman diagrams that dominate are the so-called ladder diagrams displayed in Figure \ref{fig:minkOneLoop}. All other loop diagrams (self-energy, vertex corrections) are insignificant in the eikonal regime, that is to say that their contributions are sub-leading in $s/m^2$. \\

At loop level, it is important to average over the various ways to attach the internal graviton legs, and the corresponding conserved momentum. For the one-loop case, as can be seen in \figref{fig:minkOneLoop}, there are two possibilities arising from fixing internal graviton momentum $k$ on either external leg in a direction of choice. The total amplitude at one-loop results from summing over both of these choices, and inserting a factor of a half. It is then straightforward to proceed with the calculation for the amplitude, as outlined in \cite{Kabat:1992tb}. The final expression at one-loop becomes
\begin{align}
i\mathcal{M} ~ &= ~ \kappa^2 j\left(s\right) \int \ed^4 x \gap e^{- i \left(p_1 - p_3\right) \cdot x} \Delta\left(x\right) \chi\left(x\right) \, , \\
\label{eq:defchi}
\chi ~ &\coloneqq ~ - 2 \kappa^2 j\left(s\right) \int \dfrac{\ed^4 k}{\left(2\pi\right)^4} \gap e^{i k \cdot x} \dfrac{1}{k^2 - i \epsilon} \times \left[\dfrac{1}{-2p_1\cdot k - i \epsilon} \dfrac{1}{2 p_2 \cdot k - i \epsilon} \right. \nonumber \\
&\qquad\qquad\qquad\qquad\qquad + ~ \dfrac{1}{- 2 p_1 \cdot k - i \epsilon} \dfrac{1}{- 2 p_4 \cdot k - i \epsilon} + \dfrac{1}{2 p_3 \cdot k - i \epsilon} \dfrac{1}{2 p_2 \cdot k - i \epsilon} \nonumber \\
&\qquad\qquad\qquad\qquad\qquad \left. + \dfrac{1}{2 p_3 \cdot k - i \epsilon} \dfrac{1}{- 2 p_4 \cdot k - i \epsilon}\right] \, .
\end{align}
Here, we have defined a momentum space massless Klein-Gordon propagator
\begin{eqnarray}
\dfrac{-1}{k^2 - i \epsilon} ~ = ~ \int\ed^4 x \gap e^{-i k \cdot x} \Delta\left(x\right) \, .
\end{eqnarray}
Furthermore, in arriving at the above expression for the amplitude, we approximated $(p_i + k)_{\mu}\approx p_{i\mu}$ in the vertex since $p_1$ is very large. While $k$ is integrated over, the eikonal approximation is such that the leading order contributions to the integrand are for $k\ll p_i$ allowing us to safely take only $p_i$. This approximation notwithstanding, the expression for $\chi$ does not contain any UV-divergences in the eikonal approximation. The additional $k$ from the vertex would then be countered by a $k^2$ in the matter propagator. Remarkably, the UV-divergences are all embedded in different diagrams, whose effects are subdominant in eikonal scattering. This is crucial as the theory is not renormalisable. While it may  seem counter-intuitive at first glance, the divergent diagrams when dimensionally regularized are indeed lower order in the eikonal approximation. Gravity does not allow us to do any better at this stage. 

The next step is to sum over all higher loop diagrams. Since any diagram at order $n$ would contain $n$ extra graviton legs (in comparison to the tree level diagram), there is an $n!$ that arises from symmetry. This also holds for diagrams where the graviton legs cross each other. Therefore, we find a $1/n!$ factor for each order in perturbation theory. So, non-perturbatively an exponential series ensues; this is a typical feature of eikonal scattering. The total scattering amplitude is thus written as
\begin{eqnarray}
\label{eq:amplitudeMinkowskicalc}
i\mathcal{M}=-2i\kappa^2 j(s)\int\ed^4 x\gap e^{-i(p_1-p_3)\cdot x}\Delta(x)\frac{e^{i\chi}-1}{\chi}.
\end{eqnarray}
A calculation of the non-perturbative amplitude from such an exponentiation of the one-loop result is possible owing to the fact that vertex corrections and self energy diagrams are sub-dominant, as argued in \cite{Levy:1969cr}. What remains now is the evaluation of $\chi$ by solving the integral. To this end, we now approximate $p_1\approx p_3,p_2\approx p_4$. This may interpreted as a soft limit as it restricts the virtual graviton momenta. Inserting $p_1=p_3,p_2=p_4$ into \eqref{eq:defchi} results in:
\begin{align}
\label{eq:chicalc1}
\chi ~ &= ~ - 2 \kappa^2 j\left(s\right) \int \dfrac{\ed^4 k}{(2\pi)^4} \gap e^{i k \cdot x} \dfrac{1}{k^2 - i \epsilon} \left[\dfrac{1}{2 p_1 \cdot k + i \epsilon} - \dfrac{1}{2 p_1 \cdot k - i \epsilon} \right] \nonumber\\
&\qquad\qquad\qquad\qquad\qquad\qquad\qquad \times \left[\dfrac{1}{2 p_2 \cdot k + i \epsilon} - \dfrac{1}{2 p_2 \cdot k - i \epsilon}\right] \, .
\end{align}
Since $\epsilon$ is an infinitesimal regulator,  we can now use the following delta identity:
\begin{eqnarray}
\label{eq:deltaiden}
\dfrac{1}{x + i\epsilon} - \dfrac{1}{x - i \epsilon} ~ = ~ - 2 \pi i \delta(x) \, .
\end{eqnarray}
Using this, two of the integrals can be removed from \eqref{eq:chicalc1}, allowing us to rewrite it as
\begin{align}
\label{eq:chicalc2}
\chi ~&= ~ - 2 \kappa^2 j(s) \int \dfrac{\ed^4 k}{\left(2\pi\right)^4} \gap e^{i k \cdot x} \dfrac{1}{k^2 - i \epsilon} \left(- 2 \pi i\right)^2 \delta\left(2 p_1 \cdot k\right) \delta \left(2 p_2 \cdot k\right) \\
\label{eq:chicalc3}
&= ~ \dfrac{\kappa^2 j(s)}{4 E p} \int \dfrac{\ed^2 k_{\perp}}{\left(2\pi\right)^2} \gap e^{i k_{\perp} \cdot x_{\perp}} \dfrac{1}{k_{\perp}^2 + \tilde{\mu}^2 - i \epsilon} \, ,
\end{align}
where we switched to the center of mass frame defined by $p_1=(E,0,0,p)$ and $p_2=(E,0,0,-p)$. The new parameter $\tilde{\mu}$ is an infrared regulator corresponding to the graviton mass and $x_{\perp}$ are the two remaining coordinates in the transverse directions. The solution to the integral in \eqref{eq:chicalc3} can be shown to be
\begin{eqnarray}
\chi ~ = ~ \dfrac{G_N j(s)}{E p} K_0 \left(\tilde{\mu} x_{\perp}\right) ~ \approx ~ - \dfrac{G_N j(s)}{E p} \log\left(\tilde{\mu} x_{\perp}\right) \, ,
\end{eqnarray}
where the last approximation holds for $\tilde{\mu} x_{\perp}\ll 1$ since $\tilde{\mu}$ is a regulator. In this expression, a numerical constant has been absorbed into $\tilde{\mu}$. In contrast to this result, we will see in this paper that the black hole provides for a natural infrared regulator that is not put in by hand.

We now return to solving \eqref{eq:amplitudeMinkowskicalc}; now that $\chi$ is known, only the integral remains. First we pay attention to $\Delta(x)$. In the eikonal approximation, $p_1-p_3$ only contains transverse components, i.e. only the transverse components of $q\equiv p_1-p_3$ are dominant. This allows us to write $e^{-i q\cdot x}\approx e^{-iq_{\perp}\cdot x_{\perp}}$. We can then isolate
\begin{eqnarray}
\int \ed t \ed z \gap \Delta(x) ~ = ~ \int \dfrac{\ed^2 q_{\perp}}{\left(2\pi\right)^2} e^{i q_{\perp} \cdot x_{\perp}} \dfrac{-1}{q_{\perp}^2 - i \epsilon} ~ = ~ \dfrac{-E p}{2\pi G_N j(s)} \, \chi \, .
\end{eqnarray}
Since $\chi$ only depends on $x_{\perp}$, we can insert this expression into \eqref{eq:amplitudeMinkowskicalc}. The integral to be solved simply becomes
\begin{eqnarray}
\label{eq:amplitudeMinkowskicalc2}
i \mathcal{M} ~ = ~ 8 E p\int\ed^2 x_{\perp} \gap e^{- i q_{\perp} \cdot x_{\perp}} \bigr(e^{i\chi} - 1\bigr) \, .
\end{eqnarray}
Solving this integral yields the final result for the scalar four point function in the eikonal regime
\begin{eqnarray}
\label{eq:amplitudeMinkowski}
i\mathcal{M} ~ = ~ \dfrac{2 i \kappa^2 j(s)}{-t} \dfrac{\Gamma\left(1 - i\alpha(s)\right)}{\Gamma\left(1 + i \alpha(s)\right)} \left(\dfrac{4\tilde{\mu}^2}{-t}\right)^{-i\alpha(s)} \, ,
\end{eqnarray}
where we defined
\begin{eqnarray}
\alpha(s) ~ = ~ G_N \dfrac{\left(s-2m^2\right)^2-2m^4}{\sqrt{s \left(s - 4 m^2\right)}} \, .
\end{eqnarray}
This amplitude is non-perturbative in the coupling constant $\kappa$, but is valid only to leading order in $s/m^2$. Therefore, the approximation gets better with ultra-high energy scattering. For large $s$, we notice that $\alpha(s)$ is real and the complete amplitude reduces to the tree-level amplitude in \eqref{eq:minkowskitree} up to a phase. In the limit $m\to 0$ and for $\tilde{\mu}=1$ the scattering amplitude matches the semi-classical scattering matrix derived by 't Hooft in \cite{tHooft:1987vrq} based on a first quantised description of shockwaves on an Aichelburg-Sexl metric \cite{Aichelburg:1970dh}. This corroborates the naive notion that while certain aspects of ultra-high energy scattering can be well understand, quantum gravity is generically very difficult and poorly understood because regimes of large momentum transfer (which are ignored in the eikonal approximation) are plagued with problems like UV divergences, potential non-locality at the Planck scale, etc. In what we have calculated so far, it is not clear how one may even proceed to account for these in principle.

Nevertheless, this notion is indeed naive in that quantum gravity is not only sensitive to large momentum transfers and Planckian physics, but also to strong gravitational backgrounds. Indeed, remarkable puzzles arise in strong gravitational backgrounds that are classical solutions to General Relativity. The most celebrated of them is of course the paradox of black hole information loss. 

In this paper, we will address the problem of quantum gravity in a tractable limit (that in which the virtual gravitons do not transfer momenta) in the presence of strong gravitational backgrounds where new large scales emerge dynamically. An example is the Schwarzschild solution where a new physical scale of the Schwarzschild radius emerges well before a collapsing shell has fallen past that radius. The graviton propagator is difficult to determine in generic backgrounds, and as we will see, it will take us some effort to derive the necessary Feynman rules before we are able to carry out a calculation, of the kind reviewed in this section, on the horizon of a Schwarzschild black hole.

\section{Quantum gravity in the soft limit}
\label{sec:SoftQG}
We begin by setting the scene for a putative quantum gravity path integral
\begin{equation}\label{eqn:pathIntegral}
    \mathcal{Z} ~ = ~ \int \mathcal{D} \bar{g}_{\mu\nu} \mathcal{D}\phi \exp\left(S\left[\bar{g}_{\mu\nu},\phi\right]\right) \, ,
\end{equation}
where we take the matter action to be that of a minimally coupled scalar field for simplicity:
\begin{align}
    S\left[\bar{g}_{\mu\nu,\phi}\right] ~ &= ~ S_{EH}\left[\bar{g}_{\mu\nu}\right] + S_M\left[\bar{g}_{\mu\nu},\phi\right] \nonumber \\
    &= ~ \dfrac{1}{2 \kappa^2} \int \ed^4 x\sqrt{-\bar{g}} \, \bar{R} - \dfrac{1}{2} \int \ed^4 x \sqrt{-\bar{g}} \, \bar{\nabla}_\mu \phi \bar{\nabla}^\mu \phi \, ,
\end{align}
with $\kappa^2 = 8\pi G_N$. The covariant derivatives with the bar are associated to $\bar{g}_{\mu\nu}$ and in what follows, those without will be associated to $g_{\mu\nu}$. We will work in the background field method with a vanishing matter background, 
\begin{equation}
    \bar{g}_{\mu\nu} ~ = ~ g_{\mu\nu} + \kappa h_{\mu\nu} \, ,
\end{equation}
and assume that $g_{\mu\nu}$ solves the vacuum Einstein's equations. This implies that \eqref{eqn:pathIntegral} becomes
\begin{equation}\label{eqn:pathIntegral_backgroundField}
    \mathcal{Z} ~ = ~ \int \mathcal{D} h_{\mu\nu} \mathcal{D}\phi \exp\left(S_{EH}\left[h_{\mu\nu}\right] + S_M\left[h_{\mu\nu},\phi\right]\right) \, .
\end{equation}
In principle, both terms in the exponent come with infinitely many powers of $h_{\mu\nu}$ owing to the fluctuations of the inverse metric. The matter action begins with:
\begin{align}\label{eqn:matterAction}
    S_M\left[h_{\mu\nu},\phi\right] ~ &= ~ \dfrac{1}{2} \int \ed^4 \sqrt{-g} \phi \Box \phi - \kappa \int\ed^4 x\dfrac{\delta S_M\left[\bar{g}_{\rho\sigma},\phi\right]}{\delta \bar{g}^{\mu\nu}(x)}\biggr\rvert_{\bar{g}=g}h^{\mu\nu}(x) \nonumber\\
    &= ~ \dfrac{1}{2} \int \ed^4 \sqrt{-g} \phi \Box \phi + \dfrac{1}{2} \kappa \int\ed^4 x\sqrt{-g} h^{\mu\nu}T_{\mu\nu}
\end{align}
where we defined the stress tensor 
\begin{equation}
    T_{\mu\nu} ~ = ~ \dfrac{-2}{\sqrt{-g}} \dfrac{\delta S_M}{\delta g^{\mu\nu}} \, .
\end{equation}
For most of this paper, we will restrict our attention to the quadratic kinetic terms and the cubic interaction in \eqref{eqn:matterAction}. The pure gravity action is (derived in \appref{app:quadraticFluctuations}): 
\begin{equation}\label{eqn:pureGravityAction}
    S_{EH}\left[h_{\mu\nu}\right] ~ = ~ \dfrac{1}{4} \int \ed^4 x \sqrt{-g} \left(h^{\mu\nu} - \dfrac{1}{2} g^{\mu\nu} h\right) \bigr(2 \nabla^\sigma\nabla_{(\mu}h_{\nu)\sigma} - \Box h_{\mu\nu} - \nabla_{\mu}\nabla_{\nu} h \bigr) \, ,
\end{equation}
where we defined the trace of the graviton $h = g^{\mu\nu} h_{\mu\nu}$. 

Higher order fluctuations give rise to quartic (and higher order) interaction vertices. In the Einstein-Hilbert part of the action, some of these contain terms with derivatives on $h_{\mu\nu}$. In momentum space, these contribute to graviton momenta in the Feynman diagrams. Integration over these momenta results in the familiar non-renormalisable features of gravity \cite{tHooft:1974toh, Goroff:1985sz}. These arise from large momenta in the ultraviolet. As we will show in this paper, extremely interesting physics arises in the soft limit where internal graviton momenta are taken to vanish. Although it is premature at this stage, we will argue in \secref{sec:horizon} and \secref{sec:scattering} that contributions from large virtual momenta are vanishing for the four-point function of interest; the soft limit is essentially enforced upon us when we restrict ourselves to three-point vertices. There are additional higher order interaction terms that do not contain derivatives (for instance, those coming from the expansion of the inverse metric to higher orders), which we hope to address in future work.

Therefore the complete action (upto cubic vertices) for fluctuations about a generic solution of vacuum Einstein's equations of motion is given by the sum of \eqref{eqn:pureGravityAction} and \eqref{eqn:matterAction}:
\begin{equation}\label{eqn:softAction_GenericBackground}
    S\left[h_{\mu\nu},\phi\right] ~ = ~ S_{EH}\left[h_{\mu\nu}\right] + S_M\left[h_{\mu\nu},\phi\right] \, .
\end{equation}
A recent discussion emphasising the role and importance of interactions in the context of black holes can be found in \cite{Goldberger:2020geb, Goldberger:2020wbx}.

\subsection{Spherically symmetric backgrounds}
\label{sec:SphericalBackgrounds}
In what follows, we will focus on backgrounds with spherical symmetry. Of course, Birkhoff's theorem limits the class of such allowed vacuum solutions. Nevertheless, large parts of our calculation is aimed at generic spherical backgrounds supported by matter; we will only specify the Schwarzschild metric in \secref{sec:horizon}. The general metric we will concern ourselves with is of the form:
\begin{equation}\label{eq:backgroundmetric}
    \ed s^2 ~ = ~ - 2 A(x,y) \ed x\ed y + r(x,y)^2 \ed \Omega^2 \, .
\end{equation}
With $A = e^{1-r/R} R/r$ and $r$ implicitly defined by $xy = 2 R^2(1 - r/R) e^{r/R-1}$, this reduces to the Schwarzschild solution in Kruskal-Szekeres coordinates. Perturbations around the Schwarzschild metric have long been studied \cite{Regge:1957td, Zerilli:1970se, Vishveshwara:1970cc, Teukolsky:1972my, Jensen:1995qv} with a predominant focus often laid on a study of gravitational waves, cosmic censorship, and quasi normal modes.

\paragraph{Gauge symmetry} Not all components of the fluctuations $h_{\mu\nu}$ are physical; there are gauge redundancies due to infinitesimal diffeomorphisms $x \rightarrow x + \xi(x)$, under which the graviton transforms as
\begin{equation}
h_{\mu\nu}(x) ~ \rightarrow ~ h_{\mu\nu}(x) + \nabla_{\mu} \xi_{\nu} + \nabla_{\nu}\xi_{\mu} \, .
\end{equation}
A choice of the vector $\xi_\mu$ fixes a choice of gauge. This choice is of course, not to be confused with the choice of a system of coordinates for the background. Infinitesimal diffeomorphisms acting on the graviton fluctuations are different from finite coordinate transformations of the background; as we will see, $\xi_\mu$ will be linear in the fluctuations and therefore shall not interfere with the choice of coordinates for the background. An alternative way to think about this is to note that a covariant action can be written for the fluctuations as in \eqref{eqn:softAction_GenericBackground}. A choice of background coordinates would not break that covariance, while a choice of $\xi_\mu$ would make unphysical components of the graviton $h_{\mu\nu}$ redundant.

The oft employed gauge in the literature is the harmonic (or covariant or De Donder) gauge defined by
\begin{equation}
\label{eq:harmonicgauge}
\nabla_{\mu}\left(h^{\mu\nu}-\dfrac{1}{2}g^{\mu\nu} h\right) ~ = ~ 0 \, .
\end{equation}
In this gauge, the first order variation of the Ricci tensor is of the form
\begin{equation}
R^{(1)}_{\mu\nu} ~ = ~ -\dfrac{1}{2} \left(g_{\mu\rho} g_{\nu\sigma} \Box - 2 R_{\mu\rho\nu\sigma}\right) h^{\rho\sigma} \, .
\end{equation}
The operator $g_{\mu\rho}g_{\nu\sigma}\Box+2R_{\mu\rho\nu\sigma}$ is the familiar Lichnerowicz operator. It has been studied on the Schwarzschild background in \cite{Morales:2006,Khavkine:2020ksv}. An important difficulty with the choice of the harmonic gauge is that the Lichnerowicz operator couples various spherical harmonics of the graviton. Instead, we will use an alternative gauge; the \textit{Regge-Wheeler} gauge first proposed in \cite{Regge:1957td}. This choice exploits the spherical symmetry of  the background and allows us to reduce the action \eqref{eqn:softAction_GenericBackground} to an infinite tower of decoupled\footnote{In principle interaction terms do couple the partial waves, but the coupling between the partial waves can be ignored when transverse momenta are ignored.} effective two-dimensional actions; one for each partial wave. Conveniently enough, the reduced effective action can be written covariantly in two dimensions.

\subsection{Decomposition in spherical harmonics}
The decomposition of the graviton into spherical tensor harmonics needs care. Owing to its tensorial nature, the corresponding spherical harmonic decomposition differs from the usual spherical harmonic decomposition of scalars. The angular components $h_{aA}$ and $h_{AB}$ have a nontrivial spherical harmonic expansion which must be taken into account. The lowercase indices run over the lightcone coordinates while the uppercase indices span the sphere. As was pointed out in \cite{Regge:1957td}, this can be achieved by splitting the graviton tensor harmonics into odd and even parity modes:
\begin{equation}
\label{eq:gravitonsphericalgeneral1}
h_{\mu\nu} ~ = ~ \sum\limits_{\ell,m}h^-_{\ell m,\mu\nu}+\sum\limits_{\ell,m}h^+_{\ell m,\mu\nu} \, ,
\end{equation}
where
\begin{equation}
\label{eq:gravitonodd}
h^-_{\ell m,\mu\nu}=\begin{pmatrix}
0 & 0 & -h^-_x\csc\theta\p_{\phi} & h^-_x\sin\theta\p_{\theta} \\
 0 & 0 & -h^-_y\csc\theta\p_{\phi} & h^-_y\sin\theta\p_{\theta} \\
 &  & h_{\Omega}\csc\theta\left(\p_{\theta}\p_{\phi}-\cot\theta\p_{\phi}\right) & \tfrac{1}{2}h_{\Omega}\left(\csc\theta\p_{\phi}^2+\cos\theta\p_{\theta}-\sin\theta\p_{\theta}^2\right)\\
 &  &  &  -h_{\Omega}\sin\theta\left(\p_{\theta}\p_{\phi}-\cot\theta\p_{\phi}\right)
\end{pmatrix} Y_{\ell}^{m}
\end{equation}
is the odd parity mode and
\begin{equation}
\label{eq:gravitoneven}
h^+_{\ell m,\mu\nu} = \begin{pmatrix}
H_{xx} & H_{xy} & h^+_x\p_{\theta} & h^+_x\p_{\phi} \\
 & H_{yy} & h^+_y\p_{\theta} & h^+_y\p_{\phi} \\
 &  & r^2(K+G\p_{\theta}^2) & r^2G(\p_{\theta}\p_{\phi}-\cot\theta\p_{\phi}) \\
 &  &  &  r^2(K\sin^2\theta+G(\p_{\phi}^2+\sin\theta\cos\theta\p_{\theta})) 
\end{pmatrix} Y_{\ell}^{m} 
\end{equation}
is the even parity mode, with $Y_\ell^m$ being the familiar spherical harmonics of the two-sphere. In these expressions, the missing entries are determined by the symmetry of the matrices. In \cite{Regge:1957td}, this decomposition was made for the Schwarzschild black hole, written in Schwarzschild coordinates. While we shall still work with a generic spherically symmetric background at this stage, for comparison to \cite{Regge:1957td}, the entries of the graviton modes above are given in terms of those of \cite{Regge:1957td} as:
\begin{align}
h^{\pm}_x(x,y) ~ &= ~ -\dfrac{R}{x} \left(h^{\pm}_0 + \left(1-\dfrac{R}{r}\right) h^{\pm}_1\right) \, , \\
h^{\pm}_y(x,y) ~ &= ~ -\dfrac{R}{y} \left(h^{\pm}_0 - \left(1-\dfrac{R}{r}\right) h^{\pm}_1\right) \, , \\
h_{\Omega}(x,y) ~ &= ~ h_2 \, , \\
H_{xx}(x,y) ~ &= ~ -\dfrac{y}{2x}\bigr(H_0+H_2+2H_1\bigr) \, , \\
H_{xy}(x,y) ~ &= ~ \dfrac{1}{2}\bigr(H_0-H_2\bigr)\, , \\
\label{eq:gravitonlist}
H_{yy}(x,y) ~ &= ~ -\frac{x}{2y}\bigr(H_0+H_2-2H_1\bigr) \, .
\end{align}
To arrive at these transformations, we made extensive use of the identity $\left(1-\tfrac{R}{r}\right)=\frac{xy A}{2R^2}$. It is evident that the time translation invariance of the Schwarzschild solution manifests itself as $x\to ax,y\to \tfrac{1}{a} y$, and we observe that $h_x^{\pm},h_y^{\pm}$ and $H_{xx},H_{yy}$ are not time translation invariant. 

Further on in this paper, for the sake of convenience, we will work in light-cone coordinates of the Kruskal-Szekeres kind. Nevertheless, as argued in \cite{Gaddam:2020rxb}, the covariance of the effective two-dimensional theory we shall present and the conformally flatness of the horizon indicate that the resulting amplitude is gauge invariant.

The decomposition above is general and therefore, the graviton still contains redundant gauge degrees of freedom. The $r^2$ appearing in the graviton modes depends on the light-cone coordinates $x,y$ as was noted earlier. All functions $h^{\pm}_x,h^{\pm}_y,h_{\Omega},H_{xx},H_{xy},H_{yy},K,G$ are functions of only $x,y$ with no further constraints. All of them naturally carry $\ell,m$ indices; we have dropped them to avoid clutter of notation. These account for the ten degrees of freedom of the graviton.

\subsubsection*{The Regge-Wheeler gauge}
We now perform a gauge fixing similar to Regge and Wheeler in \cite{Regge:1957td} of the form
\begin{eqnarray}
\xi_a&=&\zeta_a Y_{\ell}^{m},\\
\xi_A&=&-\tfrac{1}{2}r^2 G\p_A Y_{\ell}^{m}-\tfrac{1}{2} h_{\Omega}\epsilon_A^{\gap\gap B}\p_B Y_{\ell}^{m}.
\end{eqnarray}
Here the antisymmetric tensor is defined in Appendix \ref{app:defs} and 
\begin{eqnarray}
\zeta_a=\left(\tfrac{1}{2}r^2\p_a G-h^+_a\right).
\end{eqnarray} 
As was noted earlier, infinitesimal gauge transformations do not interfere with the choice of background Kruskal-Szekeres system of coordinates because $\xi_a,\xi_A$ are of order $h$. The above choice of gauge, along with the redefinitions of $h_a^-$ , $K$, and $H_{ab}$ as
\begin{align}
h^-_a ~ &\to ~ h_a-\tfrac{1}{2}r^2\p_a\left(\dfrac{1}{r^2}h_{\Omega}\right) \, , \\
K ~ &\to ~ K-2g^{ab}\zeta_a\p_b\log r \, , \\
H_{ab} ~ &\to ~ H_{ab}-\nabla_a\zeta_b-\nabla_b\zeta_a \, ,
\end{align}
kills $G$, $h^+_a$, and $h_\Omega$. We have dropped the minus superscript on $h^-_x$ and $h^-_y$ since $h^+_x$ and $g^+_y$ are now zero. It can be checked that the gauge transformation above, together with the field redefinitions results in the same general form for the odd and even graviton components, with $G,h_a^+,h_{\Omega}$ set to be vanishing. The gauge fixing procedure, therefore, makes four of the components redundant leaving six physical off-shell modes behind. After gauge fixing the graviton components are given as
\begin{equation}
h^-_{\ell m,\mu\nu} ~ = ~ \begin{pmatrix}
0 & 0 & -h_x\csc\theta\p_{\phi} & h_x\sin\theta\p_{\theta}\\
 0 & 0 & -h_y\csc\theta\p_{\phi} & h_y\sin\theta\p_{\theta}\\
-h_x\csc\theta\p_{\phi} & -h_y\csc\theta\p_{\phi} & 0 & 0\\
h_x\sin\theta\p_{\theta} & h_y\sin\theta\p_{\theta} & 0 & 0
\end{pmatrix} Y_{\ell}^{m} \, ,
\end{equation}
and
\begin{equation}
h^+_{\ell m,\mu\nu} ~ = ~ \begin{pmatrix}
H_{xx} & H_{xy} & 0 & 0\\
H_{xy} & H_{yy} & 0 & 0\\
0 & 0 & r^2 K  & 0\\
0 & 0 & 0 & r^2 K \sin^2\theta.
\end{pmatrix} Y_{\ell}^{m} \, .
\end{equation}
The minus sign in the odd mode $h^-_{\mu\nu}$ naturally arises when written in index notation
\begin{equation}
\label{eq:hoddscalarrep}
h^-_{aA} ~ = ~ -h_a\epsilon_A^{\gap\gap B}\p_B Y_{\ell}^{m} \, .
\end{equation}
Owing to this antisymmetric nature of the odd parity mode, the even and odd modes naturally decouple as we shall see. Furthermore, for any diffeomorphism that acts on the light-cone and angular coordinates separately, we may simply transform $h_a\to h_{a'}$ and $\epsilon_A^{\gap\gap B}\p_B Y_{\ell}^{m}\to\epsilon_{A'}^{\gap\gap B'}\p_{B'} Y_{\ell}^{m}$ accordingly. In the new coordinates then, $h^-_{\mu\nu}$ is still given by \eqref{eq:hoddscalarrep}, so that the spherical harmonics decomposition is coordinate independent for all such diffeomorphisms. Similary, for the even modes, we may write
\begin{equation}
h^+_{ab} ~ = ~ H_{ab}Y_{\ell}^m \, \quad \text{and} \quad h^+_{AB} ~ = ~ K g_{AB} Y_{\ell}^m \, ,
\end{equation}
so that coordinate invariance for the decomposition is manifest when $H_{ab}$ is transformed accordingly; $g_{AB}$ naturally transforms appropriately. Because the decomposition persists so long as we also transform the fields appropriately, we now define $H_{ab}$ to be a 2-tensor on the light-cone, $h_a$ to be a vector on the light-cone, and $K$ to be a scalar on the light-cone. Therefore, these fields transform under coordinate transformations on the light-cone but not under diffeomorphisms of the angular coordinates. In a similar fashion all quantities with indices along $A,B$ transform under angular diffeomorphisms but not under those on the light-cone. The spherical harmonics decomposition persists under any transformation that keeps the light-cone and two-sphere separate in this sense. Any coordinate transformations that break the decomposition will therefore be forbidden henceforth. 

Finally, it is worth noting that the even parity graviton carries four of the six off-shell degrees of freedom in the form of a two-tensor $H_{ab}$ and the scalar $K$, while the other two are with the odd-parity mode in the form of the two-vector $h_a$. These degrees of freedom are now defined covariantly and for arbitrary $A(r)$ in the metric. 

\subsection{Decoupling of the odd and even parity graviton modes}
In this section we will investigate the coupling between the odd and even parity modes defined by the graviton decomposition
\begin{align}
\nonumber
h^-_{aA} ~ &= ~ -h_a\epsilon_A^{\gap\gap B}\p_B Y_{\ell}^{m} \, , \\
\label{eq:decomposition1}
h^+_{ab} ~ &= ~ H_{ab}Y_{\ell}^m \, , \\
\nonumber
h^+_{AB} ~ &= ~ K g_{AB} Y_{\ell}^m \, ,
\end{align}
with the remaining terms vanishing. We expect this coupling to fall away due to spherical symmetry and will now prove this. First we see from. \eqref{eqn:softAction_GenericBackground} that the part of the Lagrangian that couples the different parity modes is of the form
\begin{equation}
\label{eq:couplingLagrangian}
\mathcal{L}_{parity-coupling} ~ = ~ -\dfrac{1}{4} h_+^{\mu\nu} G^{(1),-}_{\mu\nu} - \dfrac{1}{4} h_-^{\mu\nu} G^{(1),+}_{\mu\nu}
\end{equation}
where
\begin{align}\label{eqn:EinsteinTensorFirstOrder}
G^{(1),\pm}_{\mu\nu} ~ &= ~ \dfrac{1}{2}g^{\rho\sigma} \bigr(\nabla_{\rho}\nabla_{\mu} h^{\pm}_{\nu\sigma} + \nabla_{\rho}\nabla_{\nu} h^{\pm}_{\mu\sigma} - \nabla_{\rho}\nabla_{\sigma} h^{\pm}_{\mu\nu} - \nabla_{\mu}\nabla_{\nu} h^{\pm}_{\rho\sigma}\bigr) \nonumber \\ 
&\quad - ~ \dfrac{1}{2}g_{\mu\nu} \biggr(\nabla^{\rho}\nabla^{\sigma} h^{\pm}_{\rho\sigma} - \Box h^{\pm}\biggr) \, .
\end{align}
Evidently, the $\pm$ symbols denote the variation of the Einstein tensor corresponding to $h^{\pm}$ respectively. Exploiting the decomposition structure of \eqref{eq:decomposition1}, we find
\begin{align}
G^{(1),-}_{ab} ~ &= ~ 0 \, , \\
G^{(1),-}_{AB}g^{AB} ~ &= ~ 0 \, , \\
\label{eq:Einsteincomp1}
G_{aA}^{(1),+} ~ &= ~ \biggr(\tfrac{1}{2}\p^b H_{ab}+\tfrac{1}{A}\p_a H_{xy}-\tfrac{1}{2A}\bigr(\p_a\log(A r^2)\bigr)H_{xy}-\frac{1}{2}\p_a K\biggr)\p_A Y_{\ell}^m
\end{align}
with all other components being unnecessary owing to either the vanishing of $h^-_{ab}$, $h^-_{AB}$ and $h^+_{aA}$, or the fact that $h^+_{AB}=K g_{AB}$. The second equation $G_{AB}^{(1),-} g^{AB}=0$ results from the fact that $g^{AB}$ is even in $A,B$, while $G^{(1),-}_{AB}$ contains an $\epsilon_{AB}$ which is odd in $A,B$. We now immediately see that the first term in the coupling Lagrangian \eqref{eq:couplingLagrangian} vanishes. Therefore, we are left with
\begin{equation}
\label{eq:couplingLagrangian2}
\mathcal{L}_{parity-coupling} ~ = ~ -\dfrac{1}{2} h_-^{aA} G^{(1),+}_{aA} \, .
\end{equation}
Thus, to evaluate the coupling of the even and odd parity modes, we only need to evaluate $h^-_{aA} G_{(1),+}^{aA}$. At this point, the presence of $\ell,m$ indices that were omitted require attention and explicit re-insertion
\begin{align}
\label{eq:sphereham1}
h^-_{aA} ~ &= ~ -\sum\limits_{\ell,m} h_a^{\ell m} \epsilon_A^{\mgap B}\p_B Y_{\ell}^m \, ,\\
\label{eq:sphereham2}
G^{(1),+}_{aA} ~ &= ~ \sum\limits_{\ell,m} F_a^{\ell m} \p_A Y_{\ell}^m \, .
\end{align} 
Here we defined the new quantity
\begin{equation}
F^{\ell m}_a ~ \coloneqq ~ \dfrac{1}{2}\p^b H^{\ell m}_{ab} + \dfrac{1}{A}\p_a H^{\ell m}_{xy} - \dfrac{1}{2A} \bigr(\p_a\log(A r^2)\bigr) H^{\ell m}_{xy} - \dfrac{1}{2} \p_a K^{\ell m}.
\end{equation}
This quantity can be directly read off from \eqref{eq:Einsteincomp1}. Inserting the summations in \eqref{eq:sphereham1} and \eqref{eq:sphereham2}, the coupling Lagrangian may be rewritten as
\begin{equation}
\label{eq:app:coupling term}
\mathcal{L}_{parity-coupling} ~ = ~ -\dfrac{1}{2} \sum\limits_{\ell,m} \sum\limits_{\ell',m'} h^{\ell m}_a F^a_{\ell'm'} \epsilon^{AB}\p_B Y_{\ell}^{m}\p_A \bar{Y}_{\ell'}^{m'} \, .
\end{equation}
We inserted complex conjugation denoted by $\bar{Y}$ because the Lagrangian is required to be real and we take the imaginary representation of the spherical harmonics. The corresponding action is found by integrating \ref{eq:app:coupling term} to give
\begin{equation}
\label{eq:coupling action 1}
S_{parity-coupling} ~ = ~ - \tfrac{1}{2}\sum\limits_{\ell,m}\sum\limits_{\ell',m'}\int\ed^2 x A\left(r\right) h^{\ell m}_a F^a_{\ell'm'}\underbrace{\int\ed\phi\ed\theta\sqrt{g_{S_2}}\epsilon^{AB}\p_B Y_{\ell}^{m}\p_A \bar{Y}_{\ell'}^{m'}}_{\eqqcolon ~  C_{\ell m;\ell'm'}} \, .
\end{equation}
Here $\sqrt{g_{S_2}}$ is the volume element on $S_2$ of radius $r$, i.e. $\sqrt{g_{S_2}}=r^2\sin\theta$. The first integral is generally non-vanishing, and is the only light-cone dependent piece (the $\sqrt{g_{S_2}}$ cancels as we shall see). The second integral is only dependent on $\theta,\phi$ and contains only spherical harmonics; functions that we know exactly. This angular integrand contains contractions over $A,B$ and is integrated over the sphere. So, the quantity $ C_{\ell m;\ell'm'}$ is independent of the chosen two-sphere coordinates. 

The action in \eqref{eq:coupling action 1} explicitly splits the light-cone part depending on $x^a$, and the spherical part depending on $x^A$ in a manner that is covariant under light-cone and angular diffeomorphisms separately, without mixing. We will now explicitly calculate this coordinate independent quantity $C_{\ell m; \ell'm'}$; clearly, it can be determined exactly without the knowledge of the graviton fields. The spherical harmonics are defined by
\begin{eqnarray}
Y_{\ell}^{m}=\mathcal{N}_{\ell m}e^{i m\phi}P^m_{\ell}(\cos\theta)
\end{eqnarray}
where $\mathcal{N}_{\ell m}$ is a normalization constant and $P^m_{\ell}(x)$ the associated Legendre polynomial which are real. The antisymmetric tensor $\epsilon^{AB}$ as given in \eqref{eq:epsupper} has a prefactor $1/(r^2\sin\theta)$, so that it cancels the $\sqrt{g_{S_2}}$ in \eqref{eq:coupling action 1}. Inserting this into the definition of $C_{\ell m;\ell'm'}$ gives
\begin{align}
C_{\ell m;\ell'm'} ~ &= ~ -i\mathcal{N}_{\ell m}\mathcal{N}_{\ell' m'} \int\limits_0^{\pi} \ed\theta\int\limits_0^{2\pi} \ed\phi e^{i(m-m')\phi} \bigr[m'P^{m'}_{\ell'}(\cos\theta) \p_{\theta} P^m_{\ell}(\cos\theta) \bigr. \nonumber \\
&\qquad \qquad \qquad \qquad \qquad \qquad \bigr.+ m P^m_{\ell}(\cos\theta) \p_{\theta} P^{m'}_{\ell'}(\cos\theta)\bigr] \, .
\end{align}
Here, the $\epsilon^{AB}$ introduces a minus sign, which is then cancelled by the different signs in $\p_{\phi}e^{im\phi}$ and $\p_{\phi}e^{-im'\phi}$. Now using the orthogonality of the complex exponents
\begin{equation}
\int\limits_0^{2\pi}\ed\phi e^{i(m-m')\phi} ~ = ~ 2\pi\delta_{m m'} \, , \nonumber
\end{equation}
we may immediately write down
\begin{align}
C_{\ell m;\ell'm'} ~ &= ~ -2\pi i m\mathcal{N}_{\ell m}\mathcal{N}_{\ell' m}\delta_{m m'}\int\limits_0^{\pi}\ed\theta \bigr[P^{m}_{\ell'}(\cos\theta)\p_{\theta}P^m_{\ell}(\cos\theta) \bigr. \nonumber \\
&\qquad\qquad\qquad\qquad\qquad\qquad\qquad \bigr. + P^m_{\ell}(\cos\theta)\p_{\theta}P^{m}_{\ell'}(\cos\theta)\bigr] \, ,
\end{align}
where we used that $F_m\delta_{mm'}=F_{m'}\delta_{mm'}$ for arbitrary $F_m$. Finally this can be recognized as the product rule on derivatives of the Associated Legendre polynomials. Thus the integral becomes a total derivative, which is trivially integrated, yielding
\begin{equation}
C_{\ell m;\ell'm'} ~ = ~ -2\pi i m\mathcal{N}_{\ell m}\mathcal{N}_{\ell' m}\delta_{mm'}\bigr(P^m_{\ell}(1)P^{m}_{\ell'}(1)-P^m_{\ell}(-1)P^{m}_{\ell'}(-1)\bigr) \, .
\end{equation}
This gives the result for $C_{\ell m;\ell'm'}$ for all $\ell,m,\ell',m'$. Clearly it trivially vanishes for $m=0$, so we only need to know the values $P^m_{\ell}(\pm 1)$ for $m>0$. The definition \cite{AS} 
\begin{equation}
P_{\ell}^m(x) ~ = ~ \dfrac{(-1)^m}{2^{\ell}\gap\ell!}\left(1-x^2\right)^{m/2}\left(\dfrac{\ed}{\ed x}\right)^{\ell+m}\left(x^2-1\right)^{\ell}.
\end{equation}
shows that for $m>0$ this quantity vanishes at $x=\pm 1$. Therefore, we finally find that
\begin{equation}
C_{\ell m;\ell'm'} ~ = ~ 0
\end{equation}
in all cases. This completes the proof of the decoupling of the odd and even graviton modes. The antisymmetry arising from $\epsilon^{AB}$ is crucial. The minus sign introduced by the antisymmetry of the odd parity mode is canceled by the minus sign of the complex conjugate\footnote{However, complex conjugation is not crucial. Had we inserted two $Y_{\ell}^m$'s without complex conjugation instead, there would be no minus sign in the derivative. Nevertheless, a new minus sign would be introduced in the orthogonality relations. This allows the use of the product rule in the end.}, allowing for the use of the product rule. 

In this decoupling we find that the choice of light-cone coordinates plays no role, i.e. $r,t$ or $x,y$ yield the same result. The same holds for the choice of angular coordinates $A,B$. Both the light-cone coordinates and the angular coordinates are summed over separately. This proof also holds for any metric where $A\left(x,y\right), r\left(x,y\right)$ only depend on the light-cone coordinates, i.e. for any spherically symmetric metric of the form \eqref{eq:backgroundmetric}. In particular, of course, it is valid for the Schwarzschild metric.

\subsection{The odd parity graviton does not contribute}
We have seen that the even and odd parity gravitons decouple in the quadratic action. In what follows, we will see from studying the structure of the three-point vertex that the odd parity graviton does not contribute to the four-point function of interest. 

\subsubsection{The interaction vertex}
The interaction vertex can be inferred from \eqref{eqn:matterAction}
\begin{equation}\label{eqn:vertex}
    S_{vertex} ~ = ~ \dfrac{\kappa}{2} \int \ed^4 x \, \sqrt{- g} \,  h^{\mu\nu} \left(\partial_\mu \phi \partial_\nu\phi - \dfrac{1}{2} g_{\mu\nu} g^{\rho\sigma}\partial_\rho \phi \partial_\sigma\phi\right) \, .
\end{equation}
As discussed in the introduction, the external momenta of interest for the scalars are in-going to or out-going from the black hole horizon. We are also interested in scattering very close to the horizon. Naively, one may expect the out-going mode to carry transverse momentum after scattering in addition to the longitudinal shifts (caused by gravitational backreaction \cite{Aichelburg:1970dh, Dray:1984ha}, for instance). However, as one may have inferred from the brick-wall model \cite{tHooft:1984kcu} or the Bekenstein-Hawking entropy formula, there is roughly one degree of freedom per Planck area on the horizon. This has been more recently emphasised in the work of Dvali \textit{et.al}. \cite{Dvali:2011aa, Dvali:2014ila} by studying graviton scattering amplitudes. Therefore, loosely speaking, it is an unlikely event that particles scatter at impact parameters comparable to Planck length. This observation validates the approximation of impact parameters larger than Planck length. It is well known that transverse momentum transfer is a Planckian effect\footnote{Of course, when collision energies are extremely low for very massive bodies at large impact parameters, classical non-linear effects do not distinguish between longitudinal and transverse effects.} \cite{Dray:1984ha}. Therefore, transverse momentum transfer may safely be considered small: $p_A = 0$. Piecing this together with the Regge-Wheeler gauge, we find 
\begin{align}\label{eqn:vertexRegge-Wheeler}
    S_{vertex} ~ &= ~ \dfrac{\kappa}{2} \int \ed \Omega \int \ed^2 x \, A\left(r\right) \, r^2 \left[h^{ab} \left(\partial_a\phi \partial_b\phi - \dfrac{1}{2}g_{ab}g^{cd} \partial_c\phi\partial_d\phi\right)  + h^{AB} T_{AB}\right] \, ,
\end{align}
for the interaction vertex, where $A\left(r\right)$ and $r$ are implicitly functions of the two-dimensional coordinates $x,y$ as before: $A\left(r\right) = A\left(x,y\right)$ and $r = r\left(x,y\right)$. We see that the odd parity graviton mode $h_{aA}$ drops out of the vertex. Therefore, all scattering processes are entirely governed by the even parity graviton.

\subsection{Action for the even parity graviton}
Owing to the complete decoupling of the odd and even parity modes of the graviton, and the subsequent dropping out of the odd mode from all interactions, the Lagrangian arising from the Einstein-Hilbert action reduces to
\begin{equation}\label{eqn:actionG+}
    \mathcal{L}_{even} ~ = ~ - \dfrac{1}{4} h^+_{\mu\nu} G^{\mu\nu}_{(1),+} \, ,
\end{equation}
where $G^{\mu\nu}_{(1),+}$ is defined in \eqref{eqn:EinsteinTensorFirstOrder}. This Lagrangian can be brought into a more convenient form, exploiting the spherical symmetry of the background. To do so, we first note that the covariant tensor $h_{AB} = K g_{AB}$ and therefore depends on the angular variables. Therefore, we will now first use a raised index on the graviton to extract this angular dependence out as ${h_a}^b = {H_a}^b$ and ${h_A}^B = K \delta_A^B$. We will maintain covariance so that the indices can later be lowered or raised. This intermediate step allows us to write down a spherical harmonic expansion for the graviton as:
\begin{equation}
    {h_\nu}^\mu ~ = ~ \sum_{\ell,m} {\left(h_{lm}\right)_\nu}^\mu Y_\ell^m \, ,
\end{equation}
where $h_{lm}$ are now only dependent on the two-dimensional coordinates $x,y$. Therefore, we also have
\begin{align}
    {h_a}^b ~ &= ~ {H_a}^b Y_\ell^m \label{eqn:hdef}\\
    {h_A}^B ~ &= ~ K \delta_A^B Y_\ell^m \, . \label{eqn:kdef} 
\end{align}
The spherical harmonic expansion enables us to write the action \eqref{eqn:actionG+} as 
\begin{equation}
    \mathcal{L}_{even} ~ = ~ - \dfrac{1}{8} {h^\mu}_{\nu} G^{\nu\mgap\sigma}_{\sgap\mu\rho\sgap} {h_\sigma}^\rho \, ,
\end{equation}
where we defined 
\begin{align}\label{eq:Gdef}
    G^{\nu\mgap\sigma}_{\sgap\mu\rho\sgap} ~ &\coloneqq ~  \delta_{\mu}^{\sigma} \nabla_{\rho} \nabla^{\nu} + \delta^{\nu}_{\rho} \nabla^{\sigma} \nabla_{\mu} - \delta^{\nu}_{\mu} \nabla^{\sigma} \nabla_{\rho} - \delta^{\sigma}_{\rho} \nabla^{\nu} \nabla_{\mu} + 2 \bar{\delta}^{\nu\sigma}_{\mu\rho} \Box \, , \\
    \bar{\delta}^{\nu\sigma}_{\mu\rho} ~ &\coloneqq ~ \delta^{\nu}_{[\mu} \delta^{\rho}_{\sigma]} \, .
\end{align}
\subsection{Decoupling of the partial waves}\label{sec:partialwavedecoupling}
Given that we are working on a spherically symmetric background, it is expected that the partial waves are decoupled entirely. To see this, we first evaluate the action of the operator \eqref{eq:Gdef} on the graviton:
\begin{align}\label{eqn:GActionOnh}
    G^{\nu\mgap\sigma}_{\sgap\mu\rho\sgap} {h_\sigma}^\rho ~ &= ~ \sum_{\ell,m} G^{\nu\mgap\sigma}_{\sgap\mu\rho\sgap} (h_{\ell m})^{\rho}_{\sigma}+\sum\limits_{\ell,m}(h_{\ell m})^{\rho}_{\sigma} G^{\nu\mgap\sigma}_{\sgap\mu\rho\sgap}Y_{\ell}^m \nonumber \\
    &\qquad + ~ \sum\limits_{\ell,m}\biggr(-4\bar{\delta}^{\nu\sigma}_{\mu\rho}(\p_{\kappa}Y_{\ell}^m)\nabla^{\kappa}+\mathcal{U}^{\sigma\nu}_{\mu\rho}+\mathcal{U}^{\nu\sigma}_{\rho\mu}-\mathcal{U}^{\nu\sigma}_{\mu\rho}-\mathcal{U}^{\sigma\nu}_{\rho\mu}\biggr)(h_{\ell m})^{\rho}_{\sigma} \, ,
\end{align}
where we defined
\begin{equation}
    \mathcal{U}^{\nu\sigma}_{\mu\rho} ~ = ~ \delta^{\nu}_{\mu} \biggr[\left(\p^{\sigma}Y_{\ell}^m\right) \nabla_{\rho} + \left(\p_{\rho}Y_{\ell}^m\right) \nabla^{\sigma}\biggr] \, .
\end{equation}
The first line of \eqref{eqn:GActionOnh} is the action of all the operators in \eqref{eq:Gdef} on either the modes $h_{\ell m}$ or on the spherical harmonics, whereas the second line arises from all the mixed terms. While these terms appear complicated, noting that the spherical harmonics $Y_\ell^m$ only dependent on the angles, several indices in the second line of \eqref{eqn:GActionOnh} do not contribute. Moreover, using the tensor structure in \eqref{eqn:hdef} and \eqref{eqn:kdef}, together with the derivatives listed in \appref{app:defs}, we find
\begin{equation}
    \biggr(- 4 \bar{\delta}^{\nu\sigma}_{\mu\rho} \left(\p_{\kappa} Y_{\ell}^m\right) \nabla^{\kappa} + \mathcal{U}^{\sigma\nu}_{\mu\rho} + \mathcal{U}^{\nu\sigma}_{\rho\mu} - \mathcal{U}^{\nu\sigma}_{\mu\rho} - \mathcal{U}^{\sigma\nu}_{\rho\mu} \biggr) \left(h_{\ell m}\right)^{\rho}_{\sigma} ~ \sim ~ \p_A \left(h_{\ell m}\right)^{\rho}_{\sigma} ~ = ~ 0 \, .
\end{equation}
Therefore, the mixed terms in \eqref{eqn:GActionOnh} vanish, and we are left with 
\begin{align}
    G^{\nu\mgap\sigma}_{\sgap\mu\rho\sgap} {h_\sigma}^\rho ~ &= ~ \sum_{\ell,m} G^{\nu\mgap\sigma}_{\sgap\mu\rho\sgap} (h_{\ell m})^{\rho}_{\sigma}+\sum\limits_{\ell,m}(h_{\ell m})^{\rho}_{\sigma} G^{\nu\mgap\sigma}_{\sgap\mu\rho\sgap}Y_{\ell}^m \, .
\end{align}
The action of the operator \eqref{eq:Gdef} on the spherical harmonics is given by
\begin{align}
    G^{\nu\mgap\sigma}_{\sgap\mu\rho\sgap} Y_{\ell}^m ~ &= ~ \dfrac{\left(\Delta_{\Omega} Y_{\ell}^m\right)}{2r^2} \biggr(\delta^{\nu}_{\mu} \delta_{\rho}^{\sigma} + \delta^{\nu}_a \delta^a_{\mu} \delta^{\sigma}_b \delta^b_{\rho} - 2 \delta^{\nu}_{\rho} \delta^{\mu}_{\sigma}\biggr) \nonumber \\
    &= ~ - Y_{\ell}^m \dfrac{\ell \left(\ell + 1\right)}{2r^2} \biggr(\delta^{\nu}_{\mu} \delta_{\rho}^{\sigma} + \delta^{\nu}_a \delta^a_{\mu} \delta^{\sigma}_b \delta^b_{\rho} - 2 \delta^{\nu}_{\rho} \delta^{\mu}_{\sigma}\biggr) \, ,
\end{align}
where in the second line, we have used the property that the spherical harmonics are the eigenfunctions of the spherical Laplacian $\Delta_{\Omega} Y_{\ell}^m = \ell\left(\ell + 1\right) Y_\ell^m$. That there are no further derivatives on the $Y_\ell^m$ helps the decoupling of the partial waves. Equation \eqref{eqn:GActionOnh} can now be written as
\begin{equation}
    G^{\nu\mgap\sigma}_{\sgap\mu\rho\sgap} {h_\sigma}^\rho ~ = ~ \sum_{\ell,m} Y_\ell^m {{{\mathcal{G}_\ell}^\nu}_{\mu\rho}}^{\sigma} \left(h_{\ell m}\right)_\sigma^{\gap\rho} \, ,
\end{equation}
where we defined the modified operator
\begin{equation}\label{eqn:Inverseprop}
    \mathcal{G}^{\nu\mgap\sigma}_{\ell\gap\mu\rho\sgap} ~ \coloneqq ~ G^{\nu\mgap\sigma}_{\sgap\mu\rho\sgap} - \dfrac{\ell(\ell+1)}{2r^2} \bigr(\delta^{\mu}_{\nu} \delta_{\rho}^{\sigma} + \delta^{\mu}_a \delta^a_{\nu} \delta^{\sigma}_b \delta^b_{\rho} - 2 \delta^{\mu}_{\rho} \delta^{\sigma}_{\nu} \bigr) \, .
\end{equation}
As one many have expected, the angular parts contribute to a $1/r^2$ potential in this operation. The second order action now results in
\begin{align}\label{eqn:partialWaveAction}
    S_{even} ~ &= ~ - \dfrac{1}{8} \sum\limits_{\ell,m} \sum\limits_{\ell',m'} \int \ed\Omega \gap\bar{Y}_{\ell'}^{m'} Y_{\ell}^m \int\ed^2 x \sqrt{-g} {\left(h_{\ell'm'}\right)^{\mu}}_{\nu} \mathcal{G}^{\nu\mgap\sigma}_{\sgap\mu\rho\sgap} \left(h_{\ell m}\right)^{\gap\rho}_{\sigma} \, .
\end{align}
We now have an infinite tower of decoupled actions, one for each partial wave. The decoupling is a result of spherical symmetry of the background and may be seen as the conservation of angular momentum.

\subsection{An effective two-dimensional theory}\label{sec:effective2dTheory}
Noting that the graviton fields $h_{\ell m}$ are only functions of the light-cone coordinates since the angular pieces have been extracted out, and using. the orthogonality relations for the spherical harmonics
\begin{equation}\label{eq:sphericalharmonicsorthogonality}
    \int \ed \Omega \, \bar{Y}_{\ell '}^{m'} \, Y_\ell^m ~ = ~ \delta_{\ell \ell '} \delta_{mm'} \, ,
\end{equation}
the action \eqref{eqn:partialWaveAction} reduces to
\begin{equation}\label{eq:decoupledaction1}
    S_{even} ~ = ~ - \dfrac{1}{8} \sum\limits_{\ell,m} \int\ed^2 x A\left(r\right) \, r^2 \, {\left(h_{\ell m}\right)^{\mu}}_{\nu} \mathcal{G}^{\nu\mgap\sigma}_{\sgap\mu\rho\sgap} \left(h_{\ell m}\right)^{\gap\rho}_{\sigma} \, .
\end{equation}
While this may appear to be a two-dimensional theory, the indices on the graviton are still Greek and run over all four spacetime dimensions. Moreover, owing to the implicit presence of the angular indices, a separation of light-cone and spherical fields is no longer obvious. To appreciate this problem of separation of variables, it is instructive to evaluate the quantity $\Box h^a_{\gap b}$. Despite the presence of only the light-cone coordinates, as show in \appref{app:Identities}, we find a term of the form $\left(\partial^a r\right)\left(\partial_b r\right) K$. This arises from the two-sphere coordinates in the $\Box$ operator. The second, and perhaps more serious, problem is seen when indices are raised and lowered with the metric. Defining, for instance, $h_{\phi\phi}$ is problematic because while $h^\phi_\phi$ is independent of the angular variables, lowering the index would introduce it because $g_{\phi\phi} = r^2 \sin^2\theta$. Therefore, separation of variables along the light-cone and the two-sphere, and the spherical harmonic expansion appear to be in conflict with each other. While the two-dimensional propagator and equations of motion appear problematic, these nevertheless naturally pose no problem in the complete four-dimensional action. It is therefore evident that the price we pay for covariance on the light-cone is an inherent mixing between the fields $H_{ab}$ and $K$.

In what follows, we define differential operators with a tilde to represent those on the light-cone. So, $\tilde{\nabla}$ is the covariant derivative on the light-cone, and $\tilde{\Box} = \tilde{\nabla}^a \tilde{\nabla}_a$ is the 'd Alembertian, and the indices are raised and lowered with the light-cone metric $g_{ab}$. For consistency, we will also write the light-cone metric with a tilde as $\tilde{g}_{ab}$. In this formulation, clearly $\tilde{\Box} h^a_{\gap b}$ will never give rise to the scalar field $K$ as the lowercase Latin indices never sum over the angular coordinates. This allows us to separate variables and maintain covariance on the light-cone all the same. Notwithstanding which, the tensor $H_{ab}$ and $K$ will remain coupled. 

Symbolically we want to find an action of the following form \cite{Martel:2005ir}
\begin{eqnarray}
\label{eq:2Dlagrangian1}
S_{even} ~ = ~ \frac{1}{4}\int\ed^2 x\sqrt{-\tilde{g}}\biggr(\tilde{H}^{ab}\tilde{\Delta}^{-1}_{abcd}\tilde{H}^{cd}+\tilde{H}^{ab}\tilde{\Delta}^{-1}_{L,ab} \tilde{K}+\tilde{K}\tilde{\Delta}^{-1}_{R,ab} \tilde{H}^{ab}+\tilde{K}\tilde{\Delta}^{-1}\tilde{K}\biggr).
\end{eqnarray}
Here we defined all of the new $\tilde{\Delta}$ operators listed below. We have also defined the new lightcone fields $\tilde{H}_{ab}=r H_{ab},\tilde{K}=rK$. These definitions have been made to absorb the residual $r^2$ of the two-sphere Jacobian in the action in \eqref{eq:decoupledaction1} into the fields. This gives rise to the light-cone Jacobian $\sqrt{-\tilde{g}}$, and consequently results in the canonical form of the action \eqref{eq:2Dlagrangian1}. A detailed derivation of these operators is given in \appref{app:evenactionderivation}. The result is:
\begin{align}
\label{eq:scalarmathDpaper}
\tilde{\Delta}^{-1} ~ &= ~ - \tilde{\Box} + F^a_a \, , \\
\label{eq:mixmathDrightpaper}
\tilde{\Delta}^{-1}_{R,ab} ~ &= ~ - g_{ab} \left(\tilde{\Box} - \dfrac{1}{2} V_c \tilde{\nabla}^c + \dfrac{1}{4} V_c V^c - F_c^c - \dfrac{\ell(\ell+1)}{2r^2}\right) + \tilde{\nabla}_a \tilde{\nabla}_b - F_{ab} \, , \\
\label{eq:mixmathDleftpaper}
\tilde{\Delta}^{-1}_{L,ab} ~ &= ~ - g_{ab} \left(\tilde{\Box} + \dfrac{1}{2} V_c \tilde{\nabla}^c - \dfrac{\ell(\ell+1)}{2r^2}\right) + \tilde{\nabla}_a \tilde{\nabla}_b - F_{ab} \, , \\
\label{eq:tensormathDpaper}
\tilde{\Delta}^{-1}_{abcd} ~ &= ~ \dfrac{1}{2} g_{ac} V_{[b} \tilde{\nabla}_{d]} + \dfrac{1}{2} g_{bd} V_{[a} \tilde{\nabla}_{c]} + \dfrac{1}{2} g_{ab} \left( V_{(c} \tilde{\nabla}_{d)} + 2 F_{cd} \right) \nonumber \\
&\qquad + \dfrac{1}{2} g_{cd} \left(-V_{(a} \tilde{\nabla}_{b)} + \dfrac{1}{2} V_a V_b\right) + g_{ab} g_{cd} \left(\dfrac{1}{4} R_{2d} + \dfrac{\ell(\ell+1)}{2r^2}\right) \nonumber \\
&\qquad - g_{ac} g_{bd} \left(\dfrac{1}{2} R_{2d} + \dfrac{\ell(\ell+1)}{2r^2}\right) \, ,
\end{align}
where we defined the following tensors:
\begin{align}
V_a ~ &= ~ 2 \p_a \log r \, , \\
F_{ab} ~ &= ~ \dfrac{1}{r} \tilde{\nabla}_a \tilde{\nabla}_b r ~ = ~ \dfrac{1}{2} \tilde{\nabla}_{(a} V_{b)} + \dfrac{1}{4} V_a V_b \, , \\
R_{2d} ~ &= ~ - \dfrac{1}{A} \tilde{\Box} \log A \, .
\end{align}
Here $V_a$ is a residual curvature potential arising from the two-sphere, $F_{ab}$ its corresponding field strength, and $R_{2d}$ is the Ricci scalar of the two-dimensional background metric $\tilde{g}_{ab}$. We note that all operators are in fact symmetric in the fields; in particular, $\tilde{\Delta}^{-1}_{R,ab}$ equals $\tilde{\Delta}^{-1}_{L,ab}$ up to total derivatives. We now have a covariant description of the graviton modes on the lightcone, where the degrees of freedom $\tilde{H}_{ab}$ and $\tilde{K}$ are explicitly separated, albeit coupled. The two-sphere is now integrated out entirely; it leaves a residual $\ell,m$ dependence and the curvature potential $V_a$. The lightcone metric is at this stage still general $\tilde{g}_{ab}=A(r)\eta_{ab}$ parametrised by an arbitrary function. 

\subsubsection{Weyl rescaling}\label{subsec:WeylRescaling}
Before attempting to insert the Schwarzschild metric to find the propagators, we first observe that the effective two-dimensional metric is conformally flat:
\begin{eqnarray}
\tilde{g}_{ab} ~ = ~ A\left(r\right) \eta_{ab} \, ,
\end{eqnarray}
with $\eta_{ab}$ being the two-dimensional Minkowski metric:
\begin{equation}
\eta_{ab} ~ = ~ 
    \begin{pmatrix}
        0 & -1 \\
        -1 & 0
    \end{pmatrix} \, .
\end{equation}
This allows us to perform a Weyl transformation to reduce our theory to a two-dimensional flat theory, with modified potentials. This will allow us to trade the subtleties of curved space, and quantum field theory on the said curved space to a flat theory for additional potential terms. An important consequence is that despite being in the region within the gravitational potential, where flat space kinematics is invalid, the Weyl transformation will allow us to define centre of mass energies.

In order to migrate to a flat two-dimensional background we first make a transformation that explicitly removes the function $A(r)$ from the metric. The transformations are given by
\begin{align}
\tilde{g}_{ab} ~ &\to ~ A\left(r\right) \eta_{ab} \, , \\
\tilde{H}_{ab} ~ &\to ~ A\left(r\right) \mathfrak{h}_{ab} \, , \\
\tilde{K} ~ &\to ~ \mathcal{K} \, ,
\end{align}
where $\mathfrak{h}^{ab}$ is raised and lowered with the new flat Minkowski metric. Conveniently enough, partial derivatives are now covariant derivatives, so covariance of the physical quantities is trivially achieved. 

The Weyl transformation brings about a few noteworthy points: $\tilde{K}$ does not transform (and yet we give it a new name $\mathcal{K}$) and neither does $V_b=2\p_b\log r$ (however, $V^b=\tilde{g}^{ab}V_b=\tfrac{1}{A}\eta^{ab}V_b$ does transform). Moreover, $F_{ab}$ also does not transform. 

Inserting the Weyl rescaling, we may rewrite the action quadratic in $\mathcal{K}$, for instance, as
\begin{equation}
\label{eq:actiontransform}
\dfrac{1}{4} \int \ed^2 x A\left(r\right) \mathcal{K} \left((-\Box + F_a^a\right) \mathcal{K} ~ = ~ \dfrac{1}{4} \int\ed^2 x \mathcal{K} \left(- \p^2 + \eta^{ab}F_{ab}\right) \mathcal{K} \, ,
\end{equation}
where we included the $A\left(r\right)$ that comes from the $\sqrt{-g}$ and defined $\p^2=\eta^{ab}\p_a\p_b$. For consistency we redefine
\begin{eqnarray}
F_{ab} ~ = ~ \dfrac{1}{r} \tilde{\nabla}_a \tilde{\nabla}_b r ~ \eqqcolon ~ \dfrac{1}{r} \p_a \p_b r - \dfrac{1}{2} U_{(a} V_{b)} + \dfrac{1}{4} \eta_{ab} U^c V_c \, ,
\end{eqnarray}
where all index manipulations are now done with the flat metric, i.e. $U^c V_c=\eta^{ac}U_a V_c$ and $U^a=\eta^{ab}U_b$. This redefinition is in principle necessary, since the previous lightcone covariant derivative does not hold after the Weyl transformation. Here we defined a new potential that will embed the curvature remnants of $A\left(r\right)$:
\begin{eqnarray}
U_a ~ = ~ \p_a \log A\left(r\right) \, .
\end{eqnarray}
Then the action quadratic in $\mathcal{K}$ becomes
\begin{eqnarray}
\dfrac{1}{4} \int \ed^2 x \mathcal{K} \left(-\p^2 + F_a^a\right) \mathcal{K} \, ,
\end{eqnarray}
where $F_a^a=\eta^{ab}F_{ab}$. The Weyl transformation essentially allows us a change in viewpoint but no change in physics; we may either study free fields in a curved spacetime, or Weyl rescaled fields bounded by potentials in flat spacetime. Both descriptions are equivalent for conformally flat spacetimes. The fact that the spacetime is conformally flat is important because all curvature effects are then isotropic, allowing for an embedding into a scalar function $A\left(r\right)$. For anisotropic curvatures we naturally need a tensor potential, losing the simplicity of a Weyl transformation.

We now proceed to a Weyl rescaling of all fields and operators, with due regard for which fields transform, and which fields do not. For the following, all Christoffel symbols need to be worked out in order to safely pull the $\tfrac{1}{A}$ in $\tilde{H}^{cd}$ through. Essentially, we need to calculate:
\begin{align}
\Delta^{-1}_{R,ab} ~ &\coloneqq ~ A\left(r\right) \tilde{\Delta}^{-1}_{R,ab} \dfrac{1}{A\left(r\right)} \, , \\
\Delta^{-1}_{L,ab} ~ &\coloneqq ~ \tilde{\Delta}^{-1}_{L,ab} \, , \\
\Delta^{-1}_{abcd} ~ &\coloneqq ~ \tilde{\Delta}^{-1}_{abcd} \dfrac{1}{A\left(r\right)} \, .
\end{align}
The action is now given by
\begin{eqnarray}
\label{eq:evenactionfinal}
S ~ = ~ \dfrac{1}{4} \int\ed^2 x \biggr(\mathfrak{h}^{ab}\Delta^{-1}_{abcd}\mathfrak{h}^{cd} + \mathfrak{h}^{ab} \Delta^{-1}_{L,ab} \mathcal{K} + \mathcal{K}\Delta^{-1}_{R,ab} \mathfrak{h}^{ab} + \mathcal{K}\Delta^{-1}\mathcal{K}\biggr) \, .
\end{eqnarray}
An explicit calculation of all $\Delta$ operators shows that
\begin{subequations}\label{eqn:opers2d}
\begin{align}
\Delta^{-1} ~ &= ~ -\p^2+F_a^a \, , \label{eqn:opers2dScalar} \\
\Delta^{-1}_{R,ab} ~ &= ~ - \eta_{ab} \left(\p^2 + \dfrac{1}{2} \left(U^c - V^c\right) \p_c + \dfrac{1}{2} \eta^{cd} \left(\mathcal{W}^A_{cd} - \mathcal{W}^r_{cd}\right)\right. \nonumber \\
&\qquad\qquad\qquad - \left.A\left(r\right) \dfrac{\ell(\ell+1)}{2r^2}\right) + \p_a \p_b + U_{(a} \p_{b)} + \mathcal{W}^A_{ab} - F_{ab} \, , \label{eqn:opers2dRight} \\
\Delta^{-1}_{L,ab} ~&= ~ - \eta_{ab} \left(\p^2 - \dfrac{1}{2} \left(U^c - V^c\right) \p_c - A\left(r\right) \dfrac{\ell(\ell+1)}{2r^2} \right) + \p_a \p_b - U_{(a} \p_{b)} - F_{ab} \, , \label{eqn:opers2dLeft} \\
\Delta^{-1}_{abcd} ~ &= ~ \dfrac{1}{2} \eta_{ac} V_{[b} \p_{d]} + \dfrac{1}{2} \eta_{bd} V_{[a} \p_{c]} + \dfrac{1}{2} \eta_{ab} \left( V_{(c} \p_{d)} + \mathcal{W}^r_{cd} + \dfrac{1}{2} V_c V_d\right) \nonumber \\
&\qquad + \dfrac{1}{2} \eta_{cd} \left(- V_{(a} \p_{b)} + \dfrac{1}{2} V_a V_b\right) +  \eta_{ab} \eta_{cd} \left(\dfrac{1}{4} A\left(r\right) R_{2d} - \dfrac{1}{4} V^e U_e + A\left(r\right) \dfrac{\ell(\ell+1)}{2r^2}\right) \nonumber \\
&\qquad - \eta_{ac} \eta_{bd} \left(\dfrac{1}{2} A\left(r\right) R_{2d} - \dfrac{1}{2} V^e U_e + A\left(r\right) \dfrac{\ell(\ell+1)}{2r^2}\right) \, , \label{eqn:opers2dTensor}
\end{align}
\end{subequations}
where repeated application of the product rule on any relevant $\tfrac{1}{A}$ was necessary. We defined two new field tensors
\begin{eqnarray}
\mathcal{W}^r_{ab} ~ \coloneqq ~ \p_{(a}V_{b)} \qquad \text{and} \qquad \mathcal{W}^A_{ab} ~ \coloneqq ~ \p_{(a}U_{b)} \, .
\end{eqnarray}
While the $\Delta$ operators now appear more complicated, the background is flat, making all covariant derivatives trivial. Moreover, while coordinate transformations need care, any coordinate system may be chosen by an appropriate choice of the function $A(r)$.

This completes the calculation of the even action. The four-dimensional spherically symmetric spacetime is now reduced to a flat two dimensional Minkowski spacetime. This is largely owed to the spherical harmonics expansion and a careful exploitation of the background spherical symmetry. Curvature is embedded in two potentials $V_a,U_a$ and their respective field strengths. While we have focused our attention on vacuum solutions to Einstein's equations, extension to more general spacetimes should be straightforward. The first order term in \eqref{eq:EHactionExpansion} will be nontrivial but can be accommodated for, in this  formalism. 

To complete the effective action, we also need to reduce the scalar action in \eqref{eqn:matterAction} and the interaction vertex \eqref{eqn:vertexRegge-Wheeler} on to two-dimensional flat spacetime. We will incorporate these in the following section as we move near the horizon. Once equipped with the complete effective action, we will then proceed to determine the graviton and scalar propagators and the interaction vertex, on the Schwarzschild horizon, as advertised.


\section{Feynman rules on the horizon}\label{sec:horizon}
While the soft limit of the path integral about spherically symmetric backgrounds conveniently split into effective two-dimensional theories, it is still difficult to invert the quadratic operators \eqref{eqn:opers2d} to find the graviton propagator for general backgrounds. In what follows, we will evaluate the action \eqref{eq:evenactionfinal} on the Schwarzschild background specified by
\begin{equation}\label{eqn:background}
A\left(r\right) ~ = ~ \dfrac{R}{r}e^{1-r/R} \quad \text{and} \quad xy ~ = ~ 2 R^2\left(1-\dfrac{r}{R}\right)e^{r/R-1} \, .
\end{equation}
Therefore, the potentials defining the operators \eqref{eqn:opers2d} can now be written in the $x, y$ coordinates:
\begin{subequations}
\label{eq:potsol1}
\begin{align}
V_a ~ &= ~ \dfrac{A}{rR}x_a \, , \\
U_{a} ~ &= ~ -\dfrac{A}{2rR}\left(1+\tfrac{r}{R}\right)x_a \, , \\
\mathcal{W}^r_{ab} ~ &= ~ \dfrac{A}{rR}\eta_{ab} - \dfrac{A^2}{2R^2r^2}\left(2+\dfrac{r}{R}\right) x_a x_b \, , \\
\mathcal{W}^A_{ab} ~ &= ~ -\dfrac{A}{2rR}\left(1+\dfrac{r}{R}\right) \eta_{ab} + \dfrac{A^2}{4R^2r^2}\left(2 + 2 \dfrac{r}{R} + \dfrac{r^2}{R^2}\right) x_a x_b \, , \\
F_{ab} ~ &= ~ \dfrac{AR}{2r^3} \eta_{ab} \, , \\
R_{2d} ~ &= ~ \dfrac{2R}{r^3} \, .
\end{align}
\end{subequations}
The choice of background \eqref{eqn:background} is evidently that of an eternal Schwarzschild solution in Kruskal-Szekeres coordinates. This retains time-translational invariance in the effective two-dimensional theory: $x\to ax,y\to a^{-1} y$. In what is to follow, however, we will make a near-horizon approximation. As it turns out, the physically relevant information for the derivation of the graviton propagator is the conformal flatness of the bifurcation sphere. In a collapsing scenario, the nature of the apparent horizon may be rather different. However, in the soft limit, we will argue towards the end of the paper that the general lessons drawn may well hold. There are of course more detailed questions further away from the horizon where the near horizon propagator needs to be modified.

As an aside, the relation between the trace of the residual curvature field strength of the two-sphere $F_{ab}$ and the scalar curvature  of the two-dimensional spacetime is striking: $A\left(r\right) R_{2d} = 2 F_a^a$. They may be seen to conspire to make up the vacuum Schwarzschild solution in four dimensions.

We may now insert all curvature potentials into the $\Delta$ operators. Our interest is however, in studying scattering near the horizon. Therefore, we will first define the near-horizon approximation before writing down the simplified potentials.


\subsection{The near horizon approximation}\label{sec:horizonApprox}
The naive near-horizon approximation of the Scwharzschild solution yields Rindler spacetime. However, we would like to keep the spherical nature of the horizon intact. Moreover, as mentioned earlier, all wave fronts received on future null infinity appear to emanate from the central causal diamond in a collapsing scenario. Therefore, the natural approximation of interest is such that the two-dimensional light-cone coordinates $x, y \ll R$; the reference scale for the coordinates is the Schwarzschild radius $R$ instead of the Planck length and the near-horizon region effectively measures $1 - r/R$. In terms of the light-cone coordinates, the horizon is of course defined by $xy=0$. So, we have that 
\begin{eqnarray}
r ~ = ~ R + R \mathcal{O}\left(\dfrac{xy}{R^2}\right) \, ,
\end{eqnarray}
in this approximation. Therefore, to linear order, we simply find $r = R$ and $A\left(r\right) = 1$. There are of course other linear contributions in the form of the couplings $x_a\p_b$, while all potentials have now simply become mass terms. The $\Delta$ operators \eqref{eqn:opers2d} now simplify to
\begin{subequations}
\label{eq:operatorstemp}
\begin{align}
\Delta^{-1} ~ &= ~ -\p^2+\mu^2 \, , \\
\Delta^{-1}_{R,ab} ~ &= ~ -\eta_{ab}\left(\p^2 - \mu^2 x^c\p_c - \dfrac{1}{2}\mu^2\lambda\right) + \p_a \p_b - \mu^2 x_{(a} \p_{b)} \, , \\
\Delta^{-1}_{L,ab} ~ &= ~ -\eta_{ab}\left(\p^2 + \mu^2 x^c \p_c - \dfrac{1}{2} \mu^2\left(\lambda - 2\right)\right) + \p_a \p_b + \mu^2 x_{(a} \p_{b)} \, , \\
\Delta^{-1}_{abcd} ~&= ~ \dfrac{1}{2} \mu^2 \left(\eta_{ac} x_{[b} \p_{d]} + \eta_{bd}x_{[a} \p_{c]} + \eta_{ab} x_{(c} \p_{d)} - \eta_{cd} x_{(a} \p_{b)} \right) \nonumber \\
&\qquad +  \dfrac{\mu^2 \left(\lambda+1\right)}{2} \left(\eta_{ab} \eta_{cd} - \eta_{ac} \eta_{bd} \right) \label{eqn:tensorOperHor}\, ,
\end{align}
\end{subequations}
where we defined 
\begin{eqnarray}
\lambda ~ \coloneqq ~ \ell^2 + \ell + 1 \, ,
\end{eqnarray}
and used the inverse Schwarzschild radius $\mu=1/R$ again, which can now be understood as effective mass due to the positive curvature of the two-sphere.

It is now evident that this approximation is identical to expanding around $\mu\to 0$, which holds for large black holes. Of course the approximation is better defined in terms of the dimensionless quantity $\mu x_a$. The equations of motion are also written in terms of the $\Delta$ operators as
\begin{align}
\Delta^{-1}_{abcd}\mathfrak{h}^{cd} + \Delta^{-1}_{L,ab}\mathcal{K} ~ &= ~ 0 \, , \\
\Delta^{-1}_{R,ab}\mathfrak{h}^{ab} + \Delta^{-1}\mathcal{K} ~ &= ~ 0 \, .
\end{align}
An evident disconcerting feature of the operators \eqref{eq:operatorstemp} is the asymmetry in the `left' and `right' operators. Removing first order terms naively with $\mu x_a \rightarrow 0$ does not fix the problem either. The first derivative terms contribute to the effective mass. A symmetric representation of these operators can be achieved with a field redefinition as we will now show. We first note the following identities:
\begin{align}
\p_a \p_b e^{-\tfrac{\mu^2 x^2}{4}} ~ &= ~ e^{-\tfrac{\mu^2 x^2}{4}}\left(\p_a \p_b - \mu^2 x_{(a}\p_{b)} - \dfrac{1}{2} \mu^2 \eta_{ab} + \dfrac{1}{4} \mu^4 x_a x_b\right) \, , \\
\p_a \p_b e^{\tfrac{\mu^2 x^2}{4}} ~ &= ~ e^{\tfrac{\mu^2 x^2}{4}} \left(\p_a \p_b + \mu^2 x_{(a}\p_{b)} + \dfrac{1}{2} \mu^2 \eta_{ab} + \dfrac{1}{4} \mu^4 x_a x_b\right) \, .
\end{align}
In what follows, we will ignore the last terms quadratic in $\mu x_a$. This allows us to identify
\begin{align}
\label{eq:symmetricdelta}
e^{\tfrac{\mu^2 x^2}{4}} \Delta^{-1}_{L,ab}e^{-\tfrac{\mu^2 x^2}{4}} ~ = ~ e^{-\tfrac{\mu^2 x^2}{4}} \Delta^{-1}_{R,ab} e^{\tfrac{\mu^2 x^2}{4}} ~ = ~ -\eta_{ab}\bigr(\p^2-\tfrac{1}{2}\mu^2(\lambda-1)\bigr) + \p_a\p_b \, .
\end{align}
This gives a symmetric representation. Moreover, conveniently enough, it also removes all first derivatives and first order terms in $\mu x_a$. Therefore, the field re-definitions to achieve this symmetric representation are given by the following transformations
\begin{align}
\label{eq:transform1}
\mathcal{K} ~ &\to ~ e^{-\tfrac{\mu^2 x^2}{4}}\mathcal{K} \, , \\
\label{eq:transform2}
\mathfrak{h}_{ab} ~ &\to ~ e^{\tfrac{\mu^2 x^2}{4}}\mathfrak{h}_{ab} \, .
\end{align}
Under this field redefinition, it is worth noting that $\Delta^{-1}_{abcd}$ does not transform because of the anti-symmetry of the derivatives. What remains is $\Delta^{-1}$. We first note the part of the action quadratic in $\mathcal{K}$:
\begin{eqnarray}
S_{\mathcal{K}} ~ = ~ \dfrac{1}{4} \int\ed^2 x \mathcal{K} \Delta^{-1} \mathcal{K} ~ = ~ \dfrac{1}{4} \int \ed^2 x \gap e^{-\tfrac{\mu^2 x^2}{2}} \mathcal{K} \left(-\p^2 + \dfrac{1}{2} \mu^2 x^c \p_c + 2\mu^2\right) \mathcal{K} \, .
\end{eqnarray}
While there was no first derivative term to begin with, the field redefinition has generated one. The first derivatives in the terms coupling $\mathcal{K}$ and $\mathfrak{h}$ have essentially been traded for a first derivative in the pure $\mathcal{K}$ action. The exponent can be removed in the near-horizon approximation 
\begin{equation}
    e^{-\tfrac{\mu^2 x^2}{2}} ~ = ~ 1 + \mathcal{O}(\mu^2 xy) \, .
\end{equation} 
The use of the exponential in the field redefinition may appear to be at odds with the recurring use of the approximation $\mu x_a \ll 1$. A remark on this is in order. But before that, we will first massage the said first derivative. The integral with the first derivative can be rewritten as
\begin{eqnarray}
\int \ed^2 x \mathcal{K} \mu^2 x^c \p_c \mathcal{K} ~ = ~ \dfrac{1}{2} \mu^2 \int \ed^2 x x^c \p_c \left(\mathcal{K}^2\right) \, .
\end{eqnarray}
Integration by parts with vanishing boundary conditions yields
\begin{eqnarray}
\dfrac{1}{2} \mu^2 \int \ed^2 x x^c \p_c \left(\mathcal{K}^2\right) ~ = ~ - \dfrac{1}{2} \mu^2 \int \ed^2 x \mathcal{K}^2 \p_c x^c ~ = ~ - \mu^2 \int \ed^2 x \mathcal{K}^2 \, .
\end{eqnarray}
The first derivative is therefore traded for another mass term. The complete even parity graviton action near the horizon is now given by
\begin{eqnarray}
\label{eq:finalgravact1}
S ~ = ~ \dfrac{1}{4} \int\ed^2 x \biggr(\mathfrak{h}^{ab} \Delta^{-1}_{abcd} \mathfrak{h}^{cd} + \mathfrak{h}^{ab} \Delta^{-1}_{ab} \mathcal{K} + \mathcal{K} \Delta^{-1}_{ab} \mathfrak{h}^{ab} + \mathcal{K} \Delta^{-1} \mathcal{K}\biggr) \, ,
\end{eqnarray}
with
\begin{subequations}
\label{eq:finalop}
\begin{align}
\Delta^{-1} ~ &= ~ - \p^2 + \mu^2 \, , \\
\Delta^{-1}_{ab} ~ &= ~ - \eta_{ab} \left(\p^2 - \dfrac{1}{2} \mu^2 \left(\lambda - 1\right)\right) + \p_a \p_b \, , \\
\Delta^{-1}_{abcd} ~ &= ~ \dfrac{1}{2} \mu^2 \left(\eta_{ac} x_{[b} \p_{d]} + \eta_{bd} x_{[a} \p_{c]} + \eta_{ab} x_{(c} \p_{d)} - \eta_{cd} x_{(a} \p_{b)}\right) \nonumber \\
&\qquad + \dfrac{\mu^2 \left(\lambda + 1\right)}{2} \bigr(\eta_{ab}\eta_{cd}-\eta_{ac}\eta_{bd}\bigr) \, . \label{eqn:finalTensorOp}
\end{align}
\end{subequations}
What we see is that $\Delta^{-1}$ and $\Delta^{-1}_{abcd}$ remain unchanged. It was important that we were able to remove the exponential $\exp\left(\pm\tfrac{\mu^2 x^2}{2}\right)$ to linear order in $\mu x_a$. A word about the consistency of the approximation is now in order. In the small $\mu x_a$ limit, the fields themselves do not transform as can be seen from \eqref{eq:transform1} and \eqref{eq:transform2}. The transformations are only nontrivial at higher orders in $\mu x^a$. So we began with an infinitesimally small transformation, and nevertheless transformed the operators. Therefore, the difference between the first derivatives in $\Delta^{-1}_{L,ab}$ and $\Delta^{-1}_{R,ab}$ and the effective mass in $\Delta^{-1}_{ab}$ is vanishing to linear order in $\mu x^a$. The transformation we chose is merely one consistent choice, but the underlying structure that $\Delta^{-1}_{L,ab},\Delta^{-1}_{R,ab}$ equal $\Delta^{-1}_{ab}$ up to infinitesimal differences holds regardless of the transformation chosen.

This quadratic action is now in a convenient enough form to allow for inversion of the above operators to find the graviton propagator. All terms linear in $\mu x_a$ have dropped out, except in $\Delta^{-1}_{abcd}$. We shall see that in the following subsection that they do not contribute either.

\subsection{The graviton propagator}
\label{sec:lightconeprop}
In this subsection we derive the complete graviton propagator for the even action on the Schwarzschild horizon near the horizon. The propagator is found from the Green's function of the equation of motion. The equations of motion of the graviton are now
\begin{subequations}\label{eqn:eom}
\begin{align}
\Delta^{-1}_{abcd}\mathfrak{h}^{cd} + \Delta^{-1}_{ab}\mathcal{K} ~ &= ~ 0 \, , \\
\Delta^{-1}_{ab}\mathfrak{h}^{ab} + \Delta^{-1}\mathcal{K} ~ &= ~ 0 \, ,
\end{align}
\end{subequations}
with the operators defined in \eqref{eq:finalop}. These can be written in the following matrix form
\begin{eqnarray}
\begin{pmatrix}
\Delta^{-1}_{abcd}& \Delta^{-1}_{ab}\\
\Delta^{-1}_{cd} & \Delta^{-1}
\end{pmatrix}
\begin{pmatrix}
\mathfrak{h}^{cd} \\
\mathcal{K}
\end{pmatrix} ~ = ~ 0 \, .
\end{eqnarray} 
Then the Green's function of this matrix differential equation is defined by 
\begin{eqnarray}
\label{eq:matrixpropdef}
\begin{pmatrix}
\Delta^{-1}_{abcd}& \Delta^{-1}_{ab}\\
\Delta^{-1}_{cd} & \Delta^{-1}
\end{pmatrix}
\begin{pmatrix}
\mathcal{P}^{cdef}& \mathcal{P}^{cd}\\
\mathcal{P}^{ef} & \mathcal{P}
\end{pmatrix}
~ = ~ \begin{pmatrix}
\delta^{ef}_{ab}& 0\\
0 & 1
\end{pmatrix}
\delta^{(2)}\left(x-x'\right) \, .
\end{eqnarray}
The Green's function matrix is symmetric owing to the symmetrised operators \eqref{eq:finalop}. There are three propagators. The first is $\mathcal{P}^{cdef}$, which is the propagator of $\mathfrak{h}_{cd}\to\mathfrak{h}_{ef}$. The function $\mathcal{P}_{ab}$ corresponds to $\mathfrak{h}^{ab}\to \mathcal{K}$ whereas $\mathcal{P}$ corresponds to $\mathcal{K}\to\mathcal{K}$. From equation \eqref{eq:matrixpropdef} we find that all individual propagators are defined by
\begin{subequations}\label{eq:propdefs}
\begin{align}
\label{eq:propdef1}
\Delta^{-1}_{abcd}\mathcal{P}^{cdef} +  \Delta^{-1}_{ab}\mathcal{P}^{ef} ~ &= ~ \delta^{ef}_{ab}\delta^{(2)}\left(x-x'\right) \, , \\
\label{eq:propdef2}
\Delta^{-1}_{ab}\mathcal{P}^{abcd} + \Delta^{-1}\mathcal{P}^{cd} ~ &= ~ 0 \, , \\
\label{eq:propdef3}
\Delta^{-1}_{abcd}\mathcal{P}^{cd} + \Delta^{-1}_{ab}\mathcal{P} ~ &= ~ 0 \, , \\
\label{eq:propdef4}
\Delta^{-1}_{ab}\mathcal{P}^{ab} + \Delta^{-1}\mathcal{P} ~ &= ~ \delta^{(2)}\left(x-x'\right) \, .
\end{align}
\end{subequations}
The coupling between the propagators is naturally owed to the coupling between the tensorial and scalar fields in the quadratic action \eqref{eq:finalgravact1}. To find the propagators, we first need to find the inverses of $\Delta^{-1}$, $\Delta^{-1}_{ab}$, and $\Delta^{-1}_{abcd}$. These inverses are defined as usual:
\begin{align}
\label{eq:inversedef2}
\Delta^{-1}_{abcd}\Delta^{cdef} ~ &= ~ \delta^{ef}_{ab}\delta^{(2)}\left(x-x'\right) \, , \\
\label{eq:inversedef1}
\Delta^{-1}_{ab}\Delta^{bc} ~ &= ~ \delta^{c}_{a} \delta^{(2)}\left(x-x'\right) \, , \\
\label{eq:inversedef3}
\Delta^{-1}\Delta ~ &= ~ \delta^{(2)}\left(x-x'\right) \, .
\end{align}

\paragraph{The inverse of $\Delta^{-1}$} This is the easiest of the lot, and requires us to solve
\begin{eqnarray}
(-\p^2+\mu^2)\Delta\left(x;x'\right) ~ = ~ \delta^{(2)}\left(x-x'\right) \, .
\end{eqnarray}
Using the Fourier transforms
\begin{align}
\Delta\left(x;x'\right) ~ &= ~ \dfrac{1}{(2\pi)^2} \int \ed^2 k \gap e^{ik_a (x-x')^a} \Delta(k) \, , \\
\delta^{(2)}\left(x-x'\right) ~ &= ~ \dfrac{1}{(2\pi)^2}\int\ed^2p\gap e^{ik_a (x-x')^a} \, ,
\end{align}
the solution for the inverse is
\begin{eqnarray}
\Delta(k) ~ = ~ \dfrac{1}{k^2+\mu^2}
\end{eqnarray}
where $k^2 = \eta^{ab} k_a k_b$. This is the Klein-Gordon propagator with an effective mass $\mu^2$ arising from the two-sphere reduction of the Schwarzschild metric. In contrast to the flat space eikonal approximation, we see that the effective reduction to two dimensions near the horizon provides a natural infrared regulator given by the Schwarzschild radius. In the soft limit, this cures infrared divergences. However, the four dimensional graviton of course remains massless.

\paragraph{The inverse of $\Delta^{-1}_{abcd}$} The operator is given in \eqref{eqn:finalTensorOp}. In order to find the inverse we start with the following ansatz
\begin{eqnarray}
\label{eq:ansatz1}
\Delta^{abcd} ~ = ~ T\left(x-x'\right) \eta^{ab} \eta^{cd} + Q\left(x-x'\right) \eta^{a(c}\eta^{d)b} \, .
\end{eqnarray}
This is the most general form that respects the symmetries of the graviton $\mathfrak{h}_{ab} = \mathfrak{h}_{ba}$, and time translation symmetry $x\to ax, y\to y/a$ up to linear order in $\mu x_a$. Here we used time translation symmetry to require that $\Delta^{abcd}$ transforms at most as a tensor under time translations, so that the only allowed tensor structures are $\eta_{ab}\eta_{cd}, \mu^2 \eta_{ab}x_c x_d$ and $\mu^4 x_a x_b x_c x_d$ with any index permutation. We included the $\mu^2$ to have dimensionless tensor structures. Now the latter two can be removed up to linear order in $\mu x_a$, so that only $\eta_{ab}\eta_{cd}$ remains. Then \eqref{eq:ansatz1} gives the most general form that respects symmetries of the problem. There is another allowed dimensionless tensor as well: $\tfrac{x_a x_b}{x^2}$. This tensor is however not regular over the entire manifold. The irregularities of usual momentum space poles determine the mass-shell, whereas this tensor contains a pole that would be a coordinate singularity. In Kruskal-Szekers coordinates there is no singularity at $x^2=0$. This are not the usual $1/r$ singularity of spherical coordinates either, as that would correspond to $1/(x^2+2R^2)$ instead. Therefore, we forbid this tensor, leaving behind \eqref{eq:ansatz1} as the most general form.

Inserting \eqref{eq:ansatz1} in the definition of the inverse \eqref{eq:inversedef2} results in
\begin{align}
\nonumber
\dfrac{2}{\mu^2(\lambda+1)}\delta^{ef}_{ab}\delta^{(2)}(x-x') ~ &= ~ - Q \delta^{ab}_{ef} + \eta_{ab} \eta^{ef} \bigr(x^c \p_c T - \left(\lambda + 1\right) \left(T + Q\right) \bigr) \nonumber \\
&\qquad - \eta^{ef} x_{(a}\p_{b)} \left(2 T + Q\right) + \eta_{ab}x^{(e} \p^{f)}Q \nonumber \\ 
\label{eq:inverscalc1}
&\qquad + \dfrac{1}{2}\bigr(\delta^e_a x_{[b}\p^{f]}+\delta^f_a x_{[b}\p^{e]}+\delta^f_b x_{[a}\p^{e]}+\delta^e_b x_{[a}\p^{f]}\bigr)Q \, .
\end{align}
Neither the anti-symmetric terms nor the terms containing the metric tensor can reduce to the delta function as required by the left hand side of this equation. Therefore, we find
\begin{eqnarray}
Q ~ = ~ - \dfrac{2}{\mu^2(\lambda+1)}\delta^{(2)}\left(x-x'\right) \, .
\end{eqnarray}
This is a peculiar result; the inverse of the operator is instantaneous. Nevertheless it is allowed. The two-dimensional operator on the light-cone \eqref{eqn:finalTensorOp} does not contain any second derivatives $\p^2$. So it is not expected to have a momentum dependence. Instantaneous operators are not familiar in other gauge choices. Presumably this is a peculiarity of the choice of gauge. This operator is not the physical propagator; we denoted that by $\mathcal{P}^{abcd}$. That full propagator will of course have momentum dependence. Notice that the inclusion of the irregular $\tfrac{x_a x_b}{x^2}$ tensor would not have lead to a $\delta^{ab}_{ef}$ either and the inverse would still have been instantaneous. To proceed further, we make a guess that $T=-Q$ in \eqref{eq:ansatz1}. Inserting this into the definition of the inverse shows that \eqref{eq:ansatz1} is only a valid inverse if the following equation is satisfied:
\begin{align}
\label{eq:inverscheck}
0 ~ &\stackrel{?}{=} ~ \biggr(\eta_{ab}\eta^{ef}x^c\p_c-\eta^{ef}x_{(a}\p_{b)}-\eta_{ab}x^{(e}\p^{f)} \nonumber \\
&\qquad - \dfrac{1}{2}\bigr(\delta^e_a x_{[b}\p^{f]}+\delta^f_a x_{[b}\p^{e]}+\delta^f_b x_{[a}\p^{e]}+\delta^e_b x_{[a}\p^{f]}\bigr)\biggr)\delta^{(2)}\left(x-x'\right) \, .
\end{align}
At first glance this does not seem to vanish. However should we interpret the Dirac delta function as a distribution obeying
\begin{eqnarray}
x_a\p_b \delta^{(2)}\left(x-x'\right) ~ = ~ - \eta_{ab} \delta^{(2)} \left(x-x'\right) \, ,
\end{eqnarray}
and recognise that the right hand side of \eqref{eq:inverscheck} appears inside a two-dimensional integral, we find 
\begin{align}
\nonumber
&\eta_{ab}\eta^{ef}x^c\p_c\delta^{(2)}\left(x-x'\right) - \eta^{ef} x_{(a}\p_{b)} \delta^{(2)} \left(x-x'\right) - \eta_{ab} x^{(e}\p^{f)} \delta^{(2)} \left(x-x'\right) ~ = ~ \\
\nonumber
&\qquad= ~ - 2 \eta_{ab} \eta^{ef} \delta^{(2)} \left(x-x'\right) + \eta_{ab} \eta^{ef} \delta^{(2)}\left(x-x'\right) + \eta_{ab}\eta^{ef}\delta^{(2)}\left(x-x'\right) \nonumber \\
&\qquad= ~ 0 \, . \nonumber
\end{align}
A similar analysis for $x_{[b}\p^{f]}$ results in
\begin{align}
x_{[b}\p^{f]}\delta^{(2)}\left(x-x'\right) ~ &= ~ \dfrac{1}{2} \left(x_b \p^f - x^f \p_b\right) \delta^{(2)} \left(x-x'\right) \nonumber \\
&= ~ - \dfrac{1}{2} \left(\delta^{f}_b - \delta^f_b\right) \delta^{(2)} \left(x-x'\right) \nonumber \\
&= ~ 0 \, . \nonumber
\end{align}
Of course, all these manipulations are legitimate only under the integration over all $x,y$. Therefore, we now find that the inverse is given by
\begin{eqnarray}
\Delta_{abcd} ~ = ~ \dfrac{1}{\mu^2 \left(\lambda + 1\right)} \left(2 \eta_{ab} \eta_{cd} - \eta_{ac} \eta_{bd} - \eta_{ad} \eta_{bc}\right) \delta^{(2)}\left(x-x'\right) \, .
\end{eqnarray}
In momentum space the $\delta^{(2)}\left(x-x'\right)$ can simply be replaced by unity, giving the final tensor
\begin{eqnarray}
\label{eq:deltainv}
\Delta_{abcd} ~ = ~ \dfrac{1}{\mu^2\left(\lambda + 1\right)} \left(2 \eta_{ab} \eta_{cd} - \eta_{ac} \eta_{bd} - \eta_{ad} \eta_{bc}\right) \, .
\end{eqnarray}
With the inverses of $\Delta^{-1}$ and $\Delta^{-1}_{abcd}$, the propagators can be determined from \eqref{eq:propdefs}.

\subsubsection{The $\mathcal{K}$ propagator}
\label{sec:scalarprop}
We first address to find the easiest propagator: the scalar $\mathcal{K}$ propagator. From \eqref{eq:propdef3} and \eqref{eq:propdef4} we can find that in momentum space the propagator is given by
\begin{eqnarray}
\mathcal{P}_K ~ = ~ \dfrac{1}{\Delta^{-1} - \Delta^{-1}_{ab} \Delta^{abcd} \Delta^{-1}_{cd}} \, .
\end{eqnarray}
The second term in the denominator can be expanded as
\begin{eqnarray}
\Delta^{-1}_{ab} \Delta^{abcd} \Delta^{-1}_{cd} ~ = ~ \dfrac{\lambda - 1}{\lambda + 1} \left(\mu^2\left(\lambda - 1\right) + 2 k^2\right) \, .
\end{eqnarray}
Together with $\Delta^{-1} = k^2 + \mu^2$, we have for the $\mathcal{K}$ propagator that
\begin{eqnarray}\label{eqn:Kprop}
\mathcal{P}_K ~ = ~ - \dfrac{\lambda + 1}{\lambda - 3} \dfrac{1}{k^2 + \mu^2 \lambda} \, .
\end{eqnarray}
This obviously resembles the Klein-Gordon propagator with an effective mass $\mu\sqrt{\lambda}$. We will have more to say about the effective graviton mass at the end of this section. The prefactor, however, deserves attention. It is not well defined for $\lambda = \ell^2 + \ell + 1 = 3$ which corresponds to $\ell=1$. Moreover, there is a flip in sign for $\ell=0$, hinting at negative norm states. That the $\ell=0$ and $\ell=1$ cases are special is familiar \cite{Zerilli:1970se,Nagar:2005ea}. The equations of motion are not invertible for $\ell=1$, so a gauge redundancy is to be expected. The $\ell=0$ corresponds to a change in black hole mass since it is spherically symmetric and Birkhoff's theorem states that the $\ell=0$ perturbation is also of a Schwarzschild form \cite{Zerilli:1970se}. Therefore this also suggests leftover gauge redundancies since at $\ell=0$ we have many more degrees of freedom than just the mass perturbation. Similarly $\ell=1$ has has more degrees of freedom than can be reduced to a form with only the odd-parity graviton components $h_{aA}$ \cite{Zerilli:1970se}. So the $\ell=1$ mode may be thought of as describing a rotating gravitational field. It has to be said, however, that this intuition of \cite{Zerilli:1970se} is based on a classical analysis. Nevertheless, that these modes require additional care and indicate gauge redundancies is clear. For the four-point function of interest in this paper, fortunately, this propagator will not contribute as we will see in further sections. Therefore, we leave further analysis of the $\mathcal{K}$ propagator for future work.

\subsubsection{The $\mathfrak{h}_{ab}$ propagator}
\label{sec:flatgravprop}
For the purposes of the scalar four-point function of interest, the propagator of $\mathfrak{h}_{ab}$ will turn out to be the most important one. To find $\mathcal{P}^{abcd}$, we begin by using \eqref{eq:propdef1} and \eqref{eq:propdef2} to write
\begin{eqnarray}
\label{eq:propflatgravdef}
\mathcal{P}^{abcd} ~ &=& ~ \left(\mathcal{L}^{ab}_{\mgap ef}\right)^{-1} \Delta^{efcd} \, , 
\end{eqnarray}
where we defined
\begin{eqnarray}
\label{eq:propflatgravdef2}
\mathcal{L}^{ab}_{\mgap ef} ~ &\coloneqq& ~ \delta^{ab}_{ef}-\Delta^{abcd}\Delta^{-1}_{cd}\Delta\Delta^{-1}_{ef} \, .
\end{eqnarray}
We now write down $\Delta^{-1}_{ab}$ in momentum space:
\begin{eqnarray}
\Delta^{-1}_{ab} ~ = ~ \eta_{ab}\left(k^2 + \dfrac{1}{2} \mu^2 \left(\lambda - 1\right)\right) - k_a k_b \, .
\end{eqnarray}
Using this, $\mathcal{L}^{ab}_{\mgap ef}$ is given by its definition in \eqref{eq:propflatgravdef2}. Expanding the tensor structure results in
\begin{align}
\mathcal{L}^{ab}_{\mgap ef} ~ &= ~ \delta^{ab}_{ef} - \dfrac{\Delta\left(k\right)\left(k^2 + \frac{1}{2} \mu^2 \left(\lambda - 1\right)\right) \left(\lambda - 1\right)}{\lambda + 1} \eta^{ab} \eta_{ef} - \dfrac{2\Delta\left(k\right) \left(k^2 + \frac{1}{2} \mu^2\left(\lambda - 1\right)\right)}{\mu^2 \left(\lambda + 1\right)} k^a k^b \eta_{ef} \nonumber \\
\nonumber
&\qquad + ~ \dfrac{\Delta\left(k\right) \left(\lambda - 1\right)}{\left(\lambda + 1\right)} \eta^{ab} k_e k_f + \dfrac{2\Delta\left(k\right)}{\mu^2 \left(\lambda + 1\right)} k^a k^b k_e k_f \\
\label{eq:formL1}
&\eqqcolon ~ \delta^{ab}_{ef} - L_1 \eta^{ab} \eta_{ef} - L_2 k^a k^b \eta_{ef} + L_3 \eta^{ab} k_e k_f + L_4 k^a k^b k_e k_f \, .
\end{align}
where we defined four functions $L_i$ that can be seen by comparing the two equalities above. To invert \eqref{eq:formL1} we make the ansatz
\begin{eqnarray}
\label{eq:ansatzL1}
(\mathcal{L}^{ab}_{\mgap ef})^{-1} ~ = ~ \delta^{ab}_{ef} - Q_1 \eta^{ab} \eta_{ef} - Q_2 k^a k^b \eta_{ef} + Q_3 \eta^{ab} k_e k_f + Q_4 k^a k^b k_ek_f \, ,
\end{eqnarray}
such that
\begin{eqnarray}
\mathcal{L}^{ab}_{\mgap ef}(\mathcal{L}^{ef}_{\mgap cd})^{-1} ~ = ~ \delta^{ab}_{cd} \, .
\end{eqnarray}
Using the definition of $\mathcal{L}$ in this equation yields a neat relation between all $L_i$ and $Q_i$. This relation can be written in matrix form as follows
\begin{eqnarray}\label{eqn:matrixrelation}
\begin{pmatrix}
\mathcal{M} & 0 \\
0 & \mathcal{M}
\end{pmatrix}
\begin{pmatrix}
Q_1 \\ Q_2
\\
Q_3 \\ Q_4
\end{pmatrix}
~ = ~
\begin{pmatrix}
L_1 \\ L_2
\\
L_3 \\ L_4
\end{pmatrix} \, ,
\end{eqnarray}
where
\begin{align}
    \mathcal{M} ~ &= ~ \begin{pmatrix}
    -1+2L_1-L_3k^2 & L_1 k^2-L_3 k^4 \\
    2L_2-L_4 k^2 & -1+L_2 k^2-L_4 k^4
    \end{pmatrix} \nonumber \\
    &= ~ \frac{\Delta\left(k\right)}{\lambda+1}\begin{pmatrix}
    -2k^2+\mu^2\lambda(\lambda-3)  & \frac{1}{2}\mu^2k^2(\lambda-1)^2 \\
    \frac{2}{\mu^2}(k^2+\mu^2(\lambda-1)) & -2k^2-\mu^2(\lambda+1)\end{pmatrix} \, .
\end{align}
The matrix relation \eqref{eqn:matrixrelation} is block diagonal and therefore, an inverse of the the $2 \times 2$ matrix $\mathcal{M}$ suffices to invert the matrix relation. An explicit calculation shows that the matrix determinant is given by
\begin{align}
|\mathcal{M}| ~ &= ~ - \dfrac{\lambda-3}{\lambda+1} \left(k^2 + \mu^2\lambda\right) \Delta\left(k\right) \nonumber \\
&= ~ \dfrac{\Delta\left(k\right)}{\mathcal{P}_K} \, .
\end{align}
This determinant vanishes for $\lambda=3$; a problem similar to what was evident in the inverse propagator of $\mathcal{K}$. This shows that the equations of motion are indeed not orthogonal at $\lambda=3$ and that there is residual gauge freedom. Furthermore, the observations of the zeros and poles of the prefactor of $\mathcal{P}$ are again relevant. We can now easily write down the inverse matrix to be
\begin{eqnarray}
\mathcal{M}^{-1} ~ = ~ \dfrac{\mathcal{P}_K}{\lambda + 1} \begin{pmatrix}
- 2 k^2 - \mu^2 \left(\lambda + 1\right)  &  - \frac{1}{2} \mu^2 k^2 \left(\lambda - 1\right)^2 \\
 - \frac{2}{\mu^2} \left(k^2 + \mu^2 \left(\lambda - 1\right)\right) & - 2 k^2 + \mu^2 \lambda \left(\lambda - 3\right)
\end{pmatrix} \, .
\end{eqnarray}
We now recognize that $2L_1 = \mu^2 \left(\lambda - 1\right) L_2$, and that $2 L_3 = \mu^2 \left(\lambda - 1\right) L_4$. Thus we only need to calculate
\begin{eqnarray}
\mathcal{M}^{-1}\begin{pmatrix}
\mu^2 \left(\lambda - 1\right) \\
2
\end{pmatrix}
~ = ~ - \dfrac{\mathcal{P}_K}{\Delta} \begin{pmatrix}
\mu^2 \left(\lambda - 1\right) \\
2
\end{pmatrix} \, .
\end{eqnarray}
What we see is in fact an eigenvalue equation, so that the $2 \times 2$ matrix $\mathcal{M}$ is simply reduced to only a scalar in this particular case. The corresponding eigenvalue is exactly given by the negative of the reciprocal of the determinant. This allows us to easily identify the solution to the inverse
\begin{eqnarray}
Q_i ~ = ~ - \dfrac{\mathcal{P}_K}{\Delta} L_i \, .
\end{eqnarray}
Inserting $Q_i$ into \eqref{eq:ansatzL1} results in
\begin{align}
\label{eq:mathcalLinverse}
(\mathcal{L}^{ab}_{\mgap ef})^{-1} ~ &= ~ \delta^{ab}_{ef} + \dfrac{\mathcal{P}_K \left(k^2 + \frac{1}{2} \mu^2 \left(\lambda - 1\right)\right) \left(\lambda - 1\right)}{\lambda + 1} \eta^{ab} \eta_{ef} + \dfrac{2\mathcal{P}_K \left(k^2 + \frac{1}{2} \mu^2 \left(\lambda - 1\right)\right)}{\mu^2 \left(\lambda + 1\right)} k^a k^b \eta_{ef} \nonumber \\
&\qquad - ~ \dfrac{\mathcal{P}_K \left(\lambda - 1\right)}{\left(\lambda + 1\right)} \eta^{ab} k_e k_f - \dfrac{2 \mathcal{P}_K}{\mu^2 \left(\lambda + 1\right)} k^a k^b k_e k_f \, .
\end{align}
We can now insert the above solution for $(\mathcal{L}^{ab}_{\mgap ef})^{-1}$ into \eqref{eq:propflatgravdef} to find the propagator. It is worth noting that the detailed mathematical calculations result in what one may have expected on physical grounds based on symmetry. The underlying structure merely consists of eigentensors, elegantly displaying the symmetry structure of the graviton. 

Finally, inserting \eqref{eq:mathcalLinverse} into \eqref{eq:propflatgravdef} results in
\begin{eqnarray}
\label{eq:finalprop}
\mathcal{P}^{abcd} ~ = ~ \Delta^{abcd}+\mathcal{P}_K \proj^{ab}\proj^{cd} \, ,
\end{eqnarray}
with
\begin{eqnarray}\label{eqn:projOper}
\proj^{ab} ~ \coloneqq ~ \dfrac{\lambda-1}{\lambda+1} \eta^{ab} + \dfrac{2 k^a k^b}{\mu^2 \left(\lambda + 1\right)} \, .
\end{eqnarray}
Here $\Delta^{abcd}$ is the same quantity defined in \eqref{eq:deltainv}. The $\proj^{ab}$ operators are the projection operators for the massive graviton. In fact, observing the scalar part $\mathcal{P}_k$, there is clearly a pole at $-\mu^2\lambda$. Therefore, the tensor field $\mathfrak{h}_{ab}$ has also become massive. Note that the four-dimensional graviton is still massless, it is only the two-dimensional reduction that gains an effective mass. This is an important result, one that provides for a natural infrared regulator. This is in contrast to the flat space eikonal limit where a regulator was necessarily inserted by hand \eqref{eq:amplitudeMinkowski}.

A projection operator similar to $\proj^{ab}$ was observed for the massive graviton in \cite{Hinterbichler:2011tt}, where unlike in the present case the mass was owed to a direct insertion of mass terms in the action. While the resemblance between the propagator in \cite{Hinterbichler:2011tt} and \eqref{eq:finalprop} is notable, the most significant difference is the $\Delta^{abcd}$ term; this quantity still has no momentum dependence in the present case. It represents the effect of the mass of the spin-2 particle in two dimensions. This mass generates a curvature and $\Delta^{abcd}$ embeds this curvature into the propagator. This is also the reason that it has no momentum dependence: the curvature contribution of the tensor modes is constant in the Lagrangian, there is no $k^2$ contributing to the two-dimensional light-cone action in \eqref{eq:finalgravact1}. The second term involving the $\mathcal{K}$ mode carries momentum dependence, and implicitly also gives it to the tensor modes.

\subsubsection{Propagator for $\mathfrak{h}_{ab}\to\mathcal{K}$} The propagator that couples the tensorial mode in the graviton to the scalar one can be found from \eqref{eq:propdef2} to be given by
\begin{align}\label{eqn:habToKProp}
\mathcal{P}^{ab} ~ &= ~ - \Delta \Delta^{-1}_{cd} \mathcal{P}^{cdab} \nonumber \\
&= ~ - \mathcal{P}_K \proj^{ab} \, .
\end{align}
The operator $\proj^{ab}$ is still the same projection operator as before, defined in \eqref{eqn:projOper}. This propagator resembles that of a massive spin-1 field; given that this it couples a spin-2 field with a scalar field, this may not be a surprise. For the purposes of this article, this plays no role as we will see in the upcoming sections.

\subsection{The scalar propagator}
The propagator for the scalar field is relatively straight forward to find, but first requires a reduction of the matter action on the two sphere. We begin with the kinetic term in \eqref{eqn:matterAction}: 
\begin{eqnarray}
S ~ = ~ \dfrac{1}{2} \int\ed^4 x \sqrt{-g} \, \phi \, \Box \, \phi \, .
\end{eqnarray}
For the box operator, we write
\begin{eqnarray}
\Box ~ = ~ \tilde{\Box} + V^a \tilde{\nabla}_a + \dfrac{1}{r^2} \Delta_{\Omega} \, ,
\end{eqnarray}
where, as before, the tilde operators correspond to the light-cone. Furthermore, we used the residual curvature tensor $V_a$, which is defined to be $V_a=2\p_a\log r$. We now expand the scalar field in spherical harmonics:
\begin{eqnarray}
\phi ~ = ~ \sum\limits_{\ell,m} \phi_{\ell m} \, Y^m_{\ell} \, .
\end{eqnarray}
We again make use of $\Delta_{\Omega}Y_{\ell}^m = - \ell(\ell+1)Y_{\ell}^m$ and the orthogonality relation \eqref{eq:sphericalharmonicsorthogonality} to write the action as
\begin{eqnarray}
S_m ~ = ~ - \dfrac{1}{2} \sum\limits_{\ell,m} \int \ed^2 x \, A\left(r\right) r^2 \phi_{\ell m} \left(- \tilde{\Box} - V^a \tilde{\nabla}_a + \dfrac{\ell(\ell+1)}{r^2}\right) \phi_{\ell,m} \, .
\end{eqnarray}
We now perform the Weyl transformation as was done in \secref{subsec:WeylRescaling}, along with a field redefinition for the scalar as $\phi \to r\tilde{\phi}$. This removes the Jacobian, and yields
\begin{eqnarray}
S_m ~ = ~ - \dfrac{1}{2} \sum\limits_{\ell,m} \int \ed^2 x \gap \phi_{\ell m} \left(- \p^2 + \dfrac{1}{r} \left(\p^2 r\right) + \dfrac{A\left(r\right) \ell\left(\ell + 1\right)}{r^2}\right) \phi_{\ell,m} \, .
\end{eqnarray}
The field redefinition also removes the first derivatives, and replaces them with a potential $\sim \left(\p^2 r\right)$. Finally, we insert that on the horizon $r\sim R$ and $\p^2 r\sim -\tfrac{1}{2R}$, as derived in \appref{app:defs} in the approximation $x,y\ll R$, to find the following:
\begin{eqnarray}
\label{eq:matteraction2}
S_m ~ = ~ - \dfrac{1}{2} \sum\limits_{\ell,m} \int \ed^2 x \, \phi_{\ell m} \bigr(- \p^2 + \mu^2 \lambda\bigr) \phi_{\ell m} \, ,
\end{eqnarray}
where we defined $\lambda=\ell^2+\ell+1$ and $\mu=1/R$ as before. The propagator is easily found by the Fourier transformation
\begin{align}
\phi(x) ~ &= ~ \dfrac{1}{\left(2\pi\right)^2} \int \ed^2 p \, e^{i p_a x^a} \phi(p) \, , \\
\phi(p) ~ &= ~ \int \ed^2 x \, e^{-i p_a x^a} \phi(x) \, .
\end{align}
The inverse propagator can be read off from \eqref{eq:matteraction2} as $\p^2 - \mu^2\lambda$. Under the chosen conventions for the Fourier transform, the propagator is directly found by substituting $\p_a=ip_a$. This results in
\begin{eqnarray}
\label{eq:scalarprop1}
\mathcal{P}_{\phi} ~ = ~ - \dfrac{1}{p^2 + \mu^2 \lambda - i \epsilon} \, .
\end{eqnarray}
This is the propagator for a fixed $(\ell,m)$ wave and spherical symmetry of the background ensures that different partial waves do not interact below Planck scale. 

In what is to follow, we will assume $\p_A\to 0$ since the external particles have insignificant transverse momenta when the black hole is larger than Planck size. We will first analyse the consequence of this for the vertex, before proceeding to compute the amplitudes of interest.

\subsection{The interaction vertex}
\begin{figure}[h!]
\centering
	\includegraphics[scale=0.5]{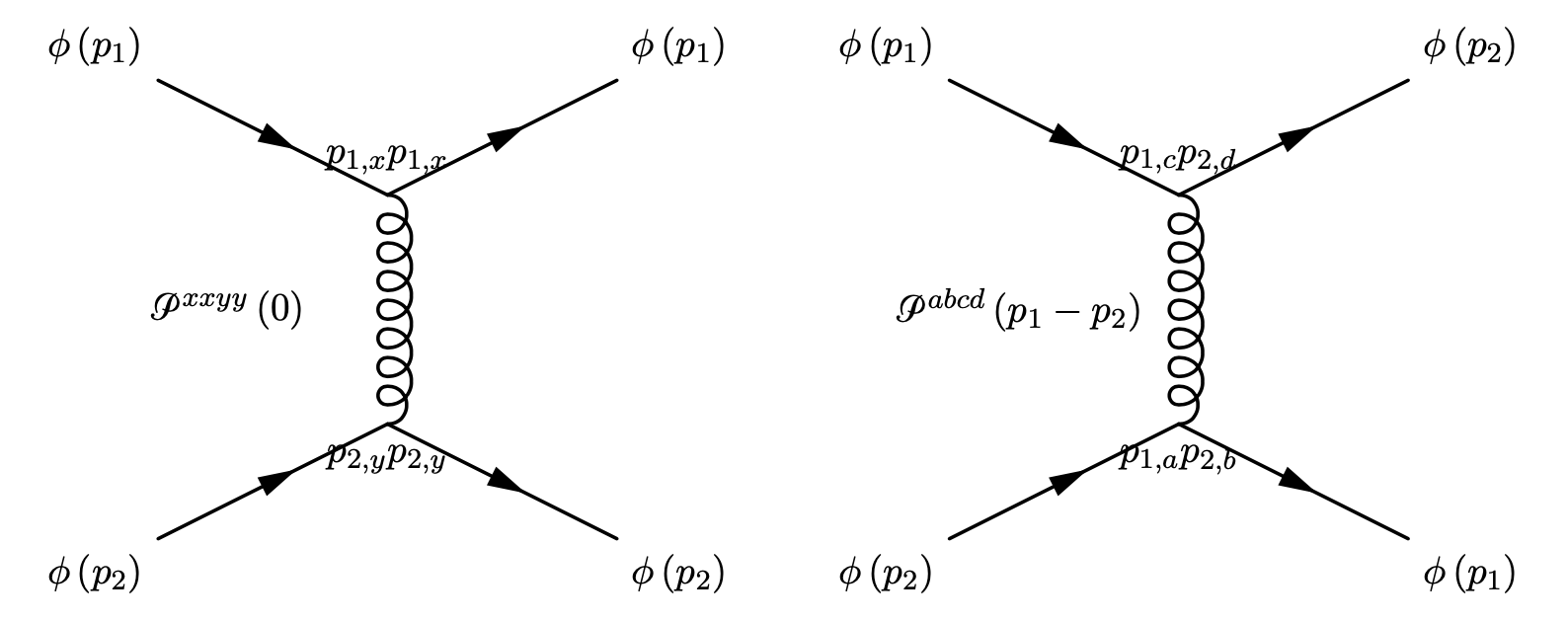}	
    \caption{The diagram on the left, say $\mathcal{M}_{soft}$, is the soft one mediated by the propagator $\mathcal{P}^{abcd}$ given in \eqref{eq:finalprop}, with no momentum exchanged. The diagram on the right yields an amplitude, say $\mathcal{M}_{hard}$, mediated by the same propagator $\mathcal{P}^{abcd}$.}
    \label{fig:diagramsWithhab}
\end{figure}

The scattering processes of interest in this article are mediated by three point vertices as we noted before. This vertex written in the Regge Wheeler gauge was given in \eqref{eqn:vertexRegge-Wheeler}. Inserting the light-cone fields taking the Weyl-rescaling into account,
\begin{equation}
    h_{ab} ~ = ~ \dfrac{1}{A\left(r\right) r} \mathfrak{h}_{ab} \, , \qquad \phi ~ = ~ \dfrac{1}{r} \phi \, ,\qquad K ~ = ~ \dfrac{1}{r} \mathcal{K} \, ,
\end{equation}
this vertex action becomes
\begin{equation}\label{eq:vertexacchorizon}
    S_{\text{vertex}} ~ = ~ \dfrac{\kappa}{2 R} \int \ed \Omega \int \ed^2 x \gap \left[\mathfrak{h}^{ab} \left(\p_{a} \phi \p_{b} \phi - \dfrac{1}{2} \eta_{ab} \eta^{cd} \p_{c} \phi \p_{d} \phi\right) + \mathcal{K} \eta^{AB} T_{AB} \right] \, .
\end{equation}
in the $x^a \ll R$ approximation. As can be seen from the second term in \eqref{eqn:vertex}, the definition of the stress tensor shows that the second term of \eqref{eq:vertexacchorizon} does not vanish with a mere assumption of vanishing transverse momenta $p_A = 0$. In fact, we see that 
\begin{equation}\label{eqn:Kvertex}
    \mathcal{K} \eta^{AB} T_{AB} ~ \sim ~ \mathcal{K} \eta^{ab} \p_a\phi \p_b \phi \, .
\end{equation}
Now, the scattering amplitude of interest has one particle, say $p_1$, with a momentum exclusively going into the black hole while the other particle, say $p_2$, is exclusively exiting the horizon. Therefore, we have that $p_1 = \left(p_{1,x}, p_{1,y}\right) = \left(p_{1,x}, 0\right)$ and $p_2 = \left(p_{2,x}, p_{2,y}\right) = \left(0, p_{2,y}\right)$. We now define the two-dimensional Mandelstam variable
\begin{equation}
    s ~ = ~ - \dfrac{1}{2} \left(p_1 + p_2\right)^2 ~ = ~ - p_{1} \cdot p_{2} ~ = ~ p_{1,x} p_{2,y} \, .
\end{equation}
The amplitudes arise from the interaction terms in \eqref{eq:vertexacchorizon}. Let us first consider the tree level diagram arising from \eqref{eqn:Kvertex} mediated by propagators involving $\mathcal{K}$:\vspace{0.5cm}

\begin{figure}[h!]
\centering
	\includegraphics[scale=0.5]{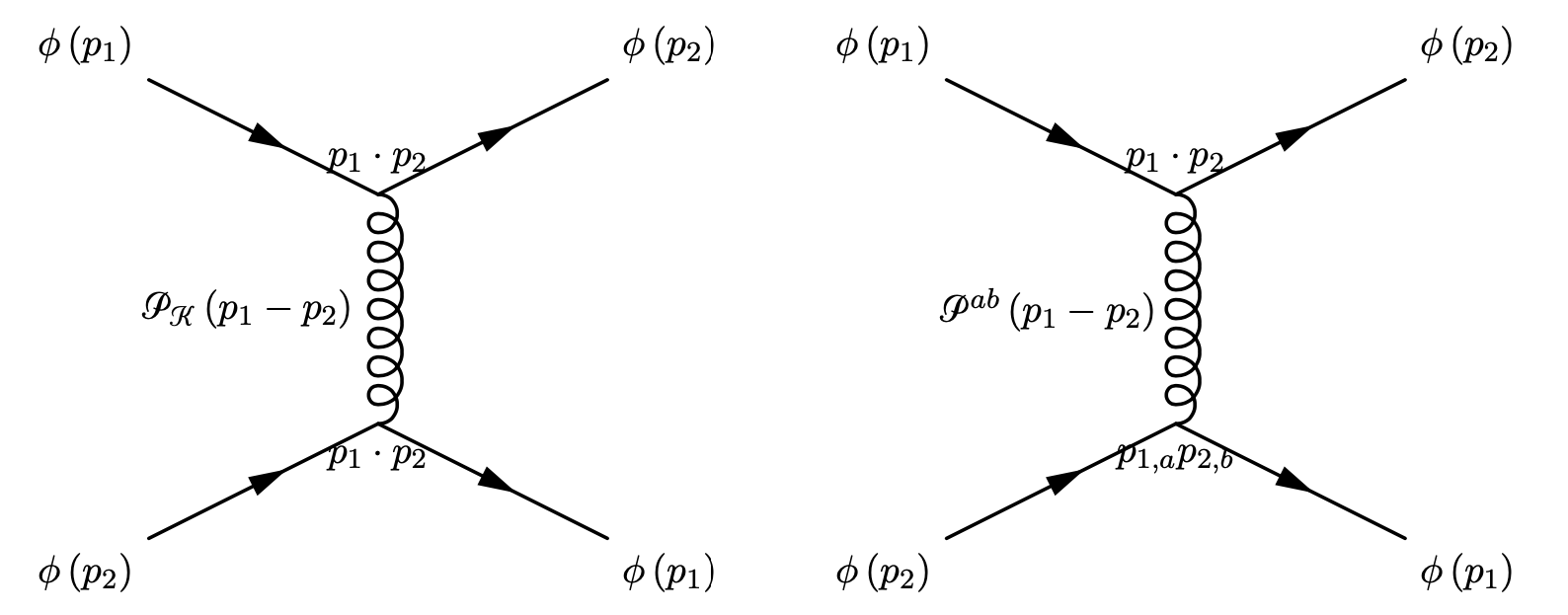}
    \caption{The diagram on the left gives rise to an amplitude, say $\mathcal{M}_\mathcal{K}$, and is mediated by the propagator $\mathcal{P}_\mathcal{K}$ given in \eqref{eqn:Kprop}. The diagram on the right yields an amplitude, say $\mathcal{M}_{\mathfrak{h}_{ab}\to\mathcal{K}}$, mediated by the propagator $\mathcal{P}^{ab}$ given in \eqref{eqn:habToKProp}.}
    \label{fig:diagramsWithK}
\end{figure}
The amplitudes in \figref{fig:diagramsWithK} can easily be calculated. The diagram on the left mediated by $\mathcal{P}_\mathcal{K}$ gives
\begin{equation}\label{eqn:Ktree}
    \mathcal{M}_\mathcal{K} ~ \sim ~ \dfrac{s^2}{2 s + \mu^2 \lambda} \, ,
\end{equation}
and the diagram on the right, mediated by $\mathcal{P}^{ab}$, gives 
\begin{align}\label{eqn:habToKtreeVanishes}
    \mathcal{M}_{\mathfrak{h}_{ab}\to\mathcal{K}} ~ &\sim ~ \left(p_1 \cdot p_2\right) \, \mathcal{P}^{ab} \left(p_{1,a}p_{2,b} - \dfrac{1}{2}\eta_{ab} \, \left(p_1 \cdot p_2\right)\right) \nonumber \\
    &= ~ \left(p_1 \cdot p_2\right) \mathcal{P}^{xy} \left(p_{1,x}p_{2,y} - p_{1,x} p_{2,y}\right) \nonumber \\ 
    &= ~ 0 \, ,
\end{align}
where we used that the light-cone metric is given by $\eta_{xy} = -1$, and that $p_1 = \left(p_{1,x}, 0\right)$ and $p_2 = \left(0, p_{2,y}\right)$. The remaining tree-level diagrams involve $\mathfrak{h}_{ab}$ exclusively, as shown in \figref{fig:diagramsWithhab}. \\

Owing to the symmetry of the propagator $\mathcal{P}^{abcd}$ under $a\leftrightarrow b$, and under $c\leftrightarrow d$, the amplitude $\mathcal{M}_{hard}$ vanishes due to an argument similar to \eqref{eqn:habToKtreeVanishes}. The soft amplitude on the other hand is again easily calculated to be
\begin{equation}\label{eqn:habtree}
    \mathcal{M}_{soft} ~ \sim ~ \left(p_{1,x} p_{2,y}\right)^2 \mathcal{P}^{xxyy}\left(0\right) ~ \sim ~ \dfrac{s^2}{\mu^2} \, .
\end{equation}

\subsection{The black hole eikonal limit: $\sqrt{s} \gg \gamma M_{Pl}$}
Using the parameters $\mu = 1/R_S$ and $\kappa = \sqrt{8 \pi G_N}$, we define
\begin{equation}\label{eqn:gammadef}
    \gamma ~ \coloneqq ~ \mu \kappa ~ \sim ~ \dfrac{M_{Pl}}{M_{BH}} \, .
\end{equation}
With this definition, we observe that
\begin{equation}
    \mu^2 ~ \sim ~ \dfrac{M^4_{Pl}}{M^2_{BH}} ~ \sim ~ \gamma^2 M^2_{Pl} \, .
\end{equation}
The only non-vanishing tree-level amplitudes were \eqref{eqn:Ktree} and \eqref{eqn:habtree}. When written in terms of $\gamma$, in the limit $s \gg \gamma^2 M^2_{Pl}$ for every fixed $\ell$, these read
\begin{align}\label{eqn:treeAmplitudes}
    \mathcal{M}_{\mathcal{K}} ~ &\sim ~ s \quad \text{and} \quad \mathcal{M}_{soft} ~ \sim ~ \dfrac{s^2}{\gamma^2 M^2_{Pl}} \, .
\end{align}
Clearly $\mathcal{M}_{\mathcal{K}}$ is sub-leading for black holes that are bigger than Planck size. In the eikonal approximation in flat space, reviewed in \secref{sec:minkEikonal}, we saw that the analogous approximation necessitated ultra-high energy scattering. A black hole background provides for a new scale that tempers this requirement dramatically. The requirement for the energy of collision is merely
\begin{equation}
    \sqrt{s} ~ \gg ~ \gamma \, M_{Pl} \, .
\end{equation}
The factor $\gamma$ is of the order of $10^{-32}$ for a black hole with a Schwarzschild radius of $1 \, cm$ (this would be a black hole with the mass of the earth). Therefore, the requirement $\sqrt{s} \gg 10^{-32} M_{Pl}$ is very easily satisfied for all known particles in the Standard Model for any black hole that is much heavier than Planck mass. In this approximation, we see that we can ignore the amplitude for $\mathcal{K}$ in \eqref{eqn:treeAmplitudes} and we are left with $\mathcal{M}_{soft}$ to leading order. Therefore, this approximation simplifies the vertex \eqref{eq:vertexacchorizon} because the field $\mathcal{K}$ drops out and we are left with
\begin{equation}
    \mathfrak{h}^{ab} \left(\p_{a} \phi\p_{b} \phi - \dfrac{1}{2} \eta_{ab} \eta^{cd} \p_{c} \phi \p_{d} \phi \right) ~ = ~ \mathfrak{h}^{xx} \p_x \phi \p_x \phi + \mathfrak{h}^{yy} \p_y \phi \p_y \phi \, .
\end{equation}
That only $\mathfrak{h}_{xx}$ and $\mathfrak{h}_{yy}$ remain was argued to be a feature of the eikonal approximation in flat space based on boundary conditions. Near the horizon of the Schwarzschild solution in the $s \gg \gamma^2 M^2_{Pl}$ approximation, we see that the same result can be explicitly derived. Therefore, with the implicit understanding that the off-diagonal components $\mathfrak{h}_{xy}$ do not contribute
\begin{equation}
    \mathfrak{h}^{ab} ~ = ~ \begin{pmatrix}
    \mathfrak{h}^{xx} & 0 \\
    0 & \mathfrak{h}^{yy}
    \end{pmatrix} \, ,
\end{equation}
the vertex action \eqref{eq:vertexacchorizon} can be written as
\begin{align}\label{eq:vertexaction1}
    S_{\text{vertex}} ~ &= ~ \dfrac{\gamma}{2} \sum\limits_{\ell,m} \sum\limits_{\ell_1,m_1} \sum\limits_{\ell_2,m_2} \int \ed \Omega \gap  Y_{\ell}^m Y_{\ell_1}^{m_1} Y_{\ell_2}^{m_2} \int \ed^2 x \gap \mathfrak{h}^{ab}_{\ell m} \p_{a} \phi_{\ell_1 m_1} \p_{b} \phi_{\ell_2 m_2} \, ,
\end{align}
where we inserted the spherical harmonic expansions
\begin{equation}
    \mathfrak{h}^{ab} ~ = ~ \sum\limits_{\ell,m} \mathfrak{h}^{ab}_{\ell m} Y_{\ell}^m \qquad \text{and} \qquad \phi ~ = ~ \sum\limits_{\ell, m} \phi_{\ell m} Y_{\ell}^m \, .
\end{equation}
Integrating over the sphere will result in a coupling between the various partial waves. We saw in \secref{sec:partialwavedecoupling} that the various partial waves decouple at quadratic order, owing to the spherical symmetry of the background. The different partial waves are then expected to only be coupled by large transverse momentum transfers \cite{Verlinde:1991iu}. This is subdominant for scattering processes on the horizon when the black hole is greater than Planck size. Therefore, we do not expect that angular momentum is distributed among external legs. This implies that one of the external scalar legs is kept with some fixed partial wave, say $\phi_0$. The vertex action therefore becomes
\begin{equation}\label{eqn:finalvertex}
    S_{\text{vertex}} ~ = ~ \gamma \sum\limits_{\ell,m} \int \ed^2 x \gap \mathfrak{h}^{ab}_{\ell m} \p_{a} \phi_0 \p_{b} \phi_{\ell m} \, ,
\end{equation}
where the additional factor of two is owed to the different possible attachments of the fixed partial wave $\phi_0$.

Since the off-diagonal components of $\mathfrak{h}^{ab}$ play no role in scattering with a massless scalar field, the graviton propagator \eqref{eq:finalprop} simplifies significantly
\begin{align}
\label{eqn:finalprop2}
\mathcal{P}^{abcd}\left(k\right) ~ &= ~ \dfrac{1}{4} f_{\ell} \biggr[ \eta_{ac} \eta_{bd} + \eta_{ad} \eta_{bc} - \eta_{ab} \eta_{cd} \nonumber \\
&\qquad \qquad + \dfrac{4}{\mu^2 \left(\lambda - 3\right)} \dfrac{1}{k^2 + \mu^2 \lambda} \left(k^a k^b - \dfrac{1}{2}k^2 \eta^{ab}\right) \left(k^c k^d - \dfrac{1}{2} k^2\eta^{cd}\right) \biggr] \, .
\end{align}
The choice of $1/4$ is arbitrary. From \eqref{eq:finalprop}, the soft form factor $f_{\ell}$ can be found to be given by
\begin{equation}\label{eqn:fzero}
f_{\ell} ~ = ~ - \dfrac{4}{\mu^2(\lambda+1)} \, .
\end{equation}

For the scattering processes to be considered, we will see in the \secref{sec:scattering} that the soft limit $k\rightarrow 0$ is imposed on us. Observing \eqref{eqn:finalprop2}, we see familiar problems in the ultraviolet. In the high energy limit, we have that $\mathcal{P}^{abcd} \sim k^4/\left(\mu^2 k^2 + \mu^4\right)$ which diverges. This is a feature familiar from massive gravity \cite{Hinterbichler:2011tt}. It is a result of the mass in the projection operator $\proj^{ab}$. For a massive spin-1 field, there is a similar feature in the UV-limit, which renders the theory non-renormalizable by the power counting theorem \cite{Weinberg:1995mt}. Massive Quantum Electrodynamics is renormalizable due to the introduction of additional scalar fields \cite{Weinberg:1996kr}. A similar trick can be applied to the spin-2 case, introducing vector fields to restore the power counting theorem for the graviton. This was also done in \cite{Hinterbichler:2011tt}, however renormalization remains a problem, as it was already a problem for the massless graviton.

Fortunately enough, owing to the soft limit that is to be enforced upon us, we may simply work with the propagator in \eqref{eqn:finalprop2}. We shall see that the divergent momentum-dependent tail is projected out in the ladder diagrams of interest. And it will be far more important that $\mathcal{P}^{abcd}$ is well defined when $k=0$. Unlike in the flat space eikonal calculation reviewed in \secref{sec:minkEikonal}, the emergent scale of the Schwarzschild radius provides for a natural regulator, with no necessity for a regulator to be put in by hand.

As a final comment, if instead of writing down the honest propagator as we have done above, had we postulated an ansatz of the form \cite{Gaddam:2020rxb}
\begin{align}\label{eqn:finalprop3}
    \mathcal{P}^{abcd}\left(k\right) ~ &= ~ \dfrac{1}{4} f_{\ell}\left(k^2\right) \left(\eta^{ac}\eta^{bd} + \eta^{ad} \eta^{bc} - \eta^{ab} \eta^{cd}\right) \, ,
\end{align}
we find that using \eqref{eq:finalprop}, the solution is given by
\begin{equation}\label{eqn:fksquared}
    f_{\ell}\left(k^2\right) ~ = ~  - \dfrac{4}{\mu^2 \left(\lambda+1\right)} - \dfrac{4}{\mu^{4} \left(\lambda + 1\right)\left(\lambda - 3\right)} \dfrac{k^{4}}{k^2 + \mu^2 \lambda} \, .
\end{equation}
In the soft limit, which we will be projected on to in \secref{sec:scattering}, either choice results in the same answer since $f_\ell\left(0\right) = f_\ell$.

\subsection{The Feynman rules}
Gathering all the results we have obtained so far in \eqref{eq:matteraction2}, \eqref{eqn:finalvertex}, and \eqref{eqn:finalprop2}, we find that the action near the horizon reduces to
\begin{align}\label{eqn:horizonaction}
    S_{\text{hor}} ~ &= ~ \dfrac{1}{4} \int \mathrm{d}^2 k \,  \mathfrak{h}^{ab} \, \mathcal{P}^{-1}_{abcd}\left(k\right) \, \mathfrak{h}^{cd} ~ - ~ \dfrac{1}{2} \int \mathrm{d}^2 p \, \phi \left(p^2 + \mu^2 \lambda\right) \phi \nonumber \\
    &\quad ~ + \gamma \int \mathrm{d}^2k \, \mathrm{d}^2p_1 \, \mathrm{d}^2p_2 \, \delta^{(2)}\left(k + p_1 + p_2\right) \mathfrak{h}_{ab}\left(k\right) p^a_1 p^b_2 \phi_0\left(p_1\right) \phi\left(p_2\right) 
\end{align}
for each partial wave, in momentum space. This allows us to read off the Feynman rules. For the propagators we find

\begin{fmffile}{feyn_propagators}
\begin{align}
  \begin{fmfgraph*}(130,0)
    \fmfleft{i1}
    \fmfright{i2}
    \fmf{phantom}{i1,i2}
    \fmfi{fermion,label=$\phi_{\ell m}(p)$,label.side=left}{vpath (__i1,__i2)}
  \end{fmfgraph*} ~~~ &= ~ \dfrac{-i}{p^2 + \frac{\lambda}{R^2_S} - i \epsilon} \nonumber \\
  \begin{fmfgraph*}(130,0)
    \fmfleft{i3}
    \fmfright{i4}
    \fmf{phantom}{i3,i4}
    \fmfi{dashes,label=$\phi_{0}(p)$,label.side=left}{vpath (__i3,__i4)}
  \end{fmfgraph*} ~~ &= ~ \dfrac{-i}{p^2 + \frac{1}{R^2_S} - i \epsilon} \nonumber \\
    \begin{fmfgraph*}(100,0)
    \fmfleft{i5}
    \fmfright{i6}
    \fmf{phantom}{i5,i6}
    \fmflabel{$\mathfrak{h}^{ab}_{\ell m}$}{i5}
    \fmflabel{$\mathfrak{h}^{cd}_{\ell m}$}{i6}
    \fmfi{curly,label=$\mathfrak{h}_{\ell m}(k)$,label.side=left}{vpath (__i5,__i6)}
  \end{fmfgraph*} ~ \qquad &= ~ 2 i \mathcal{P}^{abcd}\left(k\right) \, , \nonumber \\ \nonumber
  \end{align}
\end{fmffile}
and the vertex is given by \\
\begin{fmffile}{feyn_vertex}
\begin{align}
  \begin{fmfgraph*}(100,100)
    \fmfleft{i8,i7}
    \fmfright{o2}
    \fmf{gluon}{i7,v1}
    \fmf{dashes_arrow}{i8,v1}
    \fmf{fermion}{v1,o2}
    \fmflabel{$\mathfrak{h}^{ab}_{\ell m}$}{i7}
    \fmflabel{$p_2$}{i8}
    \fmflabel{$p_1 ~ ,$}{o2}
    \fmfdot{v1}
    \fmfv{label=$i \gamma ~ p^1_a ~ p^2_b$,label.angle=60,label.dist=20}{v1}
  \end{fmfgraph*} \nonumber
\end{align}
\end{fmffile}
with a symmetry factor of 8 for the graviton propagator accounting for the $a\leftrightarrow b$ and $c\leftrightarrow d$ symmetry and a factor of two for the external scalar legs that may be exchanged. Furthermore, we will employ the usual $i\epsilon$ prescription for the graviton propagator
\begin{equation}\label{eqn:finalgravprop}
    \mathcal{P}^{abcd}\left(k\right) ~ = ~ \dfrac{1}{4} f_{\ell} \left(\eta^{ac}\eta^{bd} + \eta^{ad} \eta^{bc} - f_\ell \left(\dfrac{\lambda + 1}{\lambda + 3}\right) \dfrac{k^a k^b k^c k^d}{4 \left(k^2 + \mu^2 \lambda - i \epsilon\right)} \right) \, ,
\end{equation}
and use the fact that the external scalar momenta dominate the momentum transferred by internal gravitons
\begin{equation}
    \dfrac{-i}{\left(p + k\right)^2 + \mu^2 \lambda - i \epsilon} ~ \approx ~ \dfrac{-i}{2 p \cdot k - i \epsilon} \, .
\end{equation}

\section{Scattering amplitudes on the horizon}\label{sec:scattering}
\begin{figure}[h!]
\centering
        \includegraphics[scale=0.6]{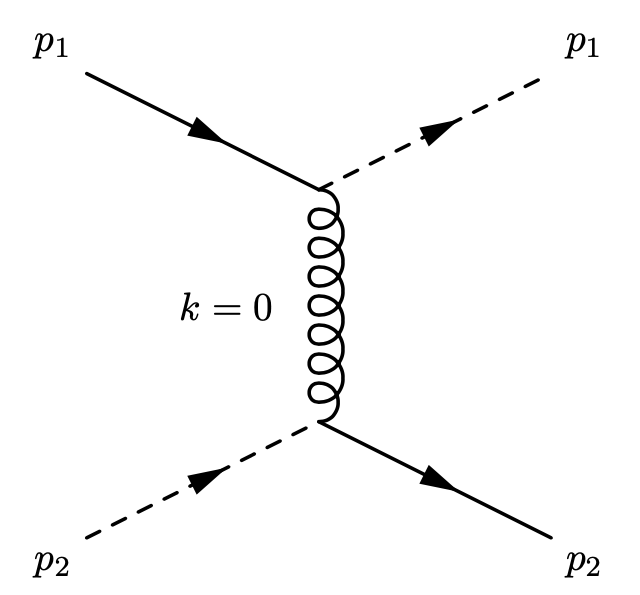}
    \caption{We call this diagram the \textit{transfer} channel as the angular momentum indices (represented by the solid lines) are transferred across the virtual graviton. Combining two such diagrams, a \textit{conserved} channel where the angular momentum stays put is generated; see \figref{fig:oneLoop}. We will discuss both these channels in this section.}
    \label{fig:treediagrams}
\end{figure}

We are now in a position to calculate all the scattering diagrams of interest. Our interest is in the four point correlator with external scalar legs: $\langle\phi_{\ell ,m}\phi_0\phi_{\ell ,m}\phi_0\rangle$. The tree level scattering diagrams are given in \figref{fig:treediagrams}. \\

This transfer channel tree amplitude can be calculated using the Feynman rules from the previous section:
\begin{align}
    i \mathcal{M}_{transfer,1} ~ &= ~ \left(i \gamma \, p^1_a \, p^1_b\right) \left(2 i \mathcal{P}^{abcd}\left(k\right)\right) \left(i \gamma \, p^2_c \, p^2_d\right) \nonumber \\
    &= ~ \gamma^2 \left(p^1_a \, p^1_b\right) \left(\dfrac{2 i}{\mu^2 \left(\lambda + 1\right)} \left(\eta^{ac}\eta^{bd} + \eta^{ad} \eta^{bc}\right)\right) \left(p^2_c \, p^2_d\right)\nonumber \\
    &= ~ \dfrac{4 i \gamma^2 s^2}{\mu^2 \left(\lambda + 1\right)} \nonumber \\
    &= ~ \dfrac{4 i \kappa^2 s^2}{\ell^2 + \ell + 2} \, ,
\end{align}
where in the first step, we used momentum conservation at each vertex and in the last step, we used the definition of $\gamma$ from \eqref{eqn:gammadef} and that $\lambda = \ell^2 + \ell + 1$. This formula is the analog of the corresponding expression for the eikonal phase in flat space \eqref{eq:minkowskitree}. In that result, all the transverse effects were embedded in the Mandelstam variable $t = - \left(p_1 - p_3\right)^2 \approx - \left(p^\perp_1 - p^\perp_3\right)^2$. Whereas here, since we have integrated the sphere out, they are now captured by the partial wave $\ell$. In contrast to the eikonal phase in flat space, where scattering is dominated by large $\ell$ modes \cite{Verlinde:1991iu}\footnote{See eqs. (6.9) and (6.10) of that reference, for instance.}, we see that it is instead dominated by the low $\ell$ ones near the horizon.

\subsection{Loop diagrams and the resummation}
\begin{figure}[h!]
\centering
	\includegraphics[scale=0.55]{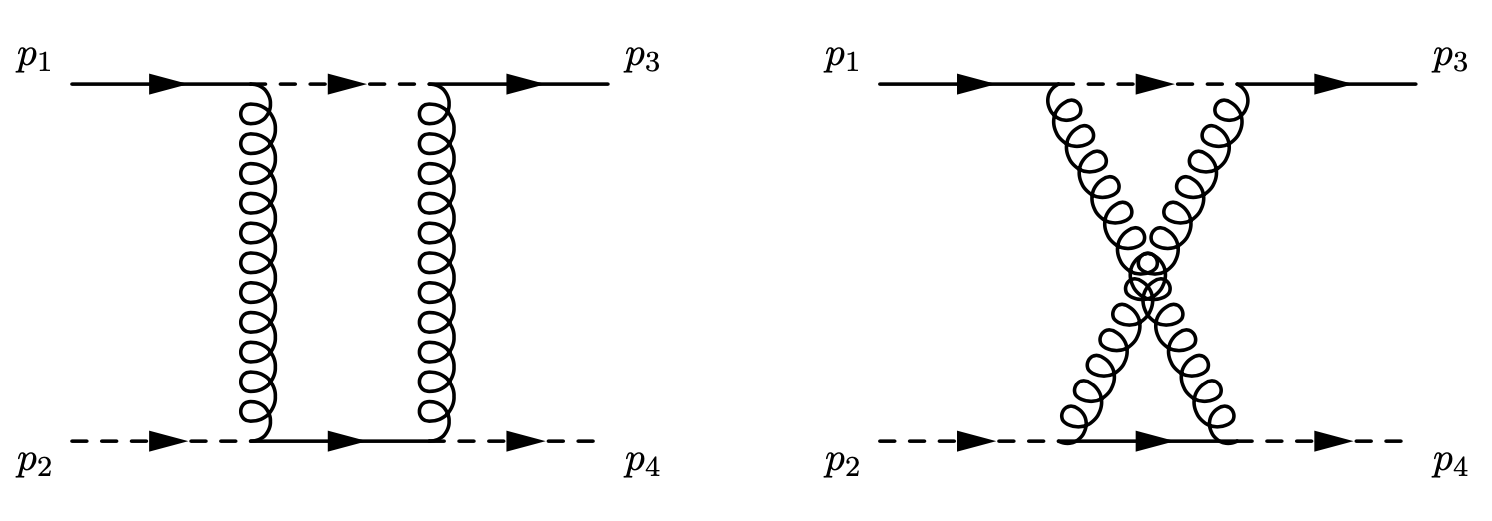}
    \caption{All leading one-loop ladder diagrams on the horizon, in the $\sqrt{s} \gg \gamma M_{Pl}$ limit. We see that the angular momentum indices (represented by the solid line) stay on top and do not get transferred to the bottom. We call this the \textit{conserved} channel. It is clear that all odd loops fall in the \textit{conserved} channel, while all even loops in the \textit{transfer} channel.}
    \label{fig:oneLoop}
\end{figure}

At one loop, much like in flat space, we have the following diagrams \figref{fig:oneLoop}. \\

\begin{figure}[h!]
\centering
    \includegraphics[scale=0.55]{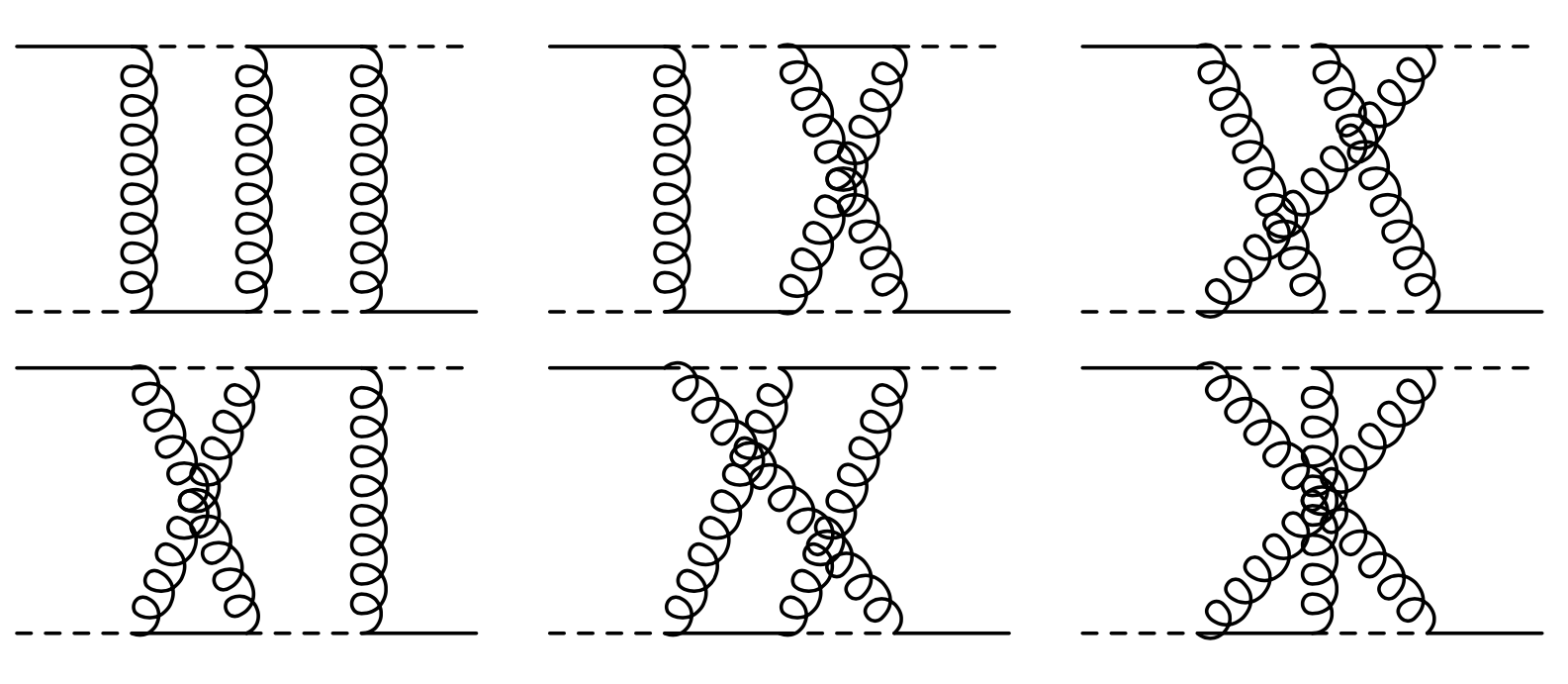}
        \caption{All leading two-loop ladder diagrams on the horizon, in the black hole eikonal limit. None of the gravitons self-interact here.}
    \label{fig:twoloop_transfer}
\end{figure}

As we see from \figref{fig:oneLoop}, the one loop diagrams only contain the \textit{conserved} channel diagrams. In fact, this is a feature of all odd loop diagrams as can be observed from the diagrams. In similar vein, all even loop diagrams transfer the angular momentum indices across the virtual gravitons to the bottom of the diagrams. For instance, all two loop diagrams in the \textit{transfer} channel, to leading order in the black hole eikonal limit are shown in \figref{fig:twoloop_transfer}.\\

In comparison to \cite{Levy:1969cr}, these general loop diagrams are largely similar, except for different external legs and the fact that we are now effectively in two-dimensions. Adapting that calculation to the present case, the general loop amplitude can be written as\footnote{Note that $n=1$ is the tree level diagram, $n=2$ is the one-loop contribution, and so on. So, $n$ counts the number of virtual gravitons exchanged. This equation is the analog of eq (3.1) of \cite{Levy:1969cr}, with gravitational vertices replacing the meson ones, and adapted to two dimensions, with $q = p_1 - p_3 = 0$.}
\begin{align}\label{eqn:generalEvenLoop}
    i \mathcal{M}_{n} ~ &= ~ \left(i \gamma\right)^{2n} \int \prod\limits_{j=1}^{n} \left[\dfrac{\ed^2 k_j}{\left(2 \pi\right)^2} \left(2 i \, p^1_a \, p^1_b \, p^2_c \, p^2_d \, \mathcal{P}^{abcd}_{\ell}\left(k_j\right)\right)\right] \times I \times \left(2 \pi\right)^2 \delta^{(2)} \left(\sum\limits_{j=1}^{n}k_j\right) \nonumber \\
    &= ~ \left(i \gamma s\right)^{2n} \int \prod\limits_{j=1}^{n} \left[\dfrac{\ed^2 k_j}{\left(2 \pi\right)^2} \left(2 i \, \mathcal{P}^{xxyy}_{\ell}\left(k_j\right) \right)\right] \times I \times \left(2 \pi\right)^2 \delta^{(2)} \left(\sum\limits_{j=1}^{n}k_j\right) \, .
\end{align}
Here, we have taken the vertices to come with a factor of $i\gamma s$ each, with all corrections arising from internal momenta $k$ being sub-leading. The quantity $I$ contains all possible matter propagators leading to diagrams of the kind shown in \figref{fig:twoloop_transfer} but at a general loop order. We also integrate over all internal graviton propagators, and insert delta functions to ensure conservation of internal momentum. The quantity $I$ represents all the matter propagators to be inserted. The matter propagators can also be derived analogously to \cite{Levy:1969cr}. The principle is to add all possible permutations of $k_j$ in the matter propagators, with the knowledge that all $f_\ell$ are the same function. These permutations are independent of the dimension of $k_j$ and therefore, the result is very similar:
\begin{align}
    i\mathcal{M}_{n} ~ &= ~ - \dfrac{\gamma^2 s^2}{n!} \int \dfrac{\ed^2 k}{\left(2\pi^2\right)} \, 2 \, i \, \mathcal{P}^{xxyy}_{\ell}\left(k\right) \int\ed^2 x \gap e^{- i k \cdot x} (i \chi)^{n - 1} \, , \nonumber \\
    \chi ~ &\coloneqq ~ - i \gamma^2 s^2\int\dfrac{\ed^2 k}{\left(2\pi\right)^2} \left(2 i \mathcal{P}^{xxyy}_{\ell}\left(k\right)\right) e^{- i k \cdot x} \left[\dfrac{1}{- 2 p_1 \cdot k - i \epsilon} \dfrac{1}{2 p_2 \cdot k - i \epsilon} \right. \nonumber \\
    &\qquad\qquad\qquad\qquad + \dfrac{1}{- 2 p_1 \cdot k - i \epsilon} \dfrac{1}{- 2 p_2 \cdot k - i \epsilon} + \dfrac{1}{2 p_1 \cdot k - i \epsilon} \dfrac{1}{2 p_2 \cdot k - i \epsilon} \nonumber \\
    &\qquad\qquad\qquad\qquad \left. + \dfrac{1}{2 p_1 \cdot k - i \epsilon} \dfrac{1}{- 2 p_2 \cdot k - i \epsilon}\right] \, .
\end{align}
The expression of $\chi$ can be massaged to find
\begin{align}
    \chi ~ &= ~ - i \gamma^2 s^2 \int \dfrac{\ed^2 k}{\left(2\pi\right)^2} \left(2 i \mathcal{P}^{xxyy}_{\ell}\left(k\right)\right) e^{- i k \cdot x} \left[\dfrac{1}{2 p_1 \cdot k + i \epsilon} - \dfrac{1}{2 p_1 \cdot k - i \epsilon}\right] \nonumber \\
    &\qquad\qquad\qquad\qquad\qquad\qquad\qquad \times \left[\dfrac{1}{2 p_2 \cdot k + i \epsilon} - \dfrac{1}{2 p_2 \cdot k - i \epsilon}\right] \, .
\end{align}
Using the identity \eqref{eq:deltaiden}, we now find
\begin{align}
    \chi ~ &= ~ - \gamma^2 s^2 \int \ed^2 k \left(2 \mathcal{P}^{xxyy}_\ell\left(k\right)\right) e^{- i k \cdot x} \delta\left(2 p_1 \cdot k\right) \delta\left(2 p_2 \cdot k\right)  \nonumber \\
    &= ~ - \dfrac{1}{2} \gamma^2 \, s \, \mathcal{P}^{xxyy}_\ell\left(0\right) \nonumber \\
    &= ~ - \dfrac{1}{4} \gamma^2 \, s \, f_\ell \, .
\end{align}
Owing to the lack of any transverse directions (they have been integrated out), we see that the soft limit is indeed forced upon us. Consequently, the ultraviolet tail in the graviton propagator is rendered inconsequential. Now, since $\chi$ is independent of space-time coordinates in this case, the amplitude can be written as
\begin{equation}
    i \mathcal{M}_n ~ = ~ - \dfrac{\gamma^2 s^2 \left(i\chi\right)^{n-1}}{n!} \int \dfrac{\ed^2 k}{(2\pi^2)}\left(2 i \mathcal{P}^{xxyy}_\ell \left(k\right)\right) \int\ed^2 x e^{- i k \cdot x} \, .
\end{equation}
Using that
\begin{equation}
    \int \ed^2 x e^{- i k \cdot x} ~ =~ \left(2 \pi\right)^2 \delta^{(2)}\left(k\right) \, ,
\end{equation}
we find for the general loop amplitude
\begin{align}
    i \mathcal{M}_n ~ &= ~ - i \dfrac{\gamma^2 s^2 \left(i\chi\right)^{n-1}}{n!} f_{\ell} \nonumber \\
    &= ~ 4 s \dfrac{\left(i\chi\right)^{n}}{n!} \, .
\end{align}
Therefore, the complete resummed non-perturbative amplitude can be written as two independent sums; one each for the \textit{transfer} and \textit{conserved} channels:
\begin{align}
    i \mathcal{M} ~ &= ~ i \mathcal{M}_{transfer} + i \mathcal{M}_{conserved} \nonumber \\
    &= ~ 4 s \left(\sum\limits_{n\text{ odd}}^{\infty}\dfrac{\left(i \chi\right)^{n}}{n!} + \sum\limits_{n\text{ even}}^{\infty}\dfrac{\left(i\chi\right)^{n}}{n!}\right) \nonumber \\
    &= ~ 4 s \left(\sum\limits_{m=0}^{\infty}\dfrac{\left(i \chi\right)^{2m+1}}{(2m+1)!} + \sum\limits_{m=1}^{\infty}\frac{(-1)^m\chi^{2m}}{(2m)!} \right) \nonumber \\
    &= ~ 4 s \left(\exp\left(i \chi\right) - 1 \right) \, .
\end{align}
Restoring factors of $\hbar$ which we have so far been negligent about, and substituting for $\chi$, we find\footnote{Analytically continuing $n$ to zero neutralises the extra factor of $1$. This would correspond to situation where no gravitons are exchanged, and the particles move without interactions. This also accounts for the normalisation, which arises from the disconnected product of two-point functions.}
\begin{align}\label{eqn:fourpointResult}
    i \mathcal{M} ~ &= ~ 4 p_{in} p_{out} \left[\exp\left(i \dfrac{\gamma^2 R^2_S}{\hbar\left(\ell^2 + \ell + 2\right)} p_{in} p_{out}\right) - 1\right] \nonumber \\
    &= ~ 4 p_{in} p_{out} \left[\exp\left(i \dfrac{8 \pi G_N}{\hbar\left(\ell^2 + \ell + 2\right)} p_{in} p_{out}\right) - 1\right] \, ,
\end{align}
where we have used the definition of $\gamma$, given in \eqref{eqn:gammadef} and that $\lambda = \ell^2 + \ell + 1$, and relabelled the momenta from $p_1$ and $p_2$ to $p_{in}$ and $p_{out}$. This amplitude is non-perturbative in $\gamma \sim M_{Pl}/M_{BH}$ and $\hbar$. Of course, it ignores effects that arise strictly from the Planck scale. Moreover, we have worked to leading order in $\sqrt{s} \gg \gamma M_{Pl}$. Unlike in flat space, this condition is easy to satisfy. In fact, it is actually difficult to violate with Standard Model particles, for large semi-classical black holes.

It is now straight forward to notice, upon Fourier transforming the right hand side, that this amplitude is only non-vanishing when 
\begin{equation}\label{eqn:shapiroResult}
    y_{out} ~ = ~ \dfrac{8 \pi G_N}{\ell^2 + \ell + 2} p_{in} \, .
\end{equation}
Similar to what was expected from a first quantised formalism \cite{Hooft:2015jea, Hooft:2016itl, Betzios:2016yaq}, this calculation suggests that near a black hole, creation and annihilation operators for in and out states, are also related by Fourier transforms in this second quantised description. However, there is a mysterious discrepancy in the denominator in comparison to the first quantised results. It is to this discrepancy that we now turn to.

\subsection{An apparent discrepancy demystified}\label{sec:discrepancy}

Analysis from first quantisation, where wavefunctions on a fixed non-linearly backreacted Dray-'t Hooft metric formed the starting point, suggested the relation
\begin{equation}\label{eqn:shapiroQM}
    y_{out} ~ = ~ \dfrac{8 \pi G_N}{\ell^2 + \ell + 1} p_{in}
\end{equation}
instead of the expression we found in \eqref{eqn:shapiroResult}. The Dray-'t Hooft metric can be readily checked to satisfy equations of motion only with \eqref{eqn:shapiroQM}. So, the first temptation is to fear the correctness of \eqref{eqn:shapiroResult}. Tracing back the origins of the factor $\ell^2 + \ell + 2 = \lambda + 1$, we see that it arises from the tensorial operator \eqref{eqn:tensorOperHor}. The only components that were relevant for the scattering in the black hole eikonal limit are:
\begin{align}
    \Delta^{-1}_{xxyy} ~ &= ~ - \dfrac{1}{2} \mu^2 \left(\lambda + 1 + x \p_x - y \p_y\right) \, \nonumber \\
    \Delta^{-1}_{yyxx} ~ &= ~ - \dfrac{1}{2} \mu^2 \left(\lambda + 1 + y \p_y - x \p_x\right) \, .
\end{align}
The on-shell metric perturbation that led to the Shapiro delay was given by (see \appref{app:onshellPert}):
\begin{equation}
    h_{xx} ~ = ~ 2 A\left(x,y\right) \delta\left(x\right) F_{\ell m} \, ,
\end{equation}
where we replaced the angular function $F\left(\Omega\right)$ by its partial wave counterpart. In particular, recall that $h_{yy}$ is vanishing. From the effective two-dimensional action \eqref{eq:finalgravact1}, we identify the first order variation of the Einstein tensor to be $G_{xx} = \Delta^{-1}_{xxyy} h_{xx}$. Writing it out, we find
\begin{align}
    G_{xx} ~ &= ~ - \mu^2 A(x,y) F_{\ell m} \delta(x) \left(\lambda + 1\right) - \mu^2 A(x,y) F_{\ell m} x \p_x \delta(x) \nonumber \\
    &\qquad - \mu^2 F_{\ell m} \delta(x) \left(x \p_x - y \p_y\right) A\left(x,y\right) \, .
\end{align}
It can now be checked that $\left(x \p_x - y \p_y\right) A\left(x,y\right) = 0$, using the derivatives from \appref{app:defs}. We are therefore left with
\begin{equation}
    G_{xx} ~ = ~ - \mu^2 F_{\ell m} \lambda \, ,
\end{equation}
which is consistent with the on-shell Dray-'t Hooft result. This confirms that the tensorial operator $\Delta^{-1}_{abcd}$ is indeed correct, the near horizon approximation in \secref{sec:horizonApprox} notwithstanding.

An important conclusion from this analysis is therefore that the discrepancy lies in the delta functions. For the Shapiro delay, we have that
\begin{equation}
    \left(x \p_x - y \p_y\right)\delta\left(x\right) ~ = ~ - \delta\left(x\right) \, .
\end{equation}
This was indeed the fact that resulted in the appropriate on-shell solution as we just saw. On the contrary, in inverting the propagator in \secref{sec:lightconeprop}, we used a double Dirac delta function, $\delta^{(2)}\left(x - x'\right) = \delta\left(x - x'\right)\delta\left(y - y'\right)$. Therefore, we will now have that
\begin{equation}
    \left(x \p_x - y \p_y\right)\delta^{(2)}\left(x - x'\right) ~ = ~ 0 \, .
\end{equation}
The term arising from the $y$ piece cancels the one arising from the $x$ piece. One may now wonder if introducing a metric perturbation $h_{yy}$ in addition to $h_{xx}$ would resolve the issue. For instance, consider splitting the metric perturbations on the future and past horizons as
\begin{equation}
    \mathfrak{h}_{ab}\left(x,y\right) ~ = ~ \mathfrak{h}^+_{ab}\left(x\right)\delta\left(y\right) + \mathfrak{h}^-_{ab}\left(y\right)\delta\left(x\right) \, ,
\end{equation}
where $\mathfrak{h}^-_{ab}$ is defined on the past horizon while $\mathfrak{h}^+_{ab}$ on the future horizon. This does not solve the problem, as can be checked; in addition to terms proportional to $\lambda$, additional $\lambda + 2$ terms arise. One may be tempted to further restrict the perturbations as (see Section 2.3 of \cite{Amati:2007ak}):
\begin{align}\label{eqn:shapiroACV}
    \mathfrak{h}_{xx}\left(x,y\right) ~ = ~ \mathfrak{h}^-_{xx}\left(y\right) \delta\left(x\right) \quad \text{and} \quad \mathfrak{h}_{yy}\left(x,y\right) ~ = ~ \mathfrak{h}^+_{yy}\left(x\right) \delta\left(y\right) \, .
\end{align}
But with this, we recover the on-shell Shapiro delay again. Therefore, if we are to restrict the possible metric perturbations, the only resulting possibility is the on-shell Dray-'t Hooft solution, leaving the different factor in \eqref{eqn:shapiroResult} still mysterious. 

In the black hole eikonal limit, we saw that the transverse mode $\mathcal{K}$ does not contribute. So, it suffices to focus on the relevant part of the action \eqref{eq:finalgravact1}:
\begin{equation}\label{eq:discacc1}
    S ~ = ~ \dfrac{1}{4} \int\ed ^2 x \left(\mathfrak{h}^{xx} \Delta^{-1}_{xxyy} \mathfrak{h}^{yy} + \left(x \leftrightarrow y\right)\right) \, .
\end{equation}
Focusing on the $\Delta^{-1}_{xxyy}$ term, and plugging in the Shapiro delay perturbations \eqref{eqn:shapiroACV} into the above action, we find
\begin{equation}
    S ~ = ~ - \dfrac{1}{8} \mu^2 \int \ed ^2 x \mathfrak{h}^+_{yy}(x) \delta(y) \bigr(\lambda + 1 + x \p_x - y \p_y\bigr) \mathfrak{h}^-_{xx}(y)\delta(x) \, .
\end{equation}
The factor $x\p_x\delta(x)$ can be integrated by parts:
\begin{align}
    \int\ed x\gap \mathfrak{h}^+_{yy}(x) x \p_x \delta(x) ~ &= ~ - \int \ed x \gap \delta(x) \p_x \bigr(x \mathfrak{h}^+_{yy}(x)\bigr) \nonumber \\
    &= ~ - \int\ed x \gap \delta(x) \bigr(1 + x \p_x \mathfrak{h}^+_{yy}(x)\bigr) \, ,
\end{align}
where the boundary terms cancel exactly because $\delta(\pm\infty)=0$. Using this relation, the above action can be rewritten as
\begin{align}
\label{eq:discinteracc1}
    S ~ &= ~ \dfrac{1}{4} \int \ed ^2 x \mathfrak{h}^+_{yy}(x) \delta(y) \left(- \dfrac{1}{2} \mu^2 \lambda\right) \mathfrak{h}^-_{xx}(y) \delta(x) \nonumber \\
    &\qquad + \dfrac{1}{8} \int \ed^2 x \delta(x) \delta(y) \bigr(x \p_x + y \p_y\bigr) \mathfrak{h}^-_{xx}(y) \mathfrak{h}^+_{yy}(x) \, .
\end{align}
The last term is now easily simplified since the delta functions do not have any derivatives on them anymore. Therefore we may simply remove the integrals and set $x=0=y$. However, in that case $x\p_x+y\p_y=0$ provided $\mathfrak{h}^-_{xx}(0)$ and $\mathfrak{h}^+_{yy}(0)$ are regular. We therefore have
\begin{eqnarray}
    S ~ = ~ \dfrac{1}{4} \int \ed ^2 x \left(\mathfrak{h}^{xx}\left(- \dfrac{1}{2} \mu^2 \lambda \right) \mathfrak{h}^{yy} + \left(x \leftrightarrow y\right)\right) \, ,
\end{eqnarray}
where we reinserted \eqref{eqn:shapiroACV}. Comparison with \eqref{eq:discacc1} now shows that we can identify $\Delta^{-1}_{xxyy}=-\tfrac{1}{2}\mu^2\lambda$. Inversion is now straightforward:
\begin{eqnarray}
\Delta_{xxyy} ~ = ~ \Delta_{yyxx} ~ = ~ - \dfrac{2}{\mu^2\lambda} ~ = ~ - \dfrac{2}{\mu^2\left(\ell^2 + \ell + 1\right)} \, .
\end{eqnarray}
This gives the on-shell Shapiro delay factor to reproduce \eqref{eqn:shapiroQM}. This proves that the discrepancy between \eqref{eqn:shapiroQM} and \eqref{eqn:shapiroResult} really arises from a choice of the specific shape of the metric perturbations made in the first quantisation calculations of \cite{Hooft:2015jea, Hooft:2016itl, Betzios:2016yaq}. Considering the $\ell=0$ mode, which is the most dominant, this results in a factor of two compared to the first quantisation calculations. To understand this better, we first note that the general operators \eqref{eqn:opers2d} can be written, using \eqref{eq:potsol1} as:
\begin{subequations}
\begin{align}
    \Delta^{-1} ~ &= ~ -\partial^2 + \dfrac{A R}{r^3} \\
    \Delta^{-1}_{R,ab} ~ &= ~ - \eta_{ab} \biggr[\partial^2 + \dfrac{A}{2 r R} \left(1 - \dfrac{r}{R} \right) x^c \p_c - \dfrac{A^2 x^2}{8 r^2 R^2} \left( 2 -\dfrac{r^2}{R^2} \right) \nonumber \\
    &\qquad \qquad \qquad - \dfrac{A}{2 r^2} \left(\ell(\ell+1) + 2 \dfrac{r}{R} - \dfrac{R}{r} \right) \biggr] \nonumber \\
    &\qquad + \p_a \p_b  - \dfrac{A}{2 r R} \left(1 - \dfrac{r}{R} \right) x_{(a} \p_{b)} - \dfrac{A^2 x^2}{4 r^2 R^2} \left(2 + 2 \dfrac{r}{R} + \dfrac{r^2}{R^2} \right) x_a x_b \\
    \Delta^{-1}_{L,ab} ~ &= ~ - \eta_{ab} \biggr[\p^2 + \dfrac{A}{2 r R} \left(1 - \dfrac{r}{R}\right) x^c \p_c - \dfrac{A}{2 r^2} \left(\ell(\ell+1) - \dfrac{R}{r}\right) \biggr] + \p_a \p_b \nonumber \\
    &\qquad + \dfrac{A}{2 r R} \left(1 - \dfrac{r}{R}\right) x_{(a}\p_{b)} \\
    \Delta^{-1}_{abcd} ~ &= ~ \dfrac{A}{4 r R} \biggr(\eta_{ac} x_{[b} \p_{d]} + \eta_{bd} x_{[a} \p_{c]} + \eta_{ad} x_{[b} \p_{c]} + \eta_{bc} x_{[a} \p_{d]} + 2 \eta_{ab} x_{(c} \p_{d)} - 2 \eta_{cd} x_{(a} \p_{b)} \biggr) \nonumber \\
    &\qquad + \eta_{ab} \eta_{cd} \biggr( \dfrac{A R}{2 r^3} + \dfrac{A \ell(\ell+1)}{2 r^2} + \dfrac{A}{2 r R} + \dfrac{A^2 x^2}{8 r^2 R^2} \left( 1 + \dfrac{r}{R}\right) \biggr) \nonumber \\
    &\qquad - \eta_{a(c} \eta_{d)b} \biggr( \dfrac{A R}{r^3} + \dfrac{A \ell(\ell+1)}{2 r^2} + \dfrac{A^2 x^2}{4 r^2 R^2} \left(1 + \dfrac{r}{R} \right) \biggr) \nonumber \\
    &\qquad - \dfrac{A^2}{4 r^2 R^2} \left(1 + \dfrac{r}{R}\right) \eta_{ab} x_c x_d + \dfrac{A^2}{4 r^2 R^2} \eta_{cd} x_a x_b \, .
\end{align}
\end{subequations}
As it turns out, there are three different approximations\footnote{In all of these, we assume that the gravitons are regular on the horizon. While the shockwave has a discontinuity on the horizon, as we explain in the following, the relevant quantity for the shockwave approximation is $x_{a} h^{a b}$ which is indeed regular.} that one may make on these operators to simplify them to an analytically invertible form. One of them is the approximation that we have described in detail in \secref{sec:horizonApprox}. Another approximation is what may be termed as the ``leading order near-horizon approximation'' where we blithely evaluate the operators on the bifurcate horizon to leading order by setting $r = R$, and $uv = x^{2} = 0$.  Moreover, to leading order, we also set all terms linear in $x^{a}$ to zero. This simplifies these operators tremendously. In particular, if the fields are regular on the horizon, we may also ignore terms with single derivatives in this approximation. The resulting operators can easily be inverted following the techniques we employed in \secref{sec:lightconeprop}. The propagators so obtained can be checked to give the eikonal amplitude we found in \eqref{eqn:shapiroResult}. In hindsight, therefore, the detailed keeping track of the single derivative terms in \secref{sec:horizonApprox} may have been unnecessary to leading order. Alternatively, another approximation that may be made on the above operators is what we may call the ``shockwave approximation'' where in addition to imposing $x^{2} = 0$, we also demand that the field configurations satisfy the condition that $x^a h_{ab} = x h_{x b} + y h_{y b}$ is negligible.\footnote{In implementing this condition, care must be taken to keep track of terms of the form $x_{a} \partial_{b} h^{a c} = - \delta_{a b} h^{a c} + \partial_{b} \left(x_{a} h^{a c}\right) =  h^{b c}$.} This condition may be understood as the demand on the horizon, say described by $x = 0$, that $h_{y b}$ is negligible. This is a covariant way to express the restriction to shockwave field configurations we found above. Carefully implementing this approximation and evaluating the eikonal amplitude following the techniques of \secref{sec:scattering}, we directly find the amplitude \eqref{eqn:shapiroQM}. Therefore, the discrepancy may finally be concluded to emerge from two possible approximations that we may make on the quadratic operators that lead to different analytic results.

\section{Discussion and conclusions}\label{sec:discussion}
\begin{figure}[h!]
    \centering
    \includegraphics[scale=0.5]{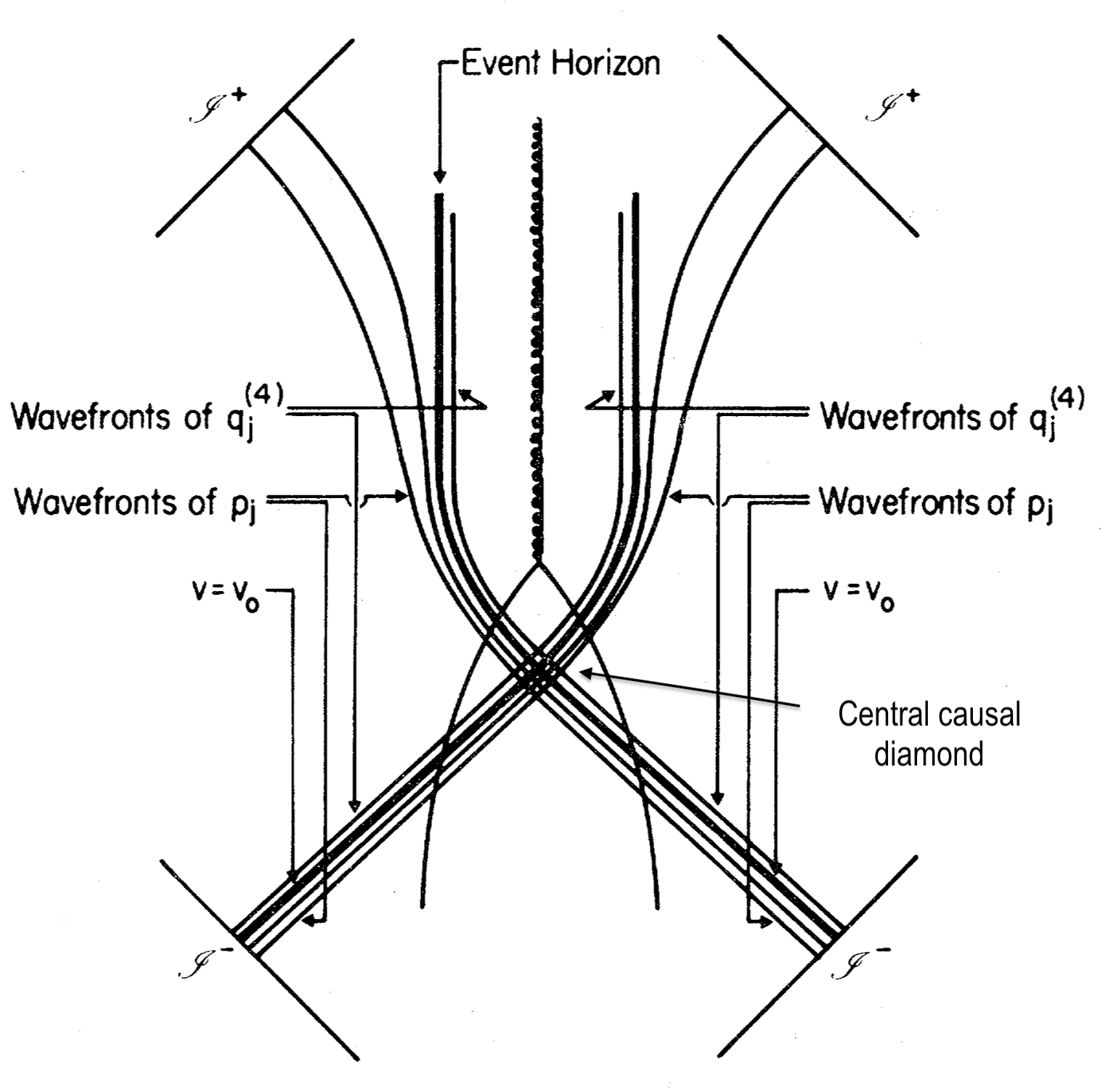}
    \caption{In Hawking's calculation of thermal radiation emanating from the black hole \cite{Hawking:1974sw, Hawking:1976ra}, wavefronts were naively extended to the past, near the central causal diamond, and further back on to the past horizon. In this article we argued that this is invalid owing to the strong gravitational interactions near the horizon that result in graviton exchange processes between in and out going quanta; the exchanged off-shell gravitons are found to be soft.}
    \label{fig:hawking1976}
\end{figure}

In this paper, we have argued that the geometric optics approximation of Hawking is invalid when the wavefronts are traced back to the region near the central causal diamond (see \figref{fig:hawking1976}) for finite sized black holes; this is owed to the strong gravitational interactions in this region. To repair this problem, we studied $2 \rightarrow 2$ scattering mediated by virtual gravitons, near this region; this near horizon region is curved and the familiar eikonal phase around flat space does not capture this physics. We found that there is a remarkable limit in quantum gravity, namely the $\sqrt{s}\gg\gamma M_{Pl}$ regime near the black hole horizon, where an infinite number of graviton exchange diagrams can be resummed exactly in $\gamma \sim M_{Pl}/M_{BH}$. We name this the \textit{black hole eikonal phase} of quantum gravity. Only upon this resummation does one recover a scattering matrix that is a pure phase.

While in this paper we have only study the elastic $2\rightarrow 2$ amplitudes which are naturally suppressed at high energies, the effective field theory we set up captures exponentially many such general $2\rightarrow N$ amplitudes. We leave the study of these amplitudes for future work. We expect that these conspire to provide for a resolution to the information paradox.  Indeed, several authors have argued that the information paradox is expected to be resolved by $1/N$ effects. It has been argued that $N$ may be interpreted as the occupation number of the number of coherent gravitons that represent the black hole state \cite{Dvali:2014ila, Dvali:2011aa}. It may be interesting to note that the coupling constant of the tree-level $2\rightarrow2$ diagram naturally comes with two powers of the coupling constant $\gamma^{2}$ which is precisely $1/N$. Both the black hole eikonal ladder (which includes loop corrections to the tree-level diagram and comes with higher powers of $\gamma$ as we explicitly computed in the present article) and (tree-level) $2\rightarrow N$ diagrams capture higher order $1/N$ corrections. 

Moreover, our calculation explicitly ignores any potential Planckian physics (by the choice that impact parameters be larger than Planck length). Ultraviolet completions are hidden in the hard sector, whereas the calculation forced us to consider soft gravitons with vanishing momenta in all the ladder diagrams. Given the universality of the result, the problem of unitarity (in effective field theory) itself does not seem to guide us to a specific ultraviolet completion of quantum gravity. Nevertheless, there is a clear path to include sub-leading corrections arising from the transverse scalar field $\mathcal{K}$, the odd-parity graviton mode, and the sub-leading contributions in the black hole eikonal limit. None of these may be analytically re-summable to all orders, but the techniques developed in this article certainly lay the path for order-by-order inclusions of these effects, in perturbation theory. It is not impossible or unreasonable that a theory such as String Theory may emerge in the ultraviolet \cite{deHaro:1998tj, deHaroOlle:1997hx}.

Of course, in addition to the shortcomings pointed out in the companion article \cite{Gaddam:2020rxb}, a natural complaint that may be raised is that the apparent horizon is radically different in nature to a Schwarzschild horizon. This is a complaint that may have been raised against Hawking's computation too. Even for collapsing black holes, provided they are large enough, this is expected to be a reliable approximation. 

\begin{figure}[h!]
    \centering
    \includegraphics[scale=0.5]{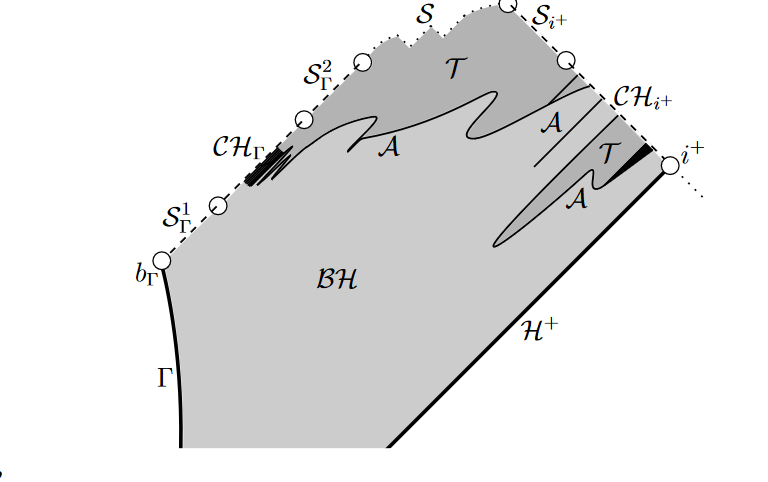}
    \caption{A depiction of a more realistic apparent horizon than the crude approximation made in this article. Figure taken from \cite{Kommemi:2011wh}.}
    \label{fig:realisticAH}
\end{figure}

A perhaps more realistic apparent horizon is shown in \figref{fig:realisticAH}, taken from \cite{Kommemi:2011wh}. While trapped surfaces were shown to dynamically form in Christodoulou's monumental work \cite{Christodoulou:2008nj}, apparent horizons (which are marginally trapped) have also been proved to dynamically form \cite{An:2017wbq}. Moreover, black hole mechanics can be established for apparent horizons should they be smooth and spacelike \cite{Ashtekar:2003hk, Ashtekar:2004cn}. In the present article, the most important ingredient that came from the horizon is the conformal flatness of the effective two-dimensional metric that allowed us to derive the effective two dimensional propagator. Given that the soft limit of the graviton propagator was the only necessary component in the end, a more realistic model of the apparent horizon may only change the soft factor, leaving behind the general lessons learnt unchanged.

Given the insensitivity of the resolution presented in this article to ultraviolet physics, it is natural to ask how the corrections we computed compare with those required for some of the existing proposals for the resolution of the information paradox. Here, we consider a few proposals and speculate on this matter.

\paragraph{Soft hair} Renewed interest in asymptotic symmetries has also provided hope that soft quantum hair could pave the way forward to information recovery from the horizon; it has also been shown that shockwaves inserted on past infinity implant soft hair on the horizon \cite{Hawking:2016msc, Hawking:2016sgy, Strominger:2017aeh, Haco:2018ske, Nomura:2019qps}. More generally, the soft sector of the gravitational scattering matrix has also received recent attention \cite{Himwich:2020rro}. In fact, the Dray-'t Hooft metric can be shown to arise from supertranlsations and to contain superrotation charges on the horizon \cite{Vleeshouwers, Gaddam}. Given that the backreacted Dray-'t Hooft solution is closely related to what can be derived from the re-summed amplitude \eqref{eqn:fourpointResult}, it is not inconceivable that the scattering matrix can be (at least partially) recast in the language of soft hair on the horizon; it is also worthwhile to note there are no infrared divergences in our calculations. Furthermore, the black hole eikonal phase may have interesting implications for (sub-dominant) soft graviton theorems near the horizon.

\paragraph{Islands} An increasingly popular suggestion is that information recovery must be seen to arise from additional saddles leading to the so called island formula that is claimed to trace the correct Page curve behaviour to restore unitarity \cite{Penington:2019npb, Almheiri:2019psf, Almheiri:2019hni}. Morally speaking, islands imply that information behind the horizon is captured by radiation. This is consistent with the present calculation and in fact, our calculation shows a dynamical process that achieves this; namely, via graviton exchange. The general lessons from the island computations may be seen as twofold: i) unitarity in black hole evaporation does not necessarily rely on the ultraviolet details of the theory, and ii) information retrieval is a non-perturbative process. The present calculation is in line with both of these expectations. An interesting caveat in the island computations is the necessity of massive gravitons for black holes in dimensions greater than three \cite{Karch:2000ct, Geng:2020qvw}. Finally, given a unitary scattering matrix that describes evolution, the notion of a Page curve is unnatural in the scattering process. Should one insist on the Page curve and generate the information paradox, one may simply set $\gamma \sim \sqrt{G_N}/R_S = 0$; this essentially ignores all the interactions that may resolve the problem. However, in our effective field theory, Page time is likely to naturally arise from longer Wigner's time delays from general $2\rightarrow N$ scattering matrices, whereas scrambling time is naturally encoded in the resummed $2\rightarrow 2$ amplitude \cite{Betzios:2016yaq} under consideration in the present paper.

\paragraph{Fuzzballs} A traditional black hole has been called by Mathur as an `information-free' one, and small corrections have been claimed to be insufficient to restore unitarity \cite{Mathur:2009hf}. Instead, new fuzzy structure is introduced, and an exponentially large number of such structures are claimed to explain the degeneracy of states. Our calculation shows that the Schwarzschild horizon is not information free in that it hosts soft gravitons. Although the fuzzball line of thought may question if a classical horizon forms at all \cite{Kraus:2015zda}, it needs to be reconciled with the expectation that apparent horizons do dynamically form \cite{An:2017wbq}. It is clear, in either case, that our calculation and the fuzzball proposal would yield different gravitational wave echoes leaking out of the classical Regge-Wheeler potential. The present calculation has no free parameters and may provide for disambiguation \cite{Betzios:2020}. Finally, it is worthwhile to note that precise statements made about fuzzballs have often been in the context of supersymmetric, extremal, or near extremal black holes inspired from string theory. In the present article, we make no assumptions about a putative ultraviolet completion.

\paragraph{Soft but strong corrections} Giddings has suggested that the resolution of the information paradox will require `soft but strong' corrections to the traditional Hawking calculation; see \cite{Giddings:2017mym} and references therein. In our calculation, the exchanged virtual gravitons near the horizon are soft.

\paragraph{`Holographic' null infinity} Another recent proposal for the resolution of the information paradox is that the information can always be recovered from an infinitesimal neighbourhood of the past boundary of future null infinity \cite{Laddha:2020kvp, Chowdhury:2020hse}. However, this proposal refers to the information present in a Cauchy slice at a specific instant in time and how this is retrievable from the boundary of the said slice. This is insufficient to draw conclusions about information preservation between Cauchy slices at different time slices, namely the S-matrix. 

\paragraph{Planckian remnants} Given that a unitary scattering matrix can be achieved with impact parameters $b\gg L_{Pl}$, the present calculation leads us to expect that one need not rely on the details of Planckian physics for the information problem.

\paragraph{Outlook and future work} Several directions for future work are listed in the companion article \cite{Gaddam:2020rxb}. Here, we list a few more.

\begin{figure}[h!]
    \centering
        \includegraphics[scale=0.6]{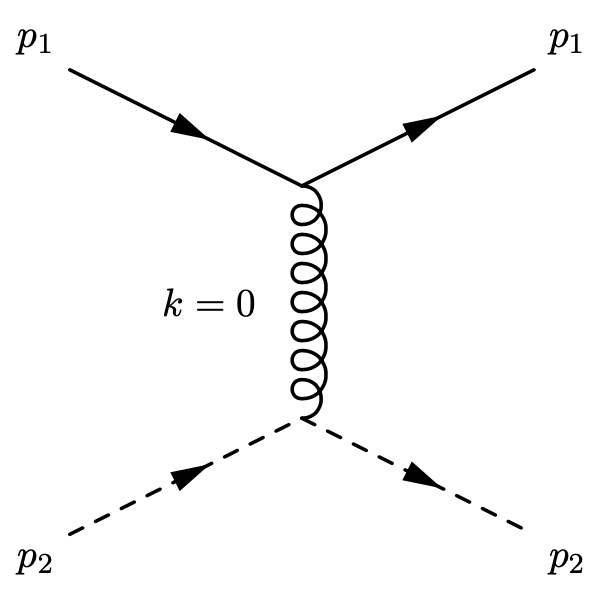}
    \caption{The vertices considered in this article had a solid and a dashed line, whereas diagrams of this kind are possibly also allowed. Given that these would exchange the $s$-wave graviton, they may be crucial in determining the change in mass of the black hole due to the scattering.}
    \label{fig:conservedtree}
\end{figure}

In the present article, we only considered three point interaction vertices that came with one solid line and one dashed line (corresponding to one field with angular momentum and another without). It would be interesting to understand the role that vertices and diagrams of the kind in \figref{fig:conservedtree} may play in the theory. 
It would also be interesting to extend our calculation to include additional external legs, and massive scalar fields. In the latter case in particular, the $\mathfrak{h}_{xy}$ component of the graviton is now expected to contribute. While the former might allow for an obvious extension of the chaos bound \cite{Maldacena:2015waa} to $n$-point functions. It is of course also interesting to explore the role of additional non-linearities arsing from graviton self-couplings, in the soft limit. For instance, a three point self-interaction would immediately allow for a study of $2\rightarrow 2$ graviton scattering \cite{Chowdhury:2019kaq}. This would allow us to ask if there are `graviballs' near the horizon, much like glueballs in QCD. Studying graviballs in the flat space eikonal regime introduces the familiar infrared divergences and unfixed parameters \cite{Blas:2020dyg}. These may be naturally addressed near the horizon using the tools developed in this article.

Owing to the Weyl rescaling we were able to perform in \secref{subsec:WeylRescaling}, we could define flat space kinematics despite being in a gravitationally interacting regime. This means that essentially every improvement done on flat space eikonal physics could potentially be repeated near the horizon using our techniques. Effective two-dimensional theories to study this region have long been sought after \cite{Lipatov:1991nf, Kirschner:1994xk, Fabbrichesi:1993kz}. Our two-dimensional effective theory may well provide for a good starting point. Some microscopic proposals for two-dimensional models may be found in \cite{Banks:2011jt, Banks:2015mya}.

It would also be interesting to explore if an emergent scale can be incorporated into Schwarzian quantum mechanics to capture the sub-leading dynamics (governed by the field $\mathcal{K}$ resulting in the tree-level amplitude \eqref{eqn:Ktree}, for instance) of four-dimensional black holes, in an analysis similar to the one carried out in \cite{Lam:2018pvp}.

The flat space eikonal phase has also been extended to Anti de-Sitter spacetimes \cite{Cornalba:2006xk, Cornalba:2006xm, Cornalba:2007fs,  Cornalba:2007zb, Parnachev:2020zbr}. The Regge-Wheeler gauge is also well understood in AdS \cite{Hui:2020xxx,  Rosen:2020crj}. All results and analyses in the present article can easily be extended to AdS. This should provide a window into studying the information paradox for small black holes in AdS. Moreover, a boundary calculation of the re-summation of the ladder diagrams on the horizon is also of significant interest. An eikonal exponentiation in the context of OTOCs in 2d holographic CFTs and their relevance to extremal BTZ black holes and corresponding microstate geometries was recently discussed in \cite{Balasubramanian:2019stt, Craps:2020ahu}.

Another avenue worth exploring is the relevance of the black hole eikonal phase introduced in this article for general constraints on graviton couplings \cite{Benincasa:2007xk, Camanho:2014apa, Kaplan:2020tdz}.

It is also interesting to explore any observational consequences of the black hole eikonal phase \cite{Betzios:2020}, for the distant observer; this requires an accounting of the classical Regge-Wheeler potential further away from the horizon.

To end, it would be a grave injustice to nature to not refer to another important emergent scale in quantum gravity, namely the size of the cosmological horizon. It shares features with the black hole horizon \cite{Gibbons:1977mu} and appears to have a large number of hitherto unexplained microstates. Although the appropriate asymptotic gauge-invariant states are far more tricky to define in quantum cosmology, the possibility of near-horizon asymptotic states giving rise to a \textit{cosmological eikonal phase} of quantum gravity shall certainly not be lost on us.

\section*{Acknowledgements}
We are grateful to Tarek Anous, Tom Banks, Panos Betzios, Fabiano Feleppa, Anne Franzen, Steve Giddings, Umut G\"{u}rsoy, Gerard 't Hooft, Olga Papadoulaki, Tomislav Prokopec, Suvrat Raju, and Lenny Susskind for various conversations, questions, comments, and criticism that helped us greatly improve this article. This work is supported by the Delta-Institute for Theoretical Physics (D-ITP) that is funded by the Dutch Ministry of Education, Culture and Science (OCW).
\begin{appendix}

\section{Metric fluctuations}
\subsection{Quadratic action}
\label{app:quadraticFluctuations}
We start with the Einstein-Hilbert action:
\begin{equation}
\label{eq:stressenergy}
S_{EH} ~ = ~ \dfrac{1}{2 \kappa^2}\int\ed^4 x\sqrt{- \bar{g}} \bar{R}.
\end{equation}
Metric fluctuations are defined in the background field method about an arbitrary background as $\bar{g}_{\mu\nu} = g_{\mu\nu} + \kappa h_{\mu\nu}$. Assuming that the background $g_{\mu\nu}$ is a vacuum solution of the equations of motion implies that the on-shell action and the variation of it to linear order in $h_{\mu\nu}$ vanish. In the soft limit, therefore, the path integral is dominated by quadratic terms in $h_{\mu\nu}$:
\begin{equation}
\label{eq:EHactionExpansion}
S_{(2)} ~ = ~ \dfrac{1}{2} \int \ed^4 x \int\ed^4 x'\frac{\delta^2 S_{EH}}{\delta \bar{g}^{\mu\nu}(x) \delta \bar{g}^{\rho\sigma}(x')}\biggr\rvert_{\bar{g}=g}h^{\mu\nu}(x)h^{\rho\sigma}(x') \, .
\end{equation}
We now expand $\sqrt{-g}$ and $R$ then
\begin{align}
\sqrt{-\bar{g}} ~ &= ~ \sqrt{-g}\left(1 + \dfrac{1}{2} \kappa h + \mathcal{O}( \kappa^2 h^2)\right),\\
\bar{R} ~ &= ~ R^{(1)} + R^{(2)} + O(h^3)
\end{align}
where we defined $h=g_{\mu\nu}h^{\mu\nu}$ and the subscripts on the Ricci scalar stand for the order in the metric fluctuations. We can then write
\begin{eqnarray}
\sqrt{-\bar{g}} \bar{R} ~ = ~ \sqrt{-g} \left[-\left(h^{\mu\nu} - \dfrac{1}{2} g^{\mu\nu}h\right)R^{(1)}_{\mu\nu} + g^{\mu\nu}R^{(2)}_{\mu\nu}\right] \, ,
\end{eqnarray}
where we have only written the quadratic terms. Therefore, we have
\begin{equation}
\label{eq:Actionvariation1}
S_2 ~ = ~ \dfrac{1}{2} \int\ed^4 x \sqrt{-g} \left[-\left(h^{\mu\nu}-\dfrac{1}{2}g_s^{\mu\nu}h\right) R^{(1)}_{\mu\nu} + g^{\mu\nu} R^{(2)}_{\mu\nu}\right].
\end{equation}

\paragraph{Variation of the Ricci tensor} The Palatini identity gives a variation of the Ricci tensor $\delta \bar{R} = R[g + \kappa h] - R[g]$:
\begin{equation}
\delta R_{\mu\nu} ~ = ~ \nabla_{\rho} \delta\tilde{\Gamma}^{\rho}_{\mu\nu} - \nabla_{\nu} \delta\tilde{\Gamma}^{\rho}_{\rho\mu} \quad \text{with} \quad  \delta\tilde{\Gamma} ~ = ~ \Gamma[g + \kappa h] - \Gamma[g] \, ,
\end{equation}
where the covariant derivatives are associated to the background metric. Since this identity holds to all orders in perturbation theory, we observe that $R_{\mu\nu}^{(2)}$ is in fact a total derivative. Therefore, the term containing $R^{(2)}_{\mu\nu}$ vanishes identically owing to vanishing boundary conditions. We therefore only need to find $\delta \bar{R}_{\mu\nu}$ up to first order since $R_{\mu\nu}=0$ for the background vacuum solution. We write the first order variation as
\begin{equation}
R^{(1)}_{\mu\nu} ~ = ~ \nabla_{\rho} \Gamma^{(1),\rho}_{\mu\nu} - \nabla_{\nu} \Gamma^{(1),\rho}_{\rho\mu}.
\end{equation}
The variation of the Christoffel symbol to first order is written as follows:
\begin{align}
\delta\bar{\Gamma}^{\rho}_{\mu\nu} ~ &= ~ -\dfrac{1}{2} h^{\rho\sigma} \left(\p_{\mu}g_{\sigma\nu} + \p_{\nu} g_{\sigma\mu} - \p_{\sigma} g_{\mu\nu}\right) + \dfrac{1}{2} g^{\rho\sigma} \left(\p_{\mu} h_{\sigma\nu} + \p_{\nu} h_{\sigma\mu} - \p_{\sigma} h_{\mu\nu}\right) \nonumber \\
&= ~ - h^{\rho\alpha} g_{\alpha\beta} \Gamma^{\beta}_{\mu\nu} + \dfrac{1}{2} g^{\rho\sigma} \left(\nabla_{\mu} h_{\sigma\nu} + \nabla_{\nu} h_{\sigma\mu} - \nabla_{\sigma} h_{\mu\nu}\right) \nonumber \\
&\lgap + \dfrac{1}{2} g^{\rho\sigma} \bigr(\cancelto{1}{\Gamma^{\alpha}_{\mu\sigma} h_{\alpha\nu}} + \Gamma^{\alpha}_{\mu\nu} h_{\alpha\sigma} + \cancelto{2}{\Gamma^{\alpha}_{\nu\sigma} h_{\alpha\mu}} + \Gamma^{\alpha}_{\mu\nu} h_{\alpha\sigma} - \cancelto{1}{\Gamma^{\alpha}_{\sigma\mu} h_{\alpha\nu}} - \cancelto{2}{\Gamma^{\alpha}_{\sigma\nu}h_{\alpha\nu}}\bigr) \nonumber \\
&= ~ -\cancelto{3}{h^{\rho\alpha}g^s_{\alpha\beta}\Gamma^{0,\beta}_{\mu\nu}}+g_s^{\rho\sigma}(\nabla^s_{(\mu}h_{\nu)\sigma}-\tfrac{1}{2}\nabla^s_{\sigma}h_{\mu\nu})+\cancelto{3}{g_s^{\rho\sigma}\Gamma^{0,\alpha}_{\mu\nu}h_{\alpha\sigma}}
\end{align}
where the identically numbered terms cancel each other in the last two lines. The Ricci tensor is thus given by
\begin{align}
\label{eq:Riccivariation}
R^{(1)}_{\mu\nu} ~ &= ~ \dfrac{1}{2} g^{\rho\sigma} \bigr(\nabla_{\rho} \nabla_{\mu} h_{\nu\sigma} + \nabla_{\rho} \nabla_{\nu} h_{\mu\sigma} - \nabla_{\rho} \nabla_{\sigma} h_{\mu\nu} - \nabla_{\mu} \nabla_{\nu} h_{\rho\sigma}\bigr) \nonumber \\
&= ~ \dfrac{1}{2} \bigr(2 \nabla^\sigma \nabla_{(\mu} h_{\nu)\sigma} - \Box h_{\mu\nu} - \nabla_{\mu} \nabla_{\nu} h \bigr)\, .
\end{align}
In addition to the kinetic term for the matter fields, we also have the interaction term given by 
\begin{equation}
    S_{\text{int}} ~ = ~ -\int\ed^4 x \dfrac{\delta S_M}{\delta \bar{g}^{\mu\nu}(x)}\biggr\rvert_{\bar{g}=g} h^{\mu\nu}(x) \, .
\end{equation}


\section{Some definitions and calculations}
\subsection{Metric and coordinates}
\label{app:defs}
We work in Kruskal-Szekeres coordinates that are defined as
\begin{eqnarray}
\label{eq:defr}
\up\um&=&2R^2\left(1-\frac{r}{R}\right)e^{\tfrac{r}{R}-1},\\
\up/\um&=&e^{2\tau}.
\end{eqnarray}
Here $\tau=\tfrac{t}{2R}$. We will often write these as a two-vector $x^a$ where $x^x=x,x^y=y$. Here $a=\{x,y\}$ denotes a light-cone index and $A=\{\theta,\phi\}$ a two-sphere index. The quantity $R$ is the Schwarzschild radius
\begin{eqnarray}
R=2GM, \lgap\lgap \mu\equiv \frac{1}{R}
\end{eqnarray}  
where we will work with the effective mass $\mu$ more often. We work in natural units where $\hbar=c=1$. The metric in these coordinates is given by
\begin{eqnarray}
\ed s^2&=&-2A(r)\ed\up\ed\um+r^2\ed\Omega^2,\\
\label{eq:defA}
A(r)&=&\tfrac{R}{r}e^{1-\tfrac{r}{R}}
\end{eqnarray}
where $r=r(x,y)$ is a function of both the light-cone coordinates. Our metric convention is the $(-,+,+,+)$ signature. This metric has the following matrix definitions:
\begin{eqnarray}
\label{eq:defmet}
g_{\mu\nu}=
\begin{pmatrix}
0 & -A & 0 & 0 \\
-A & 0 & 0 & 0 \\
0 & 0 & r^2 & 0 \\
0 & 0 & 0 & r^2\sin^2\theta
\end{pmatrix}
 & 
g^{\mu\nu}=
\begin{pmatrix}
0 & -A^{-1} & 0 & 0 \\
-A^{-1} & 0 & 0 & 0 \\
0 & 0 & r^{-2} & 0 \\
0 & 0 & 0 & r^{-2}\sin^{-2}\theta
\end{pmatrix}
\end{eqnarray}
On $r=R$ we find $A=1$ such that the metric is given by
\begin{eqnarray}
\ed s^2=-2\ed\up\ed\um+R^2\ed\Omega^2, && (r=R).
\end{eqnarray}

\paragraph{Light-cone derivatives:} Here we list the derivatives of $A$ and $r$ with respect to the light-cone coordinates. The derivative on $r$ can be found by implicit differentiation on \eqref{eq:defr} to be
\begin{eqnarray}
\label{eq:rderivs}
\p_a r = \frac{1}{2R}x_a
\end{eqnarray}
where $x_a=g_{ab} x^b$. The derivative of $A(r)$ is then:
\begin{eqnarray}
\label{eq:Aderivs}
\p_{a}A&=&\p_r A\p_{a}r =-\frac{A}{2R}\left(\frac{1}{r}+\frac{1}{R}\right)x_a.
\end{eqnarray}
When evaluated on $r=R$ these functions have the following derivatives on the horizon:
\begin{eqnarray}
\p_{a} A=\p_{a} r=0 &,& \text{  for any $a$ on $r=R$},\\
\p_{\up}\p_{\um}r\biggr\rvert_{r=R}&=&-\frac{1}{2R},\\
\p_{\up}\p_{\um}A\biggr\rvert_{r=R}&=&\frac{1}{R^2}.
\end{eqnarray}

\paragraph{Christoffel symbols:} The non-vanishing Christoffel symbols of the Schwarzschild metric in Kruskal-Szekeres coordinates are given by
\begin{eqnarray}
\Gamma^{x}_{x x}&=&\p_{x}\log A, \\
 \Gamma^{ y}_{ y y}&=&\p_{ y}\log A,\\
 \Gamma^{\theta}_{\theta x}&=& \Gamma^{\theta}_{ x\theta}= \Gamma^{\phi}_{\phi x}= \Gamma^{\phi}_{ x\phi}=\p_{ x}\log r,\\
 \Gamma^{\theta}_{\theta y}&=& \Gamma^{\theta}_{ y\theta}= \Gamma^{\phi}_{\phi y}= \Gamma^{\phi}_{ y\phi}=\p_{ y}\log r,\\
 \Gamma^{\phi}_{\theta\phi}&=&\Gamma^{\phi}_{\phi\theta}=-\sin^{-2}\theta\Gamma^{\theta}_{\phi\phi}=\cot\theta,\\
 \Gamma^{ x}_{\theta\theta}&=&\sin^{-2}\theta\Gamma^{ x}_{\phi\phi}=\frac{1}{2A}\p_{ y}r^2,\\
 \Gamma^{ y}_{\theta\theta}&=&\sin^{-2}\theta\Gamma^{ y}_{\phi\phi}=\frac{1}{2A}\p_{ x}r^2. 
\end{eqnarray}

\paragraph{The Riemann tensor:} We define the Riemann tensor as
\begin{equation}
R^{\rho}_{\gap\mu\sigma\nu} ~ = ~ \p_{\sigma}\Gamma^{\rho}_{\mu\nu}-\p_{\nu}\Gamma^{\rho}_{\mu\sigma}+\Gamma^{\rho}_{\sigma\kappa}\Gamma^{\kappa}_{\mu\nu}-\Gamma^{\rho}_{\mu\kappa}\Gamma^{\kappa}_{\sigma\nu} \, .
\end{equation}
The Ricci tensor is then given by
\begin{eqnarray}
R_{\mu\nu}=R^{\rho}_{\gap\mu\rho\nu}.
\end{eqnarray}
For the Schwarzschild metric the Ricci tensor and scalar vanish identically. Here we list all non-vanishing components of the Riemann tensor:
\begin{align}
R_{xyxy} ~ &= ~ \p_x\p_y\log A \, , \\
R_{x\theta x\theta} ~ &= ~ \dfrac{r\p_x A\p_x r-A\p_x^2 r}{A} \, , \\
R_{x\phi x\phi} ~ &= ~ \sin^2\theta R_{x\theta x\theta} \, , \\
R_{y\theta y\theta} ~ &= ~ \dfrac{r\p_y A\p_y r-A\p_y^2 r}{A} \, , \\
R_{y\phi y\phi} ~ &= ~ \sin^2\theta R_{y\theta y\theta} \, , \\
R_{x\theta y\theta} ~ &= ~ -r\p_x\p_y r \, , \\
R_{x\phi y\phi} ~ &= ~ \sin^2\theta R_{x\theta y\theta} \, , \\
R_{\theta\phi\theta\phi} ~ &= ~ r^2\sin^2\theta\left(1+\dfrac{2\p_x r\p_y r}{A}\right) \, .
\end{align}

\subsection{The antisymmetric Levi-Civita tensor}
Here we define the antisymmetric tensor on $S_2$ by
\begin{eqnarray}
\label{eq:epslower}
\epsilon_{AB}=r^2\sin\theta\begin{pmatrix}
0 & 1 \\
-1 & 0
\end{pmatrix},
\end{eqnarray}
i.e. $\epsilon_{\theta\phi}=r^2\sin\theta=-\epsilon_{\phi\theta}$. Raising and lowering goes with the metric, so for the more common form $\epsilon_A^{\gap\gap B}$ we have
\begin{eqnarray}
\label{eq:epsmix}
\epsilon_A^{\gap\gap B}=\begin{pmatrix}
0 & \sin\theta \\
-\csc\theta & 0  
\end{pmatrix}.
\end{eqnarray}
Lastly for the twice raised form we have
\begin{eqnarray}
\label{eq:epsupper}
\epsilon^{AB}=\frac{1}{r^2\sin\theta}\begin{pmatrix}
0 & 1 \\
-1 & 0
\end{pmatrix}.
\end{eqnarray}
The antisymmetric tensor on the light-cone is defined similarly, although now we work on $r=R$. It is given by
\begin{eqnarray}
\label{eq:eplightcone}
\epsilon^{ab}=\begin{pmatrix}
0 & 1 \\
-1 & 0
\end{pmatrix},\lgap\lgap\lgap \epsilon_{ab}=\begin{pmatrix}
0 & 1 \\
-1 & 0
\end{pmatrix}.
\end{eqnarray}

\subsection{Quadratic operators reduced to two dimensions}
\label{app:Identities}
In this appendix we formulate the relevant quantities in the action \eqref{eqn:partialWaveAction} in terms of the two-dimensional metric $g_{ab}$, its corresponding derivatives, and the residual curvature components arising from the two-sphere. In deriving these identities it is crucial that
\begin{align}
\label{eq:app:FirstIdentity}
\p_A K ~ &= ~ \p_A H^a_b ~ = ~ 0 \, , \\
g_{a A} ~ &= ~ 0 \, , \\
\label{eq:app:ThirdIdentity}
\Gamma^A_{ab} ~ &= ~ \Gamma^{a}_{bA} ~ = ~ 0 \, .
\end{align}
These ensure a clear split between the light-cone and two-sphere contributions. The terms proportional to $\ell(\ell+1)$ in the definition of $\mathcal{G}$ in \eqref{eqn:Inverseprop} are straightforward. Since $h^{a}_b = H^a_b$ and $h^A_B = \delta^A_B K$, the only components of interest are ${{G^A}_{A\rho}}^{\sigma}$ and ${{G^a}_{b\rho}}^{\sigma}$. The two-sphere indices are always contracted. As it turns out, checked by explicit calculation, $\Gamma^A_{BC}$ cancels in the calculation, largely owing to either \eqref{eq:app:FirstIdentity} or the fact that the two-sphere indices are contracted. Furthermore, using \eqref{eq:app:FirstIdentity} and \eqref{eq:app:ThirdIdentity} we find that the only curvature remnant of the two-sphere is neatly embedded in the following vector potential
\begin{equation}
V_a ~ \coloneqq ~ \Gamma^A_{A a} ~ = ~ -g_{ab}g^{AB}\Gamma^b_{AB} \, .
\end{equation}
We now write down all calculated expressions in terms of the light-cone metric and $V_a$. The first relevant expression is
\begin{equation}
{{G^A}_{A\rho}}^{\sigma} h^{\rho}_{\sigma} ~ = ~ 2 \nabla^{\rho} \nabla_A h^A_{\rho} - 2 \nabla^{\sigma} \nabla_{\rho} h^{\rho}_{\sigma} - \nabla_A \nabla^A h + 2 \Box h - \Box h^A_A \, .
\end{equation}
Each of these terms can be expressed in terms of the two-dimensional quantities
\begin{align}
\nabla^{\rho} \nabla_A h^A_{\rho} ~ &= ~ \tilde{\nabla}^a \left(V_b H^b_a\right) + \dfrac{3}{2} V^aV_b H^b_a - \left(\tilde{\nabla}^a V_a\right) K - \dfrac{3}{2} V^a V_a K \, , \\
\nabla^{\sigma} \nabla_{\rho} h^{\rho}_{\sigma} ~ &= ~ \tilde{\nabla}^a \tilde{\nabla}_b H^b_a + V^a \tilde{\nabla}_b H^b_a + \tilde{\nabla}^a \left(V_b H^b_a\right) + V^a V_b H^b_a - \tilde{\nabla}^a \left(V_a K\right) - V_a V^a K \, , \\
\nabla_A\nabla^A h ~ &= ~ V_a \tilde{\nabla}^a H_b^b + 2 V_a \tilde{\nabla}^a K \, , \\
\Box h ~ &= ~ \tilde{\Box} H_a^a + V_b \tilde{\nabla}^b H_a^a + 2 \tilde{\Box} K + 2 V_b \tilde{\nabla}^b K \, , \\
\Box h^A_A ~ &= ~ V_a V^b H_b^a + 2 \tilde{\Box} K + 2 V_a \tilde{\nabla}^a K - V_a V^a K \, .
\end{align}
The other relevant expression is given by
\begin{equation}
{{G^a}_{b\rho}}^{\sigma} h^{\rho}_{\sigma} ~ = ~ \nabla^{\rho} \nabla_b h^a_{\rho} + \nabla_{\rho} \nabla^a h^{\rho}_{b} - \delta^a_b \nabla^{\sigma} \nabla_{\rho} h^{\rho}_{\sigma} - \nabla^a \nabla_b h + \delta^a_b \Box h - \Box h^a_b \, .
\end{equation}
These terms separately give
\begin{align}
\nabla^{\rho} \nabla_b h^a_{\rho} ~ &= ~ \tilde{\nabla}^c \tilde{\nabla}_b H^a_c - \dfrac{1}{2} V^c V_b H^a_c + V^c \tilde{\nabla}_b H^a_c - V^a \tilde{\nabla}_b K + \dfrac{1}{2} V^a V_b K \, , \\
\nabla_{\rho} \nabla^a h^{\rho}_{b} ~ &= ~ \tilde{\nabla}_c \tilde{\nabla}^a H^c_b - \dfrac{1}{2} V_c V^a H^c_b + V_c \tilde{\nabla}^a H^c_b - V^b \tilde{\nabla}_a K + \dfrac{1}{2} V^a V_b K \, , \\
\nabla^{\sigma} \nabla_{\rho} h^{\rho}_{\sigma} ~ &= ~ \tilde{\nabla}^a \tilde{\nabla}_b H^b_a + V^a \tilde{\nabla}_b H^b_a + \tilde{\nabla}^a \left(V_b H^b_a\right) + V^a V_b H^b_a - \tilde{\nabla}^a \left(V_a K\right) - V_a V^a K \, ,\\
\nabla^a \nabla_b h ~ &= ~ \tilde{\nabla}^a \tilde{\nabla}_b H_d^d + 2 \tilde{\nabla}^a \tilde{\nabla}_b K \, , \\
\Box h ~ &= ~ \tilde{\Box} H_b^b + V_a \tilde{\nabla}^a H_b^b + 2 \tilde{\Box} K + 2 V_a \tilde{\nabla}^a K \, , \\
\Box h^a_b ~ &= ~ \tilde{\Box} H^a_b - \dfrac{1}{2} V_c V^a H^c_b - \dfrac{1}{2} V^c V_b H^a_c + V^c \tilde{\nabla}_c H^a_b + V^a V_b K \, .
\end{align}
All these expressions are covariant on the light-cone.

\subsection{Integrating the two-sphere out}
\label{app:evenactionderivation}
In this section we present the details of the derivation of the action \eqref{eq:2Dlagrangian1}:
\begin{equation}
S_{even} ~ = ~ \dfrac{1}{4} \int \ed x^2 \sqrt{-\tilde{g}} \biggr(\tilde{H}^{ab} \tilde{\Delta}^{-1}_{abcd} \tilde{H}^{cd} + \tilde{H}^{ab} \tilde{\Delta}^{-1}_{L,ab} \tilde{K} + \tilde{K} \tilde{\Delta}^{-1}_{R,ab} \tilde{H}^{ab} + \tilde{K} \tilde{\Delta}^{-1}\tilde{K}\biggr) \, .
\end{equation}
Our starting point is
\begin{align}
S_{even} ~ &= ~ -\dfrac{1}{8} \sum\limits_{\ell,m} \int\ed^2 x \, A\left(r\right) r^2 \left(h_{\ell m}\right)^{\mu}_{\nu} {{{\mathcal{G}_\ell}^\nu}_{\mu\rho}}^{\sigma} \left(h_{\ell m}\right)^{\rho}_{\sigma} \, , \\
{{{\mathcal{G}_\ell}^\nu}_{\mu\rho}}^{\sigma} ~ &= ~ {{G^\nu}_{\mu\rho}}^{\sigma} -\dfrac{\ell(\ell+1)}{2r^2} \bigr(\delta^{\mu}_{\nu}\delta_{\rho}^{\sigma} + \delta^{\mu}_a \delta^a_{\nu} \delta^{\sigma}_b \delta^b_{\rho} - 2 \delta^{\mu}_{\rho} \delta^{\sigma}_{\nu} \bigr) \, .
\end{align}
Thus we need to find the proper field transformations and operator redefinitions such that the above actions are identical. As mentioned earlier, the initial step will be to split spacetime into the light-cone $g_{ab}$ and two-sphere $g_{AB}$ pieces. The choice of field transformations will then appear naturally. For what is to come, identities in Appendix \ref{app:Identities} are crucial.

To find all relevant couplings between $H_{ab}$ and $K$ we now first split the degrees of freedom in $h^{\mu}_{\nu}G^{\nu\mgap\sigma}_{\sgap\mu\rho\sgap}h^{\rho}_{\sigma}$. To do so we are only interested in $K \mathcal{G}^{A\mgap\sigma}_{\sgap A\rho\sgap}h^{\rho}_{\sigma}$ and $H^b_a \mathcal{G}^{a\mgap\sigma}_{\sgap b\rho\sgap}h^{\rho}_{\sigma}$. Using the identities listed in Appendix \ref{app:Identities} we find that for all relevant quantities in \eqref{eq:2Dlagrangian1}
\begin{align}
\label{eq:scalarmathG}
\tilde{\mathcal{G}} ~ &= ~ 2 \tilde{\Box} + 2 V_a \tilde{\nabla}^a \, , \\
\label{eq:mixmathGprime}
\tilde{\mathcal{G}}_{R,ab} ~ &= ~ 2 g_{ab} \bigr(\tilde{\Box} + \dfrac{1}{2} V_c \tilde{\nabla}^c - \dfrac{\ell(\ell+1)}{2r^2}\bigr) - 2 \bigr(\tilde{\nabla}_a \tilde{\nabla}_b + V_{(a} \tilde{\nabla}_{b)}\bigr) \, , \\
\label{eq:mixmathG}
\tilde{\mathcal{G}}_{L,ab} ~ &= ~ 2 g_{ab} \bigr(\tilde{\Box} + V_d \tilde{\nabla}^d + \dfrac{1}{2} \tilde{\nabla}^d V_d + \dfrac{1}{2} V_d V^d - \dfrac{\ell(\ell+1)}{2r^2}\bigr) - 2 \bigr(\tilde{\nabla}_a \tilde{\nabla}_b + V_{(a} \tilde{\nabla}_{b)}\bigr) \, , \\
\label{eq:tensormathG}
\tilde{\mathcal{G}}_{abcd} ~ &= ~ g_{ac}\tilde{\nabla}_{d} \tilde{\nabla}_{b} + g_{ac} V_{d} \tilde{\nabla}_{b} + g_{bd} \tilde{\nabla}_{c} \tilde{\nabla}_{a} + g_{bd} V_{c} \tilde{\nabla}_{a} - g_{ab} \left(\tilde{\nabla}_{c} \tilde{\nabla}_{d} + 2 V_{(c} \tilde{\nabla}_{d)} \right. \nonumber \\
&\qquad \left. + V_c V_d + \left(\tilde{\nabla}_{(c} V_{d)}\right)\right) - g_{cd} \tilde{\nabla}_{a} \tilde{\nabla}_{b} + g_{ab} g_{cd} \left(\tilde{\Box} + V^e \tilde{\nabla}_e - \dfrac{\ell(\ell+1)}{r^2}\right) \nonumber \\
&\qquad - g_{ac} g_{bd} \left(\tilde{\Box} + V^e \tilde{\nabla}_e - \dfrac{\ell(\ell+1)}{r^2}\right) \, .
\end{align}
Here we have used that all quantities are now finally covariant on the lightcone; this explains the conventional covariant index form. We also defined the residual curvature tensor of the two-sphere
\begin{equation}
V_a ~ = ~ 2 \p_a \log r \, .
\end{equation}
We can now simplify \eqref{eq:tensormathG} since we are in two dimensions; explicit calculation of all double covariant derivative terms for $\tilde{\mathcal{G}}_{xxcd}H^{cd},\tilde{\mathcal{G}}_{yycd}H^{cd},\tilde{\mathcal{G}}_{xycd}H^{cd}$ shows that we can identify
\begin{align}
-\tilde{R}_{acbd} H^{cd} + g_{ac} \tilde{R}_{bd} H^{cd} ~ &= ~ g_{ac} \left[\tilde{\nabla}_d,\tilde{\nabla}_b\right] H^{cd} \nonumber \\
&= ~ \left(g_{ac} \tilde{\nabla}_{d} \tilde{\nabla}_{b} + g_{bd} \tilde{\nabla}_{c} \tilde{\nabla}_{a} - g_{ab} \tilde{\nabla}_{c} \tilde{\nabla}_{d} \right. \nonumber \\
&\qquad \left.- g_{cd} \tilde{\nabla}_{a} \tilde{\nabla}_{b} + g_{ab}g_{cd} \tilde{\Box} - g_{ac} g_{bd} \tilde{\Box} \right) H^{cd} \, .
\end{align}
The Riemann and Ricci tensors above are on the lightcone. Since the Riemann tensor has only one unique component in two dimensions (namely $\tilde{R}_{xyxy}$), we can now write:
\begin{align}
R_{acbd} ~ &= ~ \dfrac{R_{2d}}{2} \bigr(g_{ab} g_{cd} - g_{ad} g_{bc}\bigr) \, , \\
R_{bd} ~ &= ~ \dfrac{R_{2d}}{2} g_{bd} \, , \\
R_{2d} ~ &= ~ \dfrac{2}{A} \p_x \p_y \log A \, .
\end{align}
The covariant derivatives above, therefore, only contribute the following:
\begin{equation}
\dfrac{R_{2d}}{2} \bigr(g_{ad} g_{bc} - g_{ab} g_{cd} + g_{ac} g_{bd} \bigr) H^{cd} ~ = ~ \dfrac{R_{2d}}{2} \bigr(2 g_{ac} g_{bd} - g_{ab} g_{cd}\bigr) H^{cd} \, .
\end{equation}
A similar procedure for all the first derivatives shows that
\begin{align}
&\biggr(g_{ac} V_{d} \tilde{\nabla}_{b} + g_{bd} V_{c} \tilde{\nabla}_{a} - 2 g_{ab} V_{(c} \tilde{\nabla}_{d)} + g_{ab} g_{cd} V^e \tilde{\nabla}_e - g_{ac} g_{bd} V^e \tilde{\nabla}_e \biggr) H^{cd} ~ = ~ \qquad\qquad\nonumber \\
&\qquad\qquad\qquad\qquad\qquad= ~ \bigr( g_{ac} V_{[d} \tilde{\nabla}_{b]} + g_{bd} V_{[c} \tilde{\nabla}_{a]} - g_{ab} V_{(c} \tilde{\nabla}_{d)} + g_{cd} V_{(a} \tilde{\nabla}_{b)}\bigr) H^{cd} \, .
\end{align}
Finally, we can define the operator as
\begin{align}
\label{eq:tensormathGsimplified}
\tilde{\mathcal{G}}_{abcd} ~ &= ~ g_{ac} V_{[d} \tilde{\nabla}_{b]} + g_{bd} V_{[c} \tilde{\nabla}_{a]} - g_{ab} \bigr(V_{(c} \tilde{\nabla}_{d)} + V_c V_d + \left(\tilde{\nabla}_{(c} V_{d)}\right) \bigr) + g_{cd} V_{(a} \tilde{\nabla}_{b)} \nonumber \\
&\qquad- ~ g_{ab} g_{cd} \left(\dfrac{R_{2d}}{2} + \dfrac{\ell(\ell+1)}{r^2}\right) + g_{ac} g_{bd} \left(R_{2d} + \dfrac{\ell(\ell+1)}{r^2}\right) \, .
\end{align}
In what follows, we will work with this simplified form. The only fact we used to derive all these simplifications is that $g_{xx} = g_{yy}=0$, i.e. the form of the metric. Using these operators the Lagrangian is written as
\begin{equation}
\label{eq:2Dlagrangianapp}
\mathcal{L}_{even} ~ = ~ - \dfrac{r^2}{8} H^{ab} \tilde{\mathcal{G}}_{abcd} H^{cd} - \dfrac{r^2}{8} H^{ab} \tilde{\mathcal{G}}_{L,ab} K - \dfrac{r^2}{8} K \tilde{\mathcal{G}}_{R,ab} H^{ab} - \dfrac{r^2}{8} K \tilde{\mathcal{G}} K \, .
\end{equation}
Now there is a disconcerting presence of $r^2$ in this Lagrangian. Since the metric is now $g_{ab}$, not all covariant derivatives commute with $r^2$. This means that the Lagrangian is not properly symmetric between the degrees of freedom. To resolve this, we absorb the residual $r^2$ into the fields, so we redefine $\tilde{H}_{ab}=r H_{ab},\tilde{K}=r K$.

\subsubsection*{The light-cone fields}
We will now write the Lagrangian in terms of the redefined light-cone fields $\tilde{H}=r H$ and $\tilde{K}=rK$. To do so, we first inspect the quantity $V_a=2\p_a\log r$, from which we can define a derivative
\begin{equation}
\label{eq:pullrthrough}
D_a ~ \coloneqq ~ \tilde{\nabla}_a + \dfrac{1}{2} V_a ~ = \dfrac{1}{r} \tilde{\nabla}_a r \, .
\end{equation}
By replacing every covariant derivative $\tilde{\nabla}_a$ in \eqref{eq:scalarmathG}-\eqref{eq:tensormathG} by $D_a$ (using $\tilde{\nabla}_a=D_a-\tfrac{1}{2}V_a$), we can use \eqref{eq:pullrthrough} to remove one $r$ from the $r^2$ on the left hand side in \eqref{eq:2Dlagrangianapp} and introduce an $r$ on the right hand side. This automatically gives the symmetric form in $r$, from where we can redefine the fields appropriately. This gives the final effective two-dimensional theory in \eqref{eq:decoupledaction1}.

The results with the new derivatives \eqref{eq:pullrthrough} are
\begin{align}
\label{eq:scalarmathG2}
\tilde{\mathcal{G}} ~ &= ~ 2 D^2 - 2 F^a_a \, , \\
\label{eq:mixmathGprime2}
\tilde{\mathcal{G}}_{R,ab} ~ &= ~ 2 g_{ab} \left(D^2 - \dfrac{1}{2} V_c D^c + \dfrac{1}{4} V_c V^c - F_c^c - \dfrac{\ell(\ell+1)}{2r^2}\right) - 2 \left(D_a D_b - F_{ab}\right) \, , \\
\label{eq:mixmathG2}
\tilde{\mathcal{G}}_{L,ab} ~ &= ~ 2 g_{ab} \left(D^2 + \dfrac{1}{2} D^c V_c + \dfrac{1}{4} V_c V^c - F_c^c - \dfrac{\ell(\ell+1)}{2r^2}\right) - 2 \left(D_a D_b - F_{ab}\right) \, , \\
\label{eq:tensormathG2}
\tilde{\mathcal{G}}_{abcd} ~ &= ~ g_{ac} V_{[d} D_{b]} + g_{bd} V_{[c}D_{a]} - g_{ab} \left(D_{(c}V_{d)} + \dfrac{1}{2} V_c V_d\right) - g_{cd} \left(-V_{(a} D_{b)} + \dfrac{1}{2} V_a V_b\right) \nonumber \\
&\qquad - g_{ab} g_{cd} \left(\dfrac{R_{2d}}{2} + \dfrac{\ell(\ell+1)}{r^2}\right) + g_{ac} g_{bd} \left(R_{2d} + \dfrac{\ell(\ell+1)}{r^2}\right) \, .
\end{align}
Here we defined $D^2=D_aD^a$ and the new quantity
\begin{equation}
F_{ab} ~ \coloneqq ~ \dfrac{1}{2} D_{(a}V_{b)} ~ = ~ \dfrac{1}{r} \tilde{\nabla}_a \tilde{\nabla}_b r\, ,
\end{equation}
which is a symmetric tensor (not an operator). Notice that the fact that $F_{ab}$ is symmetric is a natural result from commutativity of covariant derivatives when acting on a scalar. Now, because $D_a=\tfrac{1}{r}\tilde{\nabla}_a r$, we also have that $D^2=\tfrac{1}{r}\tilde{\Box} r$. In this sense we pull a single $r$ from the left side to the right side: for the scalar field, for instance, we have
\begin{equation}
K r^2 \tilde{\mathcal{G}} K ~ = ~ 2 K r^2 \left(D^2 - F^a_a\right) K ~ \xrightarrow{D\to \tfrac{1}{r}\tilde{\nabla} r} ~ 2 K r \left(\tilde{\Box} - F^a_a\right) r K \, .
\end{equation}
We may now consistently make the field redefinitions $\tilde{H}_{ab}=r H_{ab}$, $\tilde{K}=rK$, to we end up with $2\tilde{K}(\tilde{\Box}-F^a_a)\tilde{K}$. Carrying the same procedure for all other fields, i.e. inserting $D=\tfrac{1}{r}\tilde{\nabla} r$ and the redefined fields, results in the following Lagrangian
\begin{equation}
\label{eq:2Dlagrangian4}
\mathcal{L}_{even} ~ = ~ \dfrac{1}{4} \tilde{H}^{ab} \tilde{\Delta}^{-1}_{abcd} \tilde{H}^{cd} + \dfrac{1}{4} \tilde{H}^{ab} \tilde{\Delta}^{-1}_{L,ab} \tilde{K} + \dfrac{1}{4} \tilde{K} \tilde{\Delta}^{-1}_{R,ab} \tilde{H}^{ab} + \dfrac{1}{4} \tilde{K} \tilde{\Delta}^{-1} \tilde{K} \, .
\end{equation}
Here we defined
\begin{align}
\label{eq:scalarmathD}
\tilde{\Delta}^{-1} ~ &= ~ - \tilde{\Box} + F^a_a \, , \\
\label{eq:mixmathDright}
\tilde{\Delta}^{-1}_{R,ab} ~ &= ~ - g_{ab} \left(\tilde{\Box} - \dfrac{1}{2} V_c \tilde{\nabla}^c + \dfrac{1}{4} V_c V^c - F_c^c - \dfrac{\ell(\ell+1)}{2r^2}\right) + \tilde{\nabla}_a \tilde{\nabla}_b - F_{ab} \, , \\
\label{eq:mixmathDleft}
\tilde{\Delta}^{-1}_{L,ab} ~ &= ~ - g_{ab} \left(\tilde{\Box} + \dfrac{1}{2} V_c \tilde{\nabla}^c - \dfrac{\ell(\ell+1)}{2r^2}\right) + \tilde{\nabla}_a \tilde{\nabla}_b - F_{ab} \, , \\
\label{eq:tensormathD}
\tilde{\Delta}^{-1}_{abcd} ~ &= ~ \dfrac{1}{2} g_{ac} V_{[b} \tilde{\nabla}_{d]} + \dfrac{1}{2} g_{bd} V_{[a} \tilde{\nabla}_{c]} + \dfrac{1}{2} g_{ab} \left( V_{(c} \tilde{\nabla}_{d)} + 2 F_{cd} \right) + \dfrac{1}{2} g_{cd} \left(-V_{(a} \tilde{\nabla}_{b)} + \dfrac{1}{2} V_a V_b\right) \nonumber \\
&\qquad + g_{ab} g_{cd} \left(\dfrac{1}{4} R_{2d} + \dfrac{\ell(\ell+1)}{2r^2}\right) - g_{ac} g_{bd} \left(\dfrac{1}{2} R_{2d} + \dfrac{\ell(\ell+1)}{2r^2}\right) \, .
\end{align}
A minus sign has been absorbed into all $\tilde{\Delta}$ operators for convenience. We note that all operators are in fact symmetric in the fields, since for example $\tilde{\Delta}^{-1}_{R,ab}$ equals $\tilde{\Delta}^{-1}_{L,ab}$ up to total derivatives. This completes the rewriting of the action; we have now arrived at the Lagrangian in \eqref{eq:2Dlagrangian1}.

\section{Shockwaves and on-shell perturbations}
\label{app:onshellPert}
The question of interest in this Appendix is to derive the equation of motion for the metric perturbation that results in the Dray-'t Hooft shockwave, first derived in \cite{Dray:1984ha}. Given that the source involves a delta function in the stress tensor, the perturbation is naively of the form $h_{xx}= 2 A\left(x,y\right) \delta(x) F$, where we seek to derive the equation of motion for the function $F$, such that the perturbed metric satisfies the Einstein's equations. This calculation in the harmonic gauge of course fails because the solution is not invariant under gauge transformations. So, we will instead work with the general equations of motion before gauge fixing. We begin with the Einstein's equations that are of course given by
\begin{equation}
    R_{\mu\nu} - \dfrac{1}{2} g_{\mu\nu} R ~ = ~ 8 \pi G_N T_{\mu\nu} \, ,
\end{equation}
and the variation of the Ricci tensor is derived in \eqref{eq:Riccivariation}:
\begin{eqnarray}
R_{\mu\nu} ~ = ~ \dfrac{1}{2} g^{\lambda\rho} \bigr(\nabla_{\lambda} \nabla_{\mu} h_{\rho\nu} + \nabla_{\lambda} \nabla_{\nu} h_{\rho\nu} - \nabla_{\lambda} \nabla_{\rho} h_{\mu\nu} - \nabla_{\mu} \nabla_{\nu} h_{\lambda\rho}\bigr) \, .
\end{eqnarray}
We want to find the $F$ function using the following (Schwarzschild form of the) metric:
\begin{eqnarray}
g_{\mu\nu} ~ = ~ -2 A(x,y) \ed x \ed y + r^2(x,y) \ed\Omega^2 \, ,
\end{eqnarray}
such that it still satisfies the non-linear Einstein's equations. Here, $A\left(x,y\right)$, $r\left(x,y\right)$, and the inverse metric are defined in \eqref{eq:defr}, \eqref{eq:defA}, and \eqref{eq:defmet} respectively. From the form of the source that introduces the shockwave, we may write $h_{\mu\nu}$ containing the only non-vanishing component 
\begin{eqnarray}
h_{xx} ~ = ~ 2 A(x,y) \delta(x) F\left(\Omega\right) \, .
\end{eqnarray}
Explicitly calculating all covariant derivatives in the Ricci tensor results in
\begin{align}
R_{xx} ~ &= ~ \dfrac{A\delta(x)}{r^2} \Delta_{\Omega} F + \dfrac{2 F \delta'(x)}{r} \p_y r \nonumber \\
&\qquad + \dfrac{2 F \delta(x)}{A} \left(\dfrac{\p_x A \p_y A}{A} - \p_x \p_y A - \dfrac{1}{r} \left(\p_x A \p_y r + \p_y A \p_x r\right)\right) \, .
\end{align}
Here $\Delta_{\Omega}$ is the Laplacian on the two-sphere. We now interpret the dirac delta derivative as $\delta'(x)f(x) = - \delta(x) f'(x)$. Moreover, it suffices to evaluate the equation of motion at $x=0$ owing to the $\delta(x)$; therefore, we may set $A=1$ and remove the $\p_y A,\p_y r$ terms. Finally, using the derivatives in Appendix \ref{app:defs}, the equation of motion now becomes
\begin{eqnarray}
\dfrac{\delta(x)}{R^2} \Delta_{\Omega} F - 2 \delta(x) F \left(\p_x \p_y A + \dfrac{\p_x \p_y r}{r}\right) ~ = ~ 8 \pi G_N T_{xx} \, .
\end{eqnarray}
Using the double derivative identities in Appendix \ref{app:defs} given by
\begin{eqnarray}
\p_y \p_x A \bigr\rvert_{r=R} ~ = ~ \dfrac{1}{R^2} \quad \text{and} \quad \p_y \p_x r \bigr\rvert_{r=R} ~ = ~ - \dfrac{1}{2R} \, ,
\end{eqnarray}
we therefore find
\begin{eqnarray}
\dfrac{\delta(x)}{R^2} \left(\Delta_{\Omega} - 1\right) F ~ = ~ 8 \pi G_N T_{xx} \, .
\end{eqnarray}
This shows that the Shapiro delay can be treated in linear gravity.
It is worth mentioning that all of the above was done in the discontinuous $y+\Theta(x)F$ coordinate, while the energy-momentum tensor is given in the continuous $y$ coordinate. However a quick coordinate transformation shows that on $r=R$ this has no effect on the result.

\end{appendix}

\printbibliography
\end{document}